\documentclass[11pt]{article}

\usepackage[preprint]{acl}
\usepackage{times}
\usepackage{latexsym}

\usepackage[T1]{fontenc}

\usepackage[utf8]{inputenc}

\usepackage{microtype}
\usepackage{inconsolata}

\usepackage{graphicx}

\usepackage{booktabs}
\usepackage{tabularx}
\usepackage{array}
\usepackage{multirow}
\usepackage[table]{xcolor}
\usepackage{amsmath}
\usepackage{amssymb}
\usepackage{pgfplots}
\usepackage{rotating}
\usepackage{longtable}
\usepackage{float}
\pgfplotsset{compat=1.18}

\newcolumntype{S}{>{\hsize=.88\hsize\centering\arraybackslash}X}
\newcolumntype{M}{>{\hsize=.93\hsize\centering\arraybackslash}X}
\newcolumntype{L}{>{\hsize=1.08\hsize\centering\arraybackslash}X}
\newcolumntype{R}{>{\hsize=1.14\hsize\centering\arraybackslash}X}

\usepackage[most]{tcolorbox}
\usepackage{pifont}

\newtcolorbox{insight}{
  colback=black!4, colframe=black!45, boxrule=0.6pt, arc=3pt,
  left=6pt, right=6pt, top=4pt, bottom=4pt,
  fontupper=\small
}

\usepackage{enumitem}
\usepackage{listings}
\lstdefinestyle{sayf}{
  basicstyle=\ttfamily\scriptsize,
  breaklines=true,
  breakatwhitespace=false,
  columns=fullflexible,
  keepspaces=true,
  showstringspaces=false,
  captionpos=b,
  aboveskip=4pt, belowskip=2pt,
  xleftmargin=2pt, xrightmargin=2pt,
  keywordstyle=\bfseries,
}
\newcommand{\brk}{\discretionary{}{}{}}
\newcounter{casebox}
\newcommand{\mlcase}[7]{%
  \refstepcounter{casebox}%
  \label{#1}%
  \begin{tcolorbox}[
    colback=housegreen!4,
    colframe=housegreen!85!black,
    boxrule=0.6pt,
    arc=3pt,
    left=6pt,
    right=6pt,
    top=4pt,
    bottom=4pt,
    title={\textsc{case}~\thecasebox\,\textbf{: }\,\textbf{#2}},
    fonttitle=\footnotesize\bfseries,
    coltitle=white,
    colbacktitle=housegreen!85!black
  ]
    \footnotesize
    \noindent\textbf{Prompt (truncated):} \emph{``#3''}\\[3pt]
    \noindent\textbf{Gold:} \texttt{#4}\\[1pt]
    \noindent\textbf{Model majority:} \texttt{#5}\\[3pt]
    \noindent\textbf{Verifier:} #6\\[3pt]
    \noindent\textbf{Sources:}
    \begin{itemize}[leftmargin=1.4em,topsep=2pt,parsep=0pt,itemsep=1pt]
      #7
    \end{itemize}
  \end{tcolorbox}%
}

\newcounter{examplebox}

\makeatletter
\newcommand{\evbox}[6]{%
  \refstepcounter{examplebox}%
  \protected@edef\@currentlabelname{Example \theexamplebox}%
  \label{#1}%
  \begin{tcolorbox}[
    colback=houseblue!4,
    colframe=houseblue!90!black,
    boxrule=0.6pt,
    arc=3pt,
    left=6pt,
    right=6pt,
    top=4pt,
    bottom=4pt,
    title={\textsc{example}~\theexamplebox\,\textbf{: }\,\textbf{#2}},
    fonttitle=\footnotesize\bfseries,
    coltitle=white,
    colbacktitle=houseblue!90!black
  ]
    \footnotesize
    \noindent\textbf{Prompt:} #3\\[2pt]
    \noindent\textbf{Raw output:} #4\\[2pt]
    \noindent\textbf{Extractor:} #5\\[2pt]
    \noindent\textbf{Consequence:} #6
  \end{tcolorbox}%
}
\makeatother

\newtcolorbox[auto counter]{pipelinecard}[2][]{%
  colback=houseteal!5,
  colframe=houseteal!85!black,
  boxrule=0.6pt,
  arc=3pt,
  left=6pt,
  right=6pt,
  top=4pt,
  bottom=4pt,
  title={\textsc{eval card}~\thetcbcounter\,\textbf{: }\,\textbf{#2}},
  fonttitle=\footnotesize\bfseries,
  coltitle=white,
  colbacktitle=houseteal!85!black,
  #1
}

\newtcolorbox[auto counter]{promptbox}[2][]{%
  colback=houseorange!5, colframe=houseorange!80!black, boxrule=0.6pt, arc=3pt,
  left=6pt, right=6pt, top=4pt, bottom=4pt,
  title={\textsc{prompt}~\thetcbcounter\,\textbf{: }\,\textbf{#2}},
  fonttitle=\footnotesize\bfseries,
  coltitle=white, colbacktitle=houseorange!80!black, #1}

\newtcolorbox[auto counter]{codebox}[2][]{%
  colback=housepurple!5, colframe=housepurple!85!black, boxrule=0.6pt, arc=3pt,
  left=6pt, right=6pt, top=4pt, bottom=4pt,
  title={\textsc{code}~\thetcbcounter\,\textbf{: }\,\textbf{#2}},
  fonttitle=\footnotesize\bfseries,
  coltitle=white, colbacktitle=housepurple!85!black, #1}

\title{
Benchmark Scores Are Pipeline-Dependent:\\
A Reliability Audit of Cybersecurity LLM Benchmarks
}

\author{
  \textbf{Aymene Berriche},
  \textbf{Cathrine Shalby},
  \textbf{Mohannad Alhanahnah},
  \textbf{Yazan Boshmaf}
  \\
  Qatar Computing Research Institute, HBKU
}

\definecolor{houseblue}{HTML}{1A80BB}
\definecolor{houseorange}{HTML}{EA801C}
\definecolor{housegreen}{HTML}{1F7A5C}
\definecolor{housepurple}{HTML}{6A4C93}
\definecolor{houseteal}{HTML}{2A7F8E}

\newcommand{\promptref}[1]{Prompt~\ref{#1}}
\newcommand{\coderef}[1]{Code~\ref{#1}}

\newcommand{\exampleref}[1]{Example~\ref{#1}}
\newcommand{\pipelinecardref}[1]{Eval Card~\ref{#1}}
\newcommand{\pipelinecardrefwhite}[1]{%
  Eval Card~\hyperref[#1]{\textcolor{white}{\ref*{#1}}}%
}
\newcommand{\examplerange}[2]{Examples~\ref{#1}--\ref{#2}}
\newcommand{\caserange}[2]{Cases~\ref{#1}--\ref{#2}}

\begin{document}
\maketitle

\begin{abstract}
Large language model (LLM) benchmarks are often treated as fixed datasets with stable scores, yet their outcomes depend on configurable evaluation pipelines. We audit eight cybersecurity benchmarks across 10 proprietary, open-weight, and cybersecurity-specialized LLMs. By modeling benchmarks as measurement pipelines, we identify 15 systematic failure modes and show that a single pipeline choice can change a model's score by more than 80 percentage points and substantially alter model rankings. At the cross-benchmark level, two semantically similar task pairs rank the same models differently because of incompatible evaluation conventions. Under an evaluation harness that standardizes pipeline choices while preserving task semantics, nine of 10 models shift by at least three ranks on at least one benchmark. These results show that cybersecurity LLM benchmark scores are pipeline-dependent and motivate pipeline-aware auditing as a core requirement for reliable model evaluation.
\end{abstract} 

\section{Introduction}
\label{sec:intro}
Large language model (LLM) evaluation is increasingly benchmark-driven. Benchmark scores guide model selection, support claims of state-of-the-art performance, influence deployment decisions, and shape leaderboards. Yet, these scores are often interpreted as stable measurements of model capability, even though they are produced by configurable evaluation scripts involving prompts, inference settings, output extraction, scoring, and aggregation.

Prior work shows that LLM evaluation is sensitive to prompt wording, decoding, evaluator design, answer extraction, and scoring~\citep{shi2024decoding,sun2024evaluating,wang2023fair}. These effects are amplified in generative settings, where outputs are open-ended, multiple answers may be valid, and evaluation often relies on heuristic extraction or approximate matching. Consequently, benchmark outcomes may reflect the measurement process as much as underlying model capability.

We study this problem in cybersecurity, a high-stakes domain whose benchmarks span factual recall, vulnerability analysis, threat-intelligence extraction, attacker attribution, attack-technique mapping, mitigation selection, and general security reasoning. These tasks depend on evolving, structured authoritative sources such as CVE, CWE, CVSS, and MITRE ATT\&CK~\citep{sikos2023cybersecurity}, making evaluation especially sensitive to pipeline errors and inconsistent labels. Such failures can distort claims of model specialization, obscure genuine capability differences, and mislead deployment decisions. Cybersecurity also provides a mature and heterogeneous benchmark ecosystem that has not been systematically audited, including in recent large-scale meta-evaluations~\citep{bean2025measuring}.

In this paper, we ask: \emph{to what extent do design choices of evaluation pipelines affect the reliability of cybersecurity LLM benchmark outcomes?} We argue that benchmarks should be viewed not as static datasets paired with fixed metrics, but as \emph{measurement pipelines} that transform tasks and model outputs into numerical performance estimates (\S\ref{sec:pipeline}). Under this formulation, a benchmark score is conditional on its evaluation pipeline's stages, consisting of dataset construction, prompt specification, inference, extraction and scoring, and aggregation.

We audit eight cybersecurity benchmarks comprising 48,662 questions across 23 tasks against 10 proprietary, open-weight, and cybersecurity-specialized LLMs (\S\ref{sec:setup}) and identify 15 recurring failure modes across the evaluation pipeline (\S\ref{sec:pipeline_failures}). Fixing these failures can shift a model's benchmark score by over 80 percentage points. For example, RedSage-Bench~\citep{suryanto2026redsage} uses ``\texttt{\textbackslash n}'' as a stop sequence, causing generation to halt at the first newline. For Qwen3.6, this sequence fires inside the reasoning preamble before any answer token is produced, yielding empty outputs. Generating until the end-of-sequence token, using a token budget large enough for the reasoning span to close, and stripping the reasoning span before extraction, recovers 85.9 percentage points.

Across benchmarks, broader task coverage can still yield redundant measurements and unstable model rankings (\S\ref{sec:ecosystem}). Using Principal Component Analysis (PCA), we find that the first component explains 95.25\% of the variance across the 23 task $\times$ 10 model score matrix, showing that most tasks largely capture the same broad performance dimension. However, task-induced rankings can still disagree. Two semantically similar task pairs in CTI-Bench~\citep{alam2024ctibench} and AthenaBench~\citep{alam2025athenabench} show weak rank agreement under Kendall's $\tau$-b ($0.29$ and $0.24$), with pairwise inversions revealing changes in specific model orderings. These discrepancies stem largely from incompatible scoring conventions, specifically binary versus partial-credit scoring and alias handling.

To isolate pipeline effects, we implement a harness that standardizes nine pipeline configuration fields, including prompt formatting, decoding, and extraction rules, wherever benchmark semantics permit. Under this harness, nine of 10 models shift by at least three ranks on at least one benchmark (\S\ref{sec:ecosystem}). This has an important practical implication: cybersecurity-specialized models are often selected based on leaderboard rankings, yet those rankings can reflect evaluation-pipeline choices as much as underlying domain capability. We use these findings to formulate a set of recommendations for more reliable benchmarking (\S\ref{sec:recommendations}).

Our contributions are threefold. First, we formalize LLM benchmarks as measurement pipelines and audit every pipeline stage of eight cybersecurity benchmarks across 10 LLMs under a common framework. Second, we quantify stage-level effects through controlled perturbations of pipeline configuration fields, showing that pipeline choices affect both model scores and rankings. Third, we release an evaluation harness that makes heterogeneous benchmark assumptions explicit and reproducible. Although some individual failure modes identified in our audit have been documented before, our contribution is a systematic end-to-end audit of cybersecurity benchmarks under a single formalization. Our findings motivate treating benchmark scores as pipeline-dependent measurements and adopting benchmark audits, executable reference evaluators, invalid-response reporting, and reliability statistics as standard evaluation practice.

We note that our harness is intended as a diagnostic tool for exposing and correcting evaluation-pipeline choices, rather than as a universally correct evaluator. While our audit methodology is domain-general, the observed failure rates are specific to the cybersecurity benchmarks studied. Our goal is to identify these failures and quantify their impact, rather than to propose failure-free benchmarks or a universally applicable evaluation harness. We publicly release our audit code\footnote{\url{https://github.com/qcri/cyberbench-audit}} and evaluation harness,\footnote{\url{https://github.com/qcri/sayf-eval}} and publish our meta-evaluation results on EveryEvalEver~\citep{batzner2026every}.

\section{Benchmarks as Measurement Pipelines}
\label{sec:pipeline}

LLM benchmarks are often described as datasets paired with metrics, where a metric specifies how model performance is quantified and a score is the resulting numerical measurement. In practice, that score is produced by a multi-stage evaluation procedure in which each stage transforms the output of the preceding stage and introduces choices that can affect the final measurement. We therefore model a benchmark as a \emph{measurement pipeline}, represented as a nested composition of stage-specific functions. For benchmark $b$ and model $m$, the reported score $\mathcal{S}_b(m)$ can be written as:
\begin{equation}
    \mathcal{S}_b(m) = \mathcal{A}_b\!\left(
    \mathcal{E}_b\!\left(
    \mathcal{I}_b\!\left(
    m,\mathcal{P}_b(\mathcal{D}_b)
    \right)\right)\right),
    \label{eq:pipeline}
\end{equation}
where $\mathcal{D}_b$ is the benchmark dataset, $\mathcal{P}_b$ the prompt specification, $\mathcal{I}_b$ the inference procedure, $\mathcal{E}_b$ the extraction and scoring procedure, and $\mathcal{A}_b$ the aggregation rule. Under this formulation, the benchmark score is the output of the composed evaluation pipeline rather than an intrinsic property of the model alone. As a result, $\mathcal{S}_b(m)$ is conditional on the full pipeline used to produce it.

\paragraph{Pipeline stages.}
\textit{Dataset construction ($\mathcal{D}_b$)} defines the evaluation distribution, including task coverage, label quality, and answer representation. \textit{Prompt specification ($\mathcal{P}_b$)} determines how tasks are presented to the model, including instructions, demonstrations, chat formatting, and output-format constraints. \textit{Inference configuration ($\mathcal{I}_b$)} controls how responses are generated, including decoding parameters, token budgets, stop sequences, serving backends, and backend-specific constraints. \textit{Extraction and scoring ($\mathcal{E}_b$)} converts model outputs into predictions or graded judgments. \textit{Aggregation ($\mathcal{A}_b$)} combines question- and task-level measurements into final benchmark scores. Variation at any stage can therefore change reported performance without reflecting a change in model capability.

The five stages in Eq.~\ref{eq:pipeline} are instantiated through nine \emph{pipeline configuration fields}, which we specify, perturb, and standardize throughout the paper: prompt template and chat formatting ($\mathcal{P}$); decoding, maximum new tokens, and stop sequences ($\mathcal{I}$); extraction rule, scoring rule, and denominator policy ($\mathcal{E}$); and aggregation rule ($\mathcal{A}$). App.~\ref{app:pipeline-table} records these fields for every benchmark.

\subsection{Meta-Evaluation Methodology}
\label{sec:methodology}

The pipeline view separates benchmark reliability into two levels. At the \emph{benchmark level}, we inspect each pipeline for stages where a design or implementation choice changes the reported score. We call such a pattern a \emph{failure mode}: a recurring feature of the evaluation procedure that can change reported scores without any corresponding change in model capability. Examples include prompt templates that induce unparseable outputs and aggregation rules that combine non-equivalent metrics. At the \emph{cross-benchmark level}, we ask whether the audited benchmarks support stable conclusions about which model performs better.

We denote the $i$-th failure mode at pipeline stage $\mathcal{X}_b \in \{\mathcal{D}_b,\mathcal{P}_b,\mathcal{I}_b,\mathcal{E}_b,\mathcal{A}_b\}$ for benchmark $b$ and model $m$ as $\mathcal{F}_i(m,\mathcal{X}_b)$. When discussing failure modes independent of a particular benchmark or model, we use the shorthand $\mathcal{F}_i(\mathcal{X})$ to denote the $i$-th failure mode at stage $\mathcal{X}$.

We audit every stage of each benchmark, so no stage--benchmark pair is left unexamined. For each benchmark, we separately reconstruct the documented and released pipelines, record where they disagree, and identify any choices we must supply because neither source specifies them. We log raw model outputs together with the exact prompt, decoding parameters, and extraction rules used to produce them. Where a single pipeline configuration field can be varied while holding the others fixed, we perturb that field and measure the resulting change. This perturbation step is necessarily opportunistic, and App.~\ref{app:b1} records where such isolation is not possible. Because not all failure modes admit a meaningful per-model score shift, we quantify each using a measure appropriate to its mechanism, including score deltas, invalid-response rates, extractor disagreement, denominator ratios, affected-question or benchmark shares, and rank changes under alternative but semantically equivalent implementations (App.~\ref{app:b7}).

Each failure-mode analysis is either a \emph{re-scoring} analysis, in which we reuse stored model outputs while changing a downstream pipeline configuration field (e.g., the extraction rule), or a \emph{re-generation} analysis, in which we produce new model outputs after changing an upstream pipeline configuration field (e.g., the prompt template). Re-scoring holds model outputs fixed and therefore isolates the effect of the modified downstream field exactly. Re-generation, by contrast, captures the effect of changing an upstream field through the new outputs it induces. We classify each failure mode as re-scoring or re-generation in App.~\ref{app:b2}.

At the cross-benchmark level, we use PCA to characterize redundancy in task-level scores and Kendall's $\tau$-b, together with pairwise rank inversions, to measure agreement among task-induced model rankings. Here, a pairwise inversion occurs when two models are ordered one way by one task and in the opposite order by another (App.~\ref{app:j3}). While benchmark-level analysis identifies instability within individual benchmarks, cross-benchmark analysis examines whether different tasks provide distinct evidence and support consistent comparative conclusions.

\begin{table}[t]
\centering
\scriptsize
\renewcommand{\arraystretch}{1.12}
\begin{tabular}{lrrl}
\toprule
\textbf{Benchmark} & \textbf{Tasks} & \textbf{Questions} & \textbf{Reference} \\
\midrule
MMLU-CS
    & 1 & 100
    & \citep{hendrycks2020measuring} \\
CyberMetric
    & 1 & 500
    & \citep{tihanyi2024cybermetric} \\
SecBench
    & 1 & 661
    & \citep{jing2024secbench} \\
SecEval
    & 1 & 2{,}189
    & \citep{li2023seceval} \\
SECURE
    & 3 & 2{,}502
    & \citep{bhusal2024secure} \\
CTI-Bench
    & 5 & 4{,}610
    & \citep{alam2024ctibench} \\
AthenaBench
    & 6 & 8{,}100
    & \citep{alam2025athenabench} \\
RedSage-Bench
    & 5 & 30{,}000
    & \citep{suryanto2026redsage} \\
\midrule
\textbf{Total}
    & \textbf{23}
    & \textbf{48{,}662}
    & \\
\bottomrule
\end{tabular}
\caption{Cybersecurity benchmarks included in our audit. \emph{Tasks} reports the number of scored tasks, while \emph{Questions} reports the total number of scored questions across those tasks. The counts may differ from the reported or released dataset sizes; see App.~\ref{app:c1} for details.}
\label{tab:benchmarks}
\end{table}

\section{Experimental Setup}
\label{sec:setup}

We evaluate benchmark reliability across eight cybersecurity benchmarks and 10 LLMs, measuring how pipeline choices affect reported scores and comparative conclusions.

\subsection{Benchmarks}

Table~\ref{tab:benchmarks} lists the benchmarks included in our audit. Together, they cover a broad range of cybersecurity evaluation \textit{tasks}, including multiple-choice knowledge questions, vulnerability scoring, root-cause mapping, threat-actor attribution, attack-technique extraction, mitigation selection, and general cybersecurity reasoning. These benchmarks draw on structured and semi-structured authoritative cybersecurity sources, including CVE, CWE, CVSS, MITRE ATT\&CK, vulnerability advisories, and cyber threat intelligence reports~\citep{sikos2023cybersecurity}.

We score each benchmark using its full released dataset except where the release structure or validation protocol requires a subset. For example, SecBench~\citep{jing2024secbench} reports 47,910 questions, but only 3,000 are publicly released, comprising 2,730 Multiple-Choice Questions (MCQs) and 270 Short-Answer Questions (SAQs). Of these, only 661 MCQs are in English, and none of the SAQs are. We therefore score a subset of 661 questions out of the reported 47,910. App.~\ref{app:c1} reports the reported, released, and scored sizes for every benchmark, together with the rationale for each subset, where applicable.

\subsection{Models}
We evaluate 10 LLMs spanning proprietary, open-weight, and cybersecurity-specialized models, as summarized in Table~\ref{tab:models}. This mix allows us to test whether reliability failures are model-class specific or broader properties of the benchmark ecosystem. When a benchmark specifies an inference configuration, we use it as part of the original pipeline. Otherwise, we use a fixed configuration implemented in our evaluation harness (\S\ref{sec:harness}). Open-weight models are served locally, while proprietary models are evaluated through their respective APIs.

\begin{table}[t]
\centering
\renewcommand{\arraystretch}{1.12}
\scriptsize
\setlength{\tabcolsep}{4pt}
\begin{tabular}{lll}
\toprule
\textbf{Model} & \textbf{Category} & \textbf{Reference} \\
\midrule
GPT-5.4
    & Proprietary & \citep{singh2025openai} \\
{[}Claude] Sonnet 4.6 & Proprietary & \citep{anthropic2025system} \\
\midrule
Gemma-4[-31B]
    & Open-weight & \citep{google2026gemma} \\
Qwen3.6[-35B]
    & Open-weight & \citep{yang2025qwen3} \\
Llama-3.3[-70B]
    & Open-weight & \citep{grattafiori2024llama} \\
GPT-OSS[-20B]
    & Open-weight & \citep{agarwal2025gpt} \\
\midrule
Primus-Nemotron[-70B]
    & Cybersecurity & \citep{yu2025primus} \\
Primus-Merged[-8B]
    & Cybersecurity & \citep{yu2025primus} \\
Foundation-Sec[-8B]
    & Cybersecurity & \citep{yang2026llama} \\
RedSage-Qwen3[-8B-DPO]
    & Cybersecurity & \citep{suryanto2026redsage} \\
\bottomrule
\end{tabular}
\caption{Evaluated LLMs grouped by model category. Bracketed segments are dropped when we refer to models elsewhere in the paper.}
\label{tab:models}
\end{table}

\begin{table*}[t]
\centering
\scriptsize
\setlength{\tabcolsep}{4pt}
\renewcommand{\arraystretch}{1.12}
\begin{tabular}{@{}lll l ll@{}}
\toprule
\textbf{ID} &
\textbf{Failure mode} &
\textbf{Impact measure} &
\textbf{Affected units} &
\textbf{Maximum effect} &
\textbf{Median effect} \\
\midrule
$\mathcal{F}_1(\mathcal{D})$
& Limited capability coverage
& Dominant question type
& 5 of 8 benchmarks
& \multicolumn{2}{l}{>95\% of questions are of one type} \\

$\mathcal{F}_2(\mathcal{D})$
& Gold-label correctness
& Flag precision
& 7 of 8 benchmarks
& \multicolumn{2}{l}{23.8\% of 998 checked flags} \\

\midrule

$\mathcal{F}_1(\mathcal{P})$
& Format-token leakage
& Peer score gap
& 4 of 10 models
& 90.9\,pp
& 81.3\,pp \\

$\mathcal{F}_2(\mathcal{P})$
& Prompt--question conflict
& Single-letter response rate
& 8 of 10 models
& 33.0\%
& 26.2\% \\

$\mathcal{F}_3(\mathcal{P})$
& Template incompatibility
& Peer score gap
& 1 of 10 models
& 48\,pp
& --- \\

\midrule

$\mathcal{F}_1(\mathcal{I})$
& Stop-sequence mismatch
& Score gap
& 1 of 10 models
& 85.9\,pp
& --- \\

$\mathcal{F}_2(\mathcal{I})$
& Token-budget filter
& Score gap
& 1 of 10 models
& 81\,pp
& --- \\

$\mathcal{F}_3(\mathcal{I})$
& Decoding drift
& Score gap
& 1 of 10 models
& 40\,pp
& --- \\

\midrule

$\mathcal{F}_1(\mathcal{E})$
& Extractor divergence
& Extractor score gap
& 8 of 10 models
& 79.7\,pp
& 1.3\,pp \\

$\mathcal{F}_2(\mathcal{E})$
& Denominator inflation 
& Score gap 
& 3 benchmarks 
& 99.8\,pp 
& --- \\

$\mathcal{F}_3(\mathcal{E})$
& Metric-direction mismatch
& Rank shift
& 10 of 10 models
& 5 ranks
& 1 rank \\

$\mathcal{F}_4(\mathcal{E})$
& Prompt-mode sensitivity
& Score spread
& 10 of 10 models
& 40\,pp
& 7\,pp \\

\midrule

$\mathcal{F}_1(\mathcal{A})$
& Logprob vs.\ generative scoring
& Score gap
& 8 of 10 models
& 40.9\,pp
& 2.8\,pp \\

$\mathcal{F}_2(\mathcal{A})$
& Task-level metric drift
& Aggregation score gap
& 10 of 10 models
& 70\,pp
& 33.5\,pp \\

$\mathcal{F}_3(\mathcal{A})$
& Aggregation inconsistency
& Denominator gap
& 2 of 8 benchmarks
& 90.5\,pp
& --- \\
\bottomrule
\end{tabular}
\caption{
Summary of the 15 recurring pipeline failure modes and their observed effects. \emph{Affected units} reports where each failure was observed; maximum and median effects summarize the affected units when a per-unit distribution is available, and ``---'' indicates that a median is not meaningful. More information in App.~\ref{app:b7}, App.~\ref{app:evidence}, and Table~\ref{tab:app-modes}.
}
\label{tab:failure_modes}
\end{table*}

\subsection{Evaluation Harness}
\label{sec:harness}

We implement a common harness to isolate evaluation artifacts across cybersecurity benchmarks.\footnote{\href{https://github.com/qcri/cyberbench-audit}{Audit code}, with ablations and analyses, and the \href{https://github.com/qcri/sayf-eval}{harness} are publicly available. Links are also provided in~\S\ref{sec:intro}.} The harness standardizes the nine pipeline configuration fields defined in~\S\ref{sec:pipeline} wherever benchmark semantics permit. Importantly, these standardizations modify the evaluation procedure, not the task itself: benchmark questions, gold answers, and the intended capability being measured remain unchanged (App.~\ref{app:a}). The harness records the full evaluation trace, including raw model outputs, extracted predictions, question-level and aggregate scores, invalid-response rates, and pipeline configuration, making differences between original and standardized evaluations explicit and reproducible.

App.~\ref{app:a} documents each standardization field by field, distinguishing corrections to released implementations from choices we supply when a benchmark leaves a field unspecified. Some choices are not uniquely determined by the benchmark specification. In particular, answer extraction, invalid-response handling, partial-credit scoring, and log-probability versus generative scoring admit defensible alternatives; Table~\ref{tab:app-changes} reports these alternatives and quantifies their effects. We also reviewed all eight benchmarks to determine whether output-format compliance is part of the capability being evaluated. None treats format compliance as an evaluation objective. So, the standardized pipeline uses a consistent semantic extraction rule rather than marking superficial formatting differences as incorrect. Specifically, it applies one pinned LLM-based extraction and judging policy with a fixed judge model, prompt, and temperature (App.~\ref{app:a-extract}). For example, the harness uses binary scoring while allowing the judge to recognize equivalent aliases between the gold answer and the model response.

\section{Benchmark-Level Reliability Failures}
\label{sec:pipeline_failures}

We first audit reliability within individual benchmark pipelines. Across eight cybersecurity benchmarks, we identify 15 recurring failure modes spanning all five pipeline stages. Table~\ref{tab:failure_modes} summarizes their observed effects. The main text highlights representative cases with the largest empirical impact, while Apps.~\ref{app:e}--\ref{app:i} provide additional evidence and per-model breakdowns.

\subsection{Dataset Failures}
Dataset construction determines what a benchmark can measure before any model is queried.

\begin{figure}[t]
\centering
\includegraphics[width=\columnwidth]{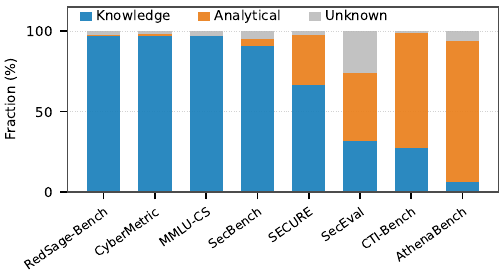}
\caption{Knowledge-oriented and analytical question composition per benchmark using 4-classifier majority vote over 2,155 stratified questions (Fleiss's $\kappa$=0.753).}
\label{fig:coverage}
\end{figure}

\paragraph{$\mathcal{F}_1(\mathcal{D})$: Limited capability coverage.}
A domain benchmark should capture a meaningful range of the capabilities it is intended to assess. We therefore examine whether each benchmark covers both knowledge recall and analytical reasoning. Using a majority vote of four LLM classifiers, we classify each \emph{question type} as either knowledge-oriented or analytical (App.~\ref{app:e1}). Figure~\ref{fig:coverage} shows that four benchmarks are highly skewed toward knowledge-oriented questions: RedSage-Bench, CyberMetric, MMLU-CS, and SecBench each contain over 95\% knowledge-oriented questions. AthenaBench shows the opposite pattern, with analytical questions comprising over 95\% of its questions. Hence, their aggregate scores capture only a narrow slice of domain capability, emphasizing either knowledge recall or analytical reasoning rather than both.

\paragraph{$\mathcal{F}_2(\mathcal{D})$: Gold-label correctness.}
Noise in gold labels is non-negligible. Model-majority disagreement flags 1,140 questions, of which an automated search-grounded verifier checks 998 (App.~\ref{app:e2}). It confirms 238 (23.8\%) as label errors and 653 (65.4\%) as false positives. Importantly, the 23.8\% figure is the precision of the flagging procedure on this disagreement-selected set, not a benchmark-wide label-error rate. Model disagreement is therefore useful for triage, but cannot by itself establish label correctness. In evolving domains such as cybersecurity, gold labels require explicit verification and uncertainty handling.

\subsection{Prompt Failures}
Prompt choices can distort evaluation by changing task interpretation, expected response format, or compatibility with extraction rules, producing score differences unrelated to model capability.

\paragraph{$\mathcal{F}_1(\mathcal{P})$: Format-token leakage.}
In CyberMetric, the answer-format template contains a literal placeholder that some models interpret incorrectly. Four of the 10 models are affected: Foundation-Sec and Primus-Nemotron produce mostly empty outputs, Gemma-4 reproduces the placeholder, and Primus-Merged continues the instruction text instead of answering. Removing the placeholder restores valid responses for all four models (App.~\ref{app:f2}).

\paragraph{$\mathcal{F}_2(\mathcal{P})$: Prompt--question conflict.}
In SecEval, benchmark-level instructions to ``select the correct answers'' sometimes conflict with question wording that suggests a single-select response, leading models to return one choice for questions that require multiple selections. Under set-exact-match scoring, these responses receive zero credit even when the selected choice is partially correct. Across the eight models that produce parseable output, single-select responses occur on 10.1\%--33.0\% of the 927 multi-select questions (App.~\ref{app:f3}).

\paragraph{$\mathcal{F}_3(\mathcal{P})$: Template incompatibility.}
In SECURE, Primus-Merged responds to MCQ questions with free-form explanations rather than the expected single- or multi-select answers. The benchmark extractor then treats the first answer-choice letter appearing in the explanation as the prediction, yielding a 48 percentage-point (pp) gap from peer models. This peer score gap reflects template incompatibility rather than task difficulty (App.~\ref{app:f4}).

\subsection{Inference Failures}
\label{sec:pipeline_failures-inference}

Inference choices determine whether a model's answer is generated and observable to the evaluator. Unlike prompt failures, which arise from task presentation, inference failures arise from generation-time pipeline configuration fields such as stop sequences, token budgets, and decoding parameters.

\paragraph{$\mathcal{F}_1(\mathcal{I})$: Stop-sequence mismatch.}
The largest inference-stage effect happens in RedSage-Bench, where the official stop sequence, the newline character ``\texttt{\textbackslash n}'', is triggered within Qwen3.6's reasoning preamble before any answer token is produced. Generating until the end-of-sequence token, allowing sufficient output tokens for the reasoning span to close, and stripping that span before extraction increases the score by 85.9 pp (App.~\ref{app:g1}).

\paragraph{$\mathcal{F}_2(\mathcal{I})$: Token-budget filter.}
In SecEval, the five-token output budget falls below Azure OpenAI's 16-token minimum, causing every GPT-5.4 request to fail and return error payloads rather than model answers. Therefore, the resulting 0.3\% score comes from accidental letter matches within these payloads. Raising the budget to 16 tokens restores valid generation and yields 81.4\% accuracy (App.~\ref{app:g2}), showing how an operational constraint can act as a hidden capability filter.

\paragraph{$\mathcal{F}_3(\mathcal{I})$: Decoding drift.}
In CyberMetric, the paper specifies sampling with temperature 1.0, top-$p$ 0.9, and top-$k$ 50, whereas the released script effectively defaults to greedy decoding. For Primus-Merged, this discrepancy changes output-format compliance and results in 40 pp score gap, where 274 of the 500 questions change correctness between the two configurations (App.~\ref{app:g3}). The effect is therefore driven largely by generation behavior and format compliance rather than knowledge. So, inference configurations should be pinned in executable code, not only described in text.

\subsection{Extraction and Scoring Failures}

Most benchmark pipelines convert free-form generations into scored predictions, making extraction a major source of measurement error.

\paragraph{$\mathcal{F}_1(\mathcal{E})$: Extractor divergence.}
Strict extractors can fail when model outputs do not match their expected format. For example, regex extractors may miss correct answers when models use free-form responses, place answers in unexpected locations, or deviate from the prescribed answer template. In CTI-Bench vulnerability scoring, the prompt asks for the CVSS vector on the final line, while the released extractor takes the last vector found anywhere in the response. These rules disagree on 30.4\% of model-item pairs where at least one extracts a vector, and on 96.5\% of Primus-Merged's pairs (App.~\ref{app:h1}). RedSage-Bench introduces a related ambiguity by reporting three metrics on the same generations without designating one as canonical. Primus-Merged scores 0.3\% under exact match but 80.0\% under prefix match, resulting in 79.7 pp score gap (App.~\ref{app:i1}). These findings show that extraction and scoring rules should be explicit and consistent.

\paragraph{$\mathcal{F}_2(\mathcal{E})$: Denominator inflation.}
In CTI-Bench's Root-Cause Mapping (CTI-RCM) task, invalid predictions, such as unparseable responses, are excluded from the denominator. Gemma-4 produces only two parseable outputs out of 1{,}000, which are both correct. Correct-over-valid scoring therefore reports 100.0\%, whereas correct-over-total scoring reports 0.2\%, a 99.8\,pp difference and $500\times$ inflation (\exampleref{example:e2-cti-rcm-gemma}; App.~\ref{app:h2}). AthenaBench's vulnerability-scoring task exhibits the same mechanism: unparseable CVSS vectors are excluded from the released denominator, affecting nine of the 10 models. Re-scoring the same generations while retaining these predictions and assigning the maximum deviation lowers scores by up to 85.4\,pp, with a median decrease of 55.9\,pp across the affected models (App.~\ref{app:h2}). Thus, invalid-response handling should be defined and reported explicitly.

\paragraph{$\mathcal{F}_3(\mathcal{E})$: Metric-direction mismatch.}
CTI-Bench and AthenaBench both evaluate CVSS-vector prediction but use different scoring conventions. CTI-Bench reports mean absolute deviation (MAD), where lower is better, while AthenaBench transforms MAD into a normalized percentage, where higher is better (App.~\ref{app:h3}). Across the 10 models, this mismatch shifts rankings by up to five positions, so similar tasks are not directly comparable unless their metric direction and scale are aligned.

\paragraph{$\mathcal{F}_4(\mathcal{E})$: Prompt-mode sensitivity.}
In AthenaBench's Attack-Technique Extraction (Athena-ATE) task, the extractor reads only the final line of the response. A Chain-of-Thought (CoT) model may identify the correct techniques in its reasoning but omit them from the final line, causing correct evidence to be ignored. Holding the task and extractor fixed, we compare zero-shot, few-shot, and CoT prompting on the same samples. The resulting Athena-ATE scores differ by up to 40 pp across the 10 models (App.~\ref{app:h4}). As each prompt mode generates new responses, we treat this as prompt-mode sensitivity rather than an extraction failure.

\subsection{Aggregation Failures}

Aggregation determines how question- and task-level scores become benchmark-level conclusions. 

\paragraph{$\mathcal{F}_1(\mathcal{A})$: Logprob vs.\ generative scoring.}
MCQ scores can change substantially depending on how answers are scored. Log-probability scoring ranks the answer choices directly without generating a response, while generative scoring extracts an answer from generated text. On RedSage-Bench, Gemma-4 scores 45.7\% under generative regex extraction but 86.6\% under log-probability scoring. Qwen3.6 shows the opposite pattern, scoring 85.9\% and 59.2\%, respectively (App.~\ref{app:i1}). These differences are not caused by the RedSage-Bench stop sequence, since the generative scores use stop-free outputs. Neither scoring method is uniformly preferable because they measure different interactions between the model and evaluator.

\paragraph{$\mathcal{F}_2(\mathcal{A})$: Task-level metric drift.}
In attacker-attribution, scores vary substantially with the scoring rules. CTI-Bench provides strict exact-match scoring and a lenient variant that additionally credits alias-connected or related threat actors (e.g., APT28 and FancyBear). AthenaBench instead uses a strict binary verdict. Across the 10 models, differences among these task-level scores reach 70~pp (App.~\ref{app:i2}). Aggregating task scores defined under different scoring rules can make benchmark-level scores and rankings depend on those rules rather than on a consistent measure of capability.

\paragraph{$\mathcal{F}_3(\mathcal{A})$: Aggregation inconsistency.}
Aggregation becomes inconsistent when scores computed over different populations are combined or compared as if they measured the same quantity. For example, \textit{accuracy} may use all questions as the denominator in one task but only valid predictions in another. Both are percentages, so this difference can disappear in benchmark-level comparisons.

Gemma-4 illustrates the effect. CTI-RCM reports 100.0\%, but only two of the model's 1,000 predictions are considered valid and both are correct (App.~\ref{app:h2}). SecEval reports 9.5\%, producing an apparent 90.5 pp gap. Its score is also affected by prompt wording that leads Gemma-4 to return single-select answers on multi-select questions, with a 33.0\% single-select rate on this subset (App.~\ref{app:f3}). Using a correct-over-total denominator reduces CTI-RCM to 0.2\%. Under the standardized pipeline, the two scores become 70.9\% and 78.3\%, respectively (App.~\ref{app:i3}). Aggregation can therefore propagate upstream pipeline inconsistencies into benchmark-level comparisons.

\begin{insight}
\textbf{Implications for Model Comparison.} Pipeline failures are widespread: every audited benchmark has at least two documented failure-mode interactions (Table~\ref{tab:failure_coverage}). Large score changes can occur without changing the dataset or model, driven instead by configuration choices spanning all pipeline stages. But does this pipeline sensitivity also change comparative conclusions about models?
\end{insight}

\newcommand{\failYes}{%
  \cellcolor{red!10}\textcolor{red!45!black}{\scriptsize\ding{51}}%
}
\newcommand{\failNo}{%
  \cellcolor{green!8}\textcolor{black!30}{$\cdot$}%
}

\begin{table*}[t]
\centering
\scriptsize
\renewcommand{\arraystretch}{1.18}
\setlength{\tabcolsep}{3.5pt}

\begin{tabularx}{\textwidth}{
@{}l
*{15}{>{\centering\arraybackslash}X}
>{\centering\arraybackslash}X@{}
}
\toprule
&
\multicolumn{2}{c}{\textbf{Dataset}}
& \multicolumn{3}{c}{\textbf{Prompt}}
& \multicolumn{3}{c}{\textbf{Inference}}
& \multicolumn{4}{c}{\textbf{Extraction}}
& \multicolumn{3}{c}{\textbf{Aggregation}}
& \\
\cmidrule(lr){2-3}
\cmidrule(lr){4-6}
\cmidrule(lr){7-9}
\cmidrule(lr){10-13}
\cmidrule(lr){14-16}

\textbf{Benchmark}
& $\mathcal{F}_1(\mathcal{D})$
& $\mathcal{F}_2(\mathcal{D})$
& $\mathcal{F}_1(\mathcal{P})$
& $\mathcal{F}_2(\mathcal{P})$
& $\mathcal{F}_3(\mathcal{P})$
& $\mathcal{F}_1(\mathcal{I})$
& $\mathcal{F}_2(\mathcal{I})$
& $\mathcal{F}_3(\mathcal{I})$
& $\mathcal{F}_1(\mathcal{E})$
& $\mathcal{F}_2(\mathcal{E})$
& $\mathcal{F}_3(\mathcal{E})$
& $\mathcal{F}_4(\mathcal{E})$
& $\mathcal{F}_1(\mathcal{A})$
& $\mathcal{F}_2(\mathcal{A})$
& $\mathcal{F}_3(\mathcal{A})$
& \textbf{Total} \\

\midrule

MMLU-CS
& \failYes & \failYes
& \failNo  & \failNo  & \failNo
& \failNo  & \failNo  & \failNo
& \failNo  & \failNo  & \failNo  & \failNo
& \failYes & \failNo  & \failNo
& 3 \\

SecEval
& \failNo  & \failYes
& \failNo  & \failYes & \failNo
& \failNo  & \failYes & \failNo
& \failNo  & \failNo  & \failNo  & \failNo
& \failNo  & \failNo  & \failYes
& 4 \\

SECURE
& \failNo  & \failYes
& \failNo  & \failNo  & \failYes
& \failNo  & \failNo  & \failNo
& \failNo  & \failYes & \failNo  & \failNo
& \failNo  & \failNo  & \failNo
& 3 \\

CTI-Bench
& \failNo  & \failYes
& \failNo  & \failNo  & \failYes
& \failYes & \failNo  & \failNo
& \failYes & \failYes & \failYes & \failNo
& \failNo  & \failYes & \failYes
& 8 \\

AthenaBench
& \failYes & \failYes
& \failNo  & \failNo  & \failNo
& \failNo  & \failNo  & \failNo
& \failNo  & \failYes & \failYes & \failYes
& \failNo  & \failYes & \failNo
& 6 \\

CyberMetric
& \failYes & \failNo
& \failYes & \failNo  & \failNo
& \failNo  & \failNo  & \failYes
& \failNo  & \failNo  & \failNo  & \failNo
& \failNo  & \failNo  & \failNo
& 3 \\

RedSage-Bench
& \failYes & \failYes
& \failNo  & \failNo  & \failNo
& \failYes & \failNo  & \failNo
& \failYes & \failNo  & \failNo  & \failNo
& \failYes & \failNo  & \failNo
& 5 \\

SecBench
& \failYes & \failYes
& \failNo  & \failNo  & \failNo
& \failNo  & \failNo  & \failNo
& \failNo  & \failNo  & \failNo  & \failNo
& \failNo  & \failNo  & \failNo
& 2 \\

\midrule

\textbf{Total}
& 5
& 7
& 1
& 1
& 2
& 2
& 1
& 1
& 2
& 3
& 2
& 1
& 2
& 2
& 2
& 34 \\
\bottomrule
\end{tabularx}
\caption{
Failure-mode incidence across the eight audited benchmarks. A red cell with a check indicates an observed incident; a green cell with a dot indicates no incident was observed. Totals report observed incidents by benchmark and failure mode. All 120 benchmark--failure-mode pairs were examined, resulting in a total of 34 incidents.
}
\label{tab:failure_coverage}
\end{table*}

\section{Cross-Benchmark Instability}
\label{sec:ecosystem}

The benchmark-level analysis shows that scores can change under different pipeline choices. We next investigate how this instability affects model comparison. We analyze the \textit{23$\times$10 task-by-model score matrix} from three perspectives: score-level redundancy, rank-level agreement, and ranking shifts under pipeline standardization.

\subsection{Score-Level Redundancy}

\begin{figure}[t]
\centering
\includegraphics[width=\columnwidth]{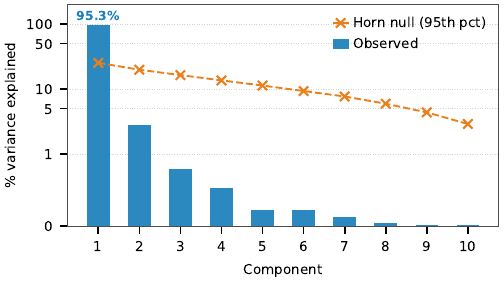}
\caption{PCA scree plot of the task-by-model accuracy matrix. Only PC1 exceeds Horn's parallel-analysis null and explains 95.25\% of the variance.}
\label{fig:pca-scree}
\end{figure}

We first ask whether the 23 tasks provide independent evidence about model capability. As shown in Figure~\ref{fig:pca-scree}, PCA of the column-standardized 23$\times$10 accuracy matrix shows that the first component explains 95.25\% of the variance (App.~\ref{app:j1}). Thus, most tasks separate stronger from weaker models along one dominant performance axis rather than capturing distinct cybersecurity capabilities. This redundancy limits what broader task coverage can establish. Adding tasks may increase benchmark scale without adding much measurement diversity. Aggregate scores may therefore provide limited evidence for claims about fine-grained cybersecurity specialization. A Mantel test finds a moderate association between task-content similarity and task-induced rank agreement ($r=0.516$; App.~\ref{app:j4}), suggesting that semantic overlap explains only part of the observed ranking similarity. This is consistent with the dominant performance axis reflecting more than repeated or similar question content.

\subsection{Rank-Level Disagreement}

Score-level redundancy does not imply stable model rankings. We measure rank agreement using Kendall's $\tau$-b, which accounts for ties, together with pairwise rank inversions (Apps.~\ref{app:j2}--\ref{app:j3}). Tasks may broadly agree on stronger and weaker models while disagreeing on specific model orderings. This pattern is clearest for semantically similar tasks. CTI-Bench and AthenaBench both evaluate vulnerability scoring and attacker attribution, yet they rank the same 10 models differently. The rank agreement is weak for vulnerability scoring ($\tau$-b=0.29) and attacker attribution ($\tau$-b=0.24). The vulnerability-scoring pair also reorders models by up to five positions (App.~\ref{app:j3}). These differences coincide with incompatible extraction rules, alias handling, and  scoring directions/rules. So, comparative claims, such as ``$m_i$ outperforms $m_j$ at vulnerability scoring,'' can depend on the pipeline.

\subsection{Ranking Shifts Under Standardization}

Using the evaluation harness (\S\ref{sec:harness}), we standardize the nine pipeline configuration fields wherever benchmark semantics allow. App.~\ref{app:k1} specifies which fields are standardized for each benchmark, and App.~\ref{app:k2} reports the underlying scores.

For benchmark $b$ and model $m$, let $r_b^{\mathrm{orig}}(m)$ and $r_b^{\mathrm{std}}(m)$ denote the model's rank under the original and standardized pipelines, respectively, with rank 1 indicating the highest-scoring model. Ties are broken deterministically using a stable sort (App.~\ref{app:k3}). We define the rank shift as
\begin{equation}
\Delta r_b(m)
=
r_b^{\mathrm{orig}}(m)
-
r_b^{\mathrm{std}}(m),
\label{eq:rank_shift}
\end{equation}
where $\Delta r_b(m)>0$ indicates upward shift after standardization and $\Delta r_b(m)<0$ indicates downward shift. Table~\ref{tab:rank_shifts} shows that nine of the 10 models shift by at least three ranks on at least one benchmark; only GPT-5.4 does not. Gemma-4 rises on every benchmark where its rank changes, consistent with the extraction and denominator effects in~\S\ref{sec:pipeline_failures}. Because standardization changes upstream fields like the prompt formatting, decoding, and token budget, these scores require re-generation rather than only re-scoring stored outputs.

\newcommand{\rankUp}[1]{\cellcolor{green!14}#1}
\newcommand{\rankDown}[1]{\cellcolor{red!14}#1}
\newcommand{\rankSame}{\cellcolor{gray!7}\textcolor{black!55}{0}}

\begin{table*}[t]
\centering
\scriptsize
\renewcommand{\arraystretch}{1.14}
\setlength{\tabcolsep}{5pt}

\resizebox{\textwidth}{!}{%
\begin{tabular}{@{}lcccccccc}
\toprule
\textbf{Model}
& \textbf{MMLU-CS}
& \textbf{SecEval}
& \textbf{SECURE}
& \textbf{CTI-Bench}
& \textbf{AthenaBench}
& \textbf{CyberMetric}
& \textbf{RedSage-Bench}
& \textbf{SecBench} \\
\midrule
GPT-5.4
& \rankUp{+1} & \rankSame & \rankSame & \rankSame
& \rankSame & \rankSame & \rankSame & \rankDown{-1} \\

Sonnet 4.6
& \rankUp{+5} & \rankDown{-5} & \rankSame & \rankSame
& \rankSame & \rankSame & \rankSame & \rankSame \\

\midrule

Gemma-4
& \rankUp{+6} & \rankUp{+7} & \rankUp{+4} & \rankUp{+3}
& \rankUp{+7} & \rankUp{+5} & \rankUp{+5} & \rankUp{+6} \\

Qwen3.6
& \rankUp{+6} & \rankUp{+3} & \rankDown{-1} & \rankDown{-2}
& \rankDown{-3} & \rankSame & \rankUp{+1} & \rankUp{+4} \\

Llama-3.3
& \rankDown{-3} & \rankDown{-5} & \rankDown{-3} & \rankDown{-2}
& \rankDown{-2} & \rankDown{-2} & \rankSame & \rankDown{-3} \\

GPT-OSS
& \rankUp{+5} & \rankUp{+4} & \rankSame & \rankDown{-5}
& \rankDown{-5} & \rankDown{-1} & \rankUp{+2} & \rankUp{+1} \\

\midrule

Primus-Nemotron
& \rankDown{-7} & \rankDown{-1} & \rankUp{+4} & \rankUp{+3}
& \rankUp{+2} & \rankUp{+2} & \rankDown{-2} & \rankDown{-1} \\

Primus-Merged
& \rankDown{-3} & \rankDown{-1} & \rankSame & \rankSame
& \rankSame & \rankDown{-1} & \rankDown{-3} & \rankDown{-1} \\

Foundation-Sec
& \rankDown{-6} & \rankDown{-5} & \rankDown{-4} & \rankSame
& \rankUp{+2} & \rankSame & \rankDown{-1} & \rankDown{-4} \\

RedSage-Qwen3
& \rankDown{-4} & \rankUp{+3} & \rankSame & \rankUp{+3}
& \rankDown{-1} & \rankDown{-3} & \rankDown{-2} & \rankDown{-1} \\

\midrule

\textbf{Spearman $\rho$}
& -0.47
& 0.03
& 0.65
& 0.64
& 0.42
& 0.73
& 0.71
& 0.50 \\

\bottomrule
\end{tabular}%
}
\caption{
Rank shifts under pipeline standardization. Each cell reports
$\Delta r_b(m)$, where positive values indicate upward movement and
negative values indicate downward movement. The bottom row reports
Spearman's $\rho$ between the original and standardized rankings;
95\% bootstrap confidence intervals are shown in
Figure~\ref{fig:rank-shift-corr}.
}
\label{tab:rank_shifts}
\end{table*}

\begin{figure}[t]
\centering
\resizebox{\columnwidth}{!}{%
\begin{tikzpicture}
\begin{axis}[
  width=12cm, height=6.8cm,
  xbar, bar width=7.2pt,
  xmin=-1.05, xmax=1.05, xtick={-1,-0.5,0,0.5,1},
  ytick={1,2,3,4,5,6,7,8},
  yticklabels={MMLU-CS,SecEval,AthenaBench,SecBench,CTI-Bench,SECURE,RedSage-Bench,CyberMetric},
  ymin=0.4, ymax=8.6, y dir=reverse,
  xlabel={Spearman $\rho$},
  tick label style={font=\sffamily\fontsize{12}{11}\selectfont},
  yticklabel style={font=\sffamily\fontsize{12}{11}\selectfont, xshift=1.5pt},
  label style={font=\sffamily\fontsize{12}{12}\selectfont},
  axis line style={line width=0.86pt, draw=black!70},
  every tick/.style={draw=black!70, line width=0.72pt},
  tick align=outside, tick pos=left,
  xmajorgrids, grid style={dotted, line width=0.65pt, draw=black!50},
  enlarge y limits=false,
]
\addplot[
  fill=houseblue, fill opacity=0.92, draw=none,
  error bars/.cd, x dir=both, x explicit,
  error bar style={draw=black!70, line width=0.6pt},
  error mark options={draw=black!70, mark size=2pt, line width=0.6pt}]
  table[row sep=\\, x=rho, y=yy, x error plus=ep, x error minus=em]{
  rho & yy & ep & em \\
  -0.47 & 1 & 0.22 & 0.24 \\
  0.03 & 2 & 0.12 & 0.11 \\
  0.42 & 3 & 0.05 & 0.03 \\
  0.50 & 4 & 0.08 & 0.18 \\
  0.64 & 5 & 0.12 & 0.08 \\
  0.65 & 6 & 0.07 & 0.09 \\
  0.71 & 7 & 0.01 & 0.11 \\
  0.73 & 8 & 0.08 & 0.26 \\
  };
\draw[black!55, dashed, line width=0.72pt]
  (axis cs:0,0.4) -- (axis cs:0,8.6);
\end{axis}
\end{tikzpicture}}
\caption{Spearman $\rho$ between original and standardized model rankings for each benchmark. Whiskers show question-level bootstrap 95\% confidence intervals (App.~\ref{app:k4}). Lower $\rho$ indicates greater reordering.}
\label{fig:rank-shift-corr}
\end{figure}
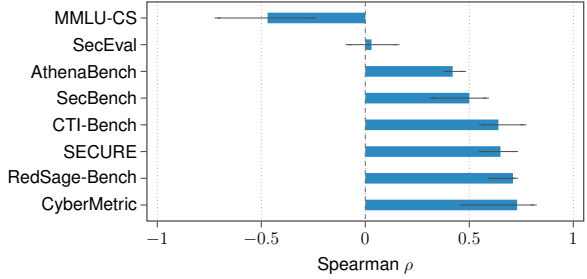

As shown in Figure~\ref{fig:rank-shift-corr}, the original and standardized rankings also show substantial disagreement. Spearman's $\rho$ is -0.47 for MMLU-CS and 0.03 for SecEval, indicating strong reordering, and ranges from 0.42 to 0.73 on the other benchmarks. On MMLU-CS, all four cybersecurity-specialized models shift downward by three to seven positions, while Gemma-4 shifts upward by six.

These shifts are robust to question-sampling variability. Using 5,000 paired bootstrap samples, every rank shift of at least three positions keeps the same direction in at least 97.7\% of samples. The 95\% confidence intervals for Spearman's $\rho$ all exclude 1, indicating that the original and standardized rankings are not statistically consistent with being identical. Thus, the ranking changes cannot be explained by question-sampling noise alone.

\begin{insight}
\textbf{Implications for Benchmark Design.}
Broader task coverage does not guarantee more informative or stable model comparisons. Many tasks provide redundant score-level evidence, while model rankings remain sensitive to pipeline choices. Reliable benchmarking therefore requires both diverse capability coverage and explicit, consistent evaluation pipelines.
\end{insight}

\section{Toward Reliable Benchmarking}
\label{sec:recommendations}

Benchmark reliability deserves the same care as task design. A benchmark release should specify the full measurement pipeline used to produce its scores, not only the dataset and metric. This includes dataset and label provenance, the exact prompt and chat template, inference settings, the executable extractor or judge, denominator policy, scoring and aggregation rules, and reliability statistics such as invalid-response rates and rank stability. App.~\ref{app:card} provides a complete field list.

Evaluation harnesses should also minimize avoidable nondeterminism. Open-weight models should use pinned model and serving configurations. Closed-source evaluations should report the exact model ID and run timestamp. Judge-based evaluation should fix the judge model, prompt, and decoding parameters. Storing raw outputs further improves reproducibility: changes to extraction, scoring, denominators, or aggregation can then be re-scored without querying the evaluated model again. An analysis that requires new generations should report the model and serving configuration used for those runs.

Benchmarks that evaluate similar tasks should align metric direction, scoring rules, alias handling, and aggregation conventions. If not, their scores should be reported as distinct measurements rather than treated as directly comparable. In evolving domains such as cybersecurity, gold labels should also be periodically audited against authoritative sources, with uncertainty reported explicitly.

Not every reliability failure can be fixed automatically. Some arise from underspecified benchmark intent, such as whether partial credit should be awarded or which semantic equivalences (e.g., aliases) should count as correct. These choices cannot be recovered reliably from the released artifact alone. Evaluation harnesses should therefore make such decisions explicit and encode them in executable form. App.~\ref{app:m} classifies the identified failure modes as automatically detectable, automatically fixable, or requiring manual judgment.

\section{Related Work}
\label{sec:related}

Large-scale frameworks such as BIG-bench~\citep{srivastava2022beyond}, HELM~\citep{liang2022helm}, and DynaBench~\citep{kiela2021dynabench} have made benchmark-based comparison central to LLM evaluation. Recent work shows that reported performance can be sensitive to prompt wording, decoding parameters, evaluator design, and extraction rules~\citep{shi2024decoding,sun2024evaluating,wang2023fair}. We build on this work by modeling benchmarks as end-to-end measurement pipelines and auditing failures across all five stages. Prior work also shows that aggregate score correlations can obscure disagreement in model rankings~\citep{perlitz2024benchmark}. We extend this perspective to cybersecurity, where even semantically similar tasks can rank the same models differently when their evaluation conventions differ.

Bean et al.~\citep{bean2025measuring} study the complementary problem of \emph{construct validity}: whether benchmarks measure the phenomena they claim to measure and support the resulting claims. Their review spans multiple benchmark domains but does not include cybersecurity benchmarks. We instead focus on \emph{measurement reliability} of executable evaluation pipelines and provide a systematic audit of this problem in cybersecurity.

\section{Conclusion}
\label{sec:conclusion}

We audited cybersecurity LLM benchmarks as end-to-end measurement pipelines. Across eight benchmarks, 23 tasks, and 10 LLMs, we identified 15 recurring failure modes spanning all five pipeline stages. Individual pipeline choices can shift scores by more than 80 pp, and nine of the 10 models shift by at least three ranks under standardization. Also, broader task coverage does not guarantee more informative or stable comparisons: many tasks provide redundant evidence, while similar tasks can rank models differently. Thus, benchmark scores should be treated as pipeline-dependent measurements, supported by meaningful capability coverage and explicit, consistent evaluation pipelines.

\section{Software}
\label{sec:software}
The harness is released on PyPI as an open-source Python package for model-agnostic evaluation of cybersecurity LLMs.\footnote{\url{https://pypi.org/project/sayf-eval}} It provides a common interface for hosted APIs and locally served models. Open-ended responses can be scored using a configurable LLM judge, allowing the evaluated model and judge model to be selected independently. For transparency, an adapter exports aggregate scores and pipeline metadata to EveryEvalEver~\citep{batzner2026every}, while withholding question-level content in accordance with our Ethical Considerations.

\section*{Limitations}
\label{sec:limitations}

Our empirical findings are specific to cybersecurity. We chose this domain because it makes pipeline failures unusually observable. Much of its ground truth can be checked against authoritative sources such as CVE, CWE, CVSS, and MITRE ATT\&CK (App.~\ref{app:e2}). Many answers are also structured, so certain failures such as an invalid CVSS vector can be detected mechanically. Finally, the benchmark ecosystem is heterogeneous: 44 of 72 pipeline configuration fields cannot be reproduced from documentation alone (App.~\ref{app:a-underspec}). Mechanical failures involving token budgets, stop sequences, extraction, denominators, and aggregation may transfer to other domains, but we do not test this directly.

Our audit covers eight benchmarks, 23 tasks, and 10 LLMs. Other benchmarks, models, serving backends, and evaluators may exhibit failures outside our taxonomy. Some benchmark artifacts are also underspecified, requiring us to make implementation choices. We document these choices in the paper and appendix, but alternative interpretations may be defensible.

We study measurement reliability, not construct validity~\citep{bean2025measuring}. A benchmark can be reproducible and internally consistent while still failing to measure the real-world capability it claims to represent. Reliability is therefore necessary for valid evaluation, but it is not sufficient.

Finally, some conclusions depend on contestable standardization choices. Table~\ref{tab:app-changes} identifies four such choices: answer extraction, invalid-response denominators, partial credit, and log-probability versus generative scoring. These choices are defensible but not unique, so comparative conclusions remain conditional on them. Our standardized extraction also uses a pinned LLM judge. An independent judge agrees on 99.6\% of verdicts and produces nearly identical rankings (App.~\ref{app:a-judgeindep}), reducing but not eliminating concerns about judge dependence. We quantify question-sampling variability with a question-level bootstrap (App.~\ref{app:k4}), but do not estimate variability across repeated model generations. Gold-label and judge decision verification are also LLM-assisted, with manual validation on random stratified samples of questions drawn across each benchmark's tasks (Apps.~\ref{app:a-judgestats} and~\ref{app:e2}).

\section*{Ethical Considerations}
\label{sec:ethics}

This work studies the reliability of cybersecurity benchmarks rather than the development of new offensive capabilities. The audited tasks cover vulnerabilities, attack techniques, malware behavior, and threat intelligence, but are drawn from existing public benchmarks and authoritative cybersecurity sources. We do not introduce new exploit procedures, malware implementations, offensive datasets, or attack automation.

The main ethical concern is the harm caused by unreliable evaluation. In a high-stakes domain such as cybersecurity, unstable benchmark pipelines can support misleading claims about model capability, safety, or specialization. Such claims may influence model selection, deployment decisions, and trust in automated security systems. By exposing pipeline failures and making evaluation choices explicit, our goal is to improve transparency, reproducibility, and scientific reliability.

A secondary concern is dual use. Detailed descriptions of benchmark failure modes could facilitate benchmark-specific optimization or gaming. We therefore focus on evaluation mechanisms rather than methods for exploiting security systems, and our public reporting emphasizes aggregate results and pipeline metadata rather than unnecessary question-level content.

\bibliography{refs}

@article{ghosh2026evaluation,
  title={Evaluation Cards: An Interpretive Layer for AI Evaluation Reporting},
  author={Ghosh, Avijit and Reuel, Anka and Chim, Jenny and Kennedy, Wm Matthew and Yadav, Srishti and Mickel, Jennifer and Long, Yanan and Tran, Andrew and Kornilova, Anastassia and Stachura, Damian and others},
  journal={arXiv preprint arXiv:2606.09809},
  year={2026}
}

@article{batzner2026every,
  title={Every Eval Ever: A Unifying Schema and Community Repository for AI Evaluation Results},
  author={Batzner, Jan and Nelaturu, Sree Harsha and Stachura, Damian and Kornilova, Anastassia and Crall, Jon and Cerruti, Tommaso and Long, Yanan and Mai, Yifan and Ahuja, Sanchit and Yehudai, Asaf and others},
  journal={arXiv preprint arXiv:2606.14516},
  year={2026}
}

@article{bean2025measuring,
  title={Measuring what Matters: Construct Validity in Large Language Model Benchmarks},
  author={Bean, Andrew M and Kearns, Ryan Othniel and Romanou, Angelika and Hafner, Franziska Sofia and Mayne, Harry and Batzner, Jan and Foroutan Eghlidi, Negar and Schmitz, Chris and Korgul, Karolina and Batra, Hunar and others},
  journal={Advances in Neural Information Processing Systems 38},
  pages={19868--19949},
  year={2025},
  publisher={Neural Information Processing Systems Foundation, Inc.(NeurIPS)}
}

@inproceedings{shi2024decoding,
  title={{A Thorough Examination of Decoding Methods in the Era of Large Language Models}},
  author={Shi, Cheng and Zhang, Zhisheng and Yang, Yujiu and others},
  booktitle={Proceedings of EMNLP},
  year={2024}
}

@inproceedings{sun2024evaluating,
title={{Evaluating the Zero-Shot Robustness of Instruction-Tuned Language Models}},
  author={Sun, Jiuding and Shaib, Chantal and Wallace, Byron},
  booktitle={International Conference on Learning Representations},
  volume={2024},
  pages={48103--48141},
  year={2024}
}

@article{wang2023fair,
  title={{Large Language Models are not Fair Evaluators}},
  author={Wang, Peiyi and Li, Lei and Chen, Liang and others},
  journal={arXiv preprint arXiv:2305.17926},
  year={2023}
}

@article{sikos2023cybersecurity,
  title={{Cybersecurity Knowledge Graphs}},
  author={Sikos, Leslie F.},
  journal={Knowledge and Information Systems},
  volume={65},
  number={9},
  pages={3511--3531},
  year={2023},
  publisher={Springer}
}

@inproceedings{suryanto2026redsage,
  title={{RedSage: A Cybersecurity Generalist LLM}},
  author={Suryanto, Naufal and Naseer, Muzammal and Li, Pengfei and Wasim, Syed Talal and Yi, Jinhui and Gall, Juergen and Ceravolo, Paolo and Damiani, Ernesto},
  booktitle={The Fourteenth International Conference on Learning Representations},
  year={2026},
  url={https://openreview.net/forum?id=W4FAenIrQ2}
}

@article{alam2025athenabench,
  title   = {{AthenaBench: A Dynamic Benchmark for Evaluating LLMs in Cyber Threat Intelligence}},
  author  = {Alam, Md Tanvirul and Bhusal, Dipkamal and Ahmad, Salman and Rastogi, Nidhi and Worth, Peter},
  journal = {arXiv preprint arXiv:2511.01144},
  year    = {2025}
}

@article{hendrycks2020measuring,
  title={{Measuring Massive Multitask Language Understanding}},
  author={Hendrycks, Dan and Burns, Collin and Basart, Steven and Zou, Andy and Mazeika, Mantas and Song, Dawn and Steinhardt, Jacob},
  journal={arXiv preprint arXiv:2009.03300},
  year={2020}
}

@inproceedings{tihanyi2024cybermetric,
  title={{CyberMetric: A Benchmark Dataset Based on Retrieval-Augmented Generation for Evaluating LLMs in Cybersecurity Knowledge}},
  author={Tihanyi, Norbert and Ferrag, Mohamed Amine and Jain, Ridhi and Bisztray, Tamas and Debbah, Merouane},
  booktitle={2024 IEEE International Conference on Cyber Security and Resilience (CSR)},
  pages={296--302},
  year={2024},
  organization={IEEE}
}

@article{li2023seceval,
  title={{SecEval: A Comprehensive Benchmark for Evaluating Cybersecurity Knowledge of Foundation Models}},
  author={Li, Guancheng and Li, Yifeng and Guannan, Wang and Yang, Haoyu and Yu, Yang},
  journal={GitHub},
  year={2023}
}

@article{alam2024ctibench,
  title={{CTIBench: A Benchmark for Evaluating LLMs in Cyber Threat Intelligence}},
  author={Alam, Md Tanvirul and Bhusal, Dipkamal and Nguyen, Le and Rastogi, Nidhi},
  journal={Advances in Neural Information Processing Systems},
  volume={37},
  pages={50805--50825},
  year={2024}
}

@article{jing2024secbench,
  title   = {{SecBench}: A Comprehensive Multi-Dimensional Benchmarking Dataset for {LLM}s in Cybersecurity},
  author  = {Jing, Pengfei and Tang, Mengyun and Shi, Xiaorong and Zheng, Xing and Nie, Sen and Wu, Shi and Yang, Yong and Luo, Xiapu},
  journal = {arXiv preprint arXiv:2412.20787},
  year    = {2024}
}

@inproceedings{bhusal2024secure,
  title={Secure: Benchmarking large language models for cybersecurity},
  author={Bhusal, Dipkamal and Alam, Md Tanvirul and Nguyen, Le and Mahara, Ashim and Lightcap, Zachary and Frazier, Rodney and Fieblinger, Romy and Torales, Grace Long and Blakely, Benjamin A and Rastogi, Nidhi},
  booktitle={2024 Annual Computer Security Applications Conference (ACSAC)},
  pages={15--30},
  year={2024},
  organization={IEEE}
}

@article{singh2025openai,
  title={{OpenAI GPT-5 System Card}},
  author={Singh, Aaditya and Fry, Adam and Perelman, Adam and Tart, Adam and Ganesh, Adi and El-Kishky, Ahmed and McLaughlin, Aidan and Low, Aiden and Ostrow, AJ and Ananthram, Akhila and others},
  journal={arXiv preprint arXiv:2601.03267},
  year={2025}
}

@misc{anthropic2025system,
  title={{System Card: Claude Sonnet 4.6}},
  author={Anthropic},
  url={https://www.anthropic.com/news/claude-sonnet-4-6},
  year={2025}
}

@misc{google2026gemma,
  title={{Gemma 4 Model Card}},
  author={Google},
  url={https://ai.google.dev/gemma/docs/core/model_card_4},
  year={2026}
}

@article{yang2025qwen3,
  title={Qwen3 Technical Report},
  author={Yang, An and Li, Anfeng and Yang, Baosong and Zhang, Beichen and Hui, Binyuan and Zheng, Bo and Yu, Bowen and Gao, Chang and Huang, Chengen and Lv, Chenxu and others},
  journal={arXiv preprint arXiv:2505.09388},
  year={2025}
}

@article{grattafiori2024llama,
  title={{The Llama 3 Herd of Models}},
  author={Grattafiori, Aaron and Dubey, Abhimanyu and Jauhri, Abhinav and Pandey, Abhinav and Kadian, Abhishek and Al-Dahle, Ahmad and Letman, Aiesha and Mathur, Akhil and Schelten, Alan and Vaughan, Alex and others},
  journal={arXiv preprint arXiv:2407.21783},
  year={2024}
}

@article{agarwal2025gpt,
  title={{GPT-OSS-120B \& GPT-OSS-20B Model Card}},
  author={Agarwal, Sandhini and Ahmad, Lama and Ai, Jason and Altman, Sam and Applebaum, Andy and Arbus, Edwin and Arora, Rahul K and Bai, Yu and Baker, Bowen and Bao, Haiming and others},
  journal={arXiv preprint arXiv:2508.10925},
  year={2025}
}

@inproceedings{yu2025primus,
  title={{Primus: A Pioneering Collection of Open-Source Datasets for Cybersecurity LLM Training}},
  author={Yu, Yao-Ching and Chiang, Tsun-Han and Tsai, Cheng-Wei and Huang, Chien-Ming and Tsao, Wen-Kwang},
  booktitle={Proceedings of the 2025 Conference on Empirical Methods in Natural Language Processing},
  pages={10402--10424},
  year={2025}
}

@article{yang2026llama,
  title={{Llama-3.1-FoundationAI-SecurityLLM-Reasoning-8B Technical Report}},
  author={Yang, Zhuoran and Li, Ed and He, Jianliang and Priyanshu, Aman and Saglam, Baturay and Kassianik, Paul and Weerawardhena, Sajana and Vellore, Anu and Nelson, Blaine and Javidnia, Neusha and others},
  journal={arXiv preprint arXiv:2601.21051},
  year={2026}
}

@article{liang2022helm,
  title={{Holistic Evaluation of Language Models}},
  author={Percy Liang and Rishi Bommasani and Tony Lee and Dimitris Tsipras and Dilara Soylu and Michihiro Yasunaga and Yian Zhang and Deepak Narayanan and Yuhuai Wu and Ananya Kumar and Benjamin Newman and Binhang Yuan and Bobby Yan and Ce Zhang and Christian Cosgrove and Christopher D Manning and Christopher Re and Diana Acosta-Navas and Drew A. Hudson and Eric Zelikman and Esin Durmus and Faisal Ladhak and Frieda Rong and Hongyu Ren and Huaxiu Yao and Jue WANG and Keshav Santhanam and Laurel Orr and Lucia Zheng and Mert Yuksekgonul and Mirac Suzgun and Nathan Kim and Neel Guha and Niladri S. Chatterji and Omar Khattab and Peter Henderson and Qian Huang and Ryan Andrew Chi and Sang Michael Xie and Shibani Santurkar and Surya Ganguli and Tatsunori Hashimoto and Thomas Icard and Tianyi Zhang and Vishrav Chaudhary and William Wang and Xuechen Li and Yifan Mai and Yuhui Zhang and Yuta Koreeda},
  journal={Transactions on Machine Learning Research},
  issn={2835-8856},
  year={2023},
  url={https://openreview.net/forum?id=iO4LZibEqW},
}

@article{srivastava2022beyond,
  title={{Beyond the Imitation Game: Quantifying and Extrapolating the Capabilities of Language Models}},
  author={Srivastava, Aarohi and Rastogi, Abhinav and Rao, Abhishek and others},
  journal={Transactions on Machine Learning Research},
  year={2022}
}

@inproceedings{perlitz2024benchmark,
  title={{Benchmark Agreement Testing Done Right: A Guide for LLM Benchmark Evaluation}},
  author={Perlitz, Yotam and Gera, Ariel and Arviv, Ofir and Yehudai, Asaf and Bandel, Elron and Shnarch, Eyal and Shmueli-Scheuer, Michal and Choshen, Leshem},
  booktitle={NeurIPS 2025 Workshop on Evaluating the Evolving LLM Lifecycle: Benchmarks, Emergent Abilities, and Scaling},
  year={2024}
}

@inproceedings{kiela2021dynabench,
  title={{Dynabench: Rethinking Benchmarking in NLP}},
  author={Kiela, Douwe and Bartolo, Max and Nie, Yixin and others},
  booktitle={Proceedings of NAACL},
  year={2021}
}

\appendix

\section{Where to Find Each Artifact}
\label{app:map}

Table~\ref{tab:app-map} provides a roadmap to the artifacts supporting our pipeline audit and standardization decisions. Materials that are impractical to include in the paper, such as raw generations, run logs, plotting code, and interactive notebooks, are released in the audit repository, which is also linked in the table.

Four box types recur throughout the appendix. A \textit{prompt} box reproduces model or judge instructions. An \textit{example} box shows a concrete failure-mode instance, including the prompt, raw output, evaluator behavior, and consequence. A \textit{case} box presents a verified gold-label case grounded in authoritative sources. A \textit{code} box shows an executable harness artifact. Prompt, example, case, and code boxes are numbered by type.

\section{Evaluation Setup}
\label{app:c}

This section provides the implementation details needed to reproduce our
evaluation. We first define the scored benchmark subsets and task inventory.
We then describe the serving environment, prompt construction, and the
configuration records produced by the evaluation harness. Benchmark-specific
pipeline choices and their standardization are documented separately in
App.~\ref{app:a}.

\subsection{Evaluation Scale and Sampling}
\label{app:c1}

Table~\ref{tab:app-scale} distinguishes three dataset sizes for each benchmark.
\emph{Reported} is the number of questions stated by the benchmark paper or
release documentation for the corresponding evaluation scope.
\emph{Released} is the number of questions available in the public artifact.
\emph{Scored} is the subset evaluated in our audit. The scored subsets sum to
48{,}662 questions across 23 tasks, which is the evaluation scope reported in
the main presentation. Per-task counts are given in Table~\ref{tab:app-inventory}.

Only three benchmarks have matching reported and released sizes. Several of
the remaining differences arise from counting or release conventions rather
than from our sampling. SecEval reports an overall total of 2{,}126 in its
README, although its per-topic counts match the 2{,}189 distinct questions in
the released file. CyberMetric reports 10{,}000 questions, while its release
contains 10{,}180. CTI-Bench does not report a benchmark-wide total. Summing
the paper's per-task figures gives 4{,}947, partly because CTI-ATE is described
using 397 unique ATT\&CK techniques rather than its 60 evaluation questions.
For RedSage-Bench, the paper reports 30{,}240 questions, of which 240 are open ended Q\&A. The released artifact also contains a separate 50-question
validation split.

\begin{table}[t]
\centering
\scriptsize
\setlength{\tabcolsep}{4pt}
\renewcommand{\arraystretch}{1.18}
\begin{tabular}{@{}
  >{\raggedright\arraybackslash}p{0.70\columnwidth}
  >{\raggedright\arraybackslash}p{0.25\columnwidth}
@{}}
\toprule
\textbf{Artifact} & \textbf{Location} \\
\midrule

\multicolumn{2}{@{}l}{\textbf{Pipeline specification}} \\[1pt]
Original and standardized pipelines, by benchmark
    & App.~\ref{app:pipeline-table}; Table~\ref{tab:pipeline-summary} \\
Specification status of all 72 pipeline fields
    & App.~\ref{app:a-underspec}; Table~\ref{tab:spec-matrix} \\
Standardizations with defensible alternatives
    & App.~\ref{app:d2}, Table~\ref{tab:app-changes} \\
Field-level pipeline ledger and measured effects
    & App.~\ref{app:pipeline-table}, Table~\ref{tab:pipeline-ledger} \\
Harness configuration
    & App.~\ref{app:a} \\

\midrule

\multicolumn{2}{@{}l}{\textbf{Audit protocol}} \\[1pt]
Audit procedure
    & \S\ref{sec:methodology}; App.~\ref{app:b1} \\
Stage-by-benchmark audit coverage
    & App.~\ref{app:b5}; Table~\ref{tab:app-coverage} \\
Re-scoring, re-generation, and automation limits
    & App.~\ref{app:b2}; Table~\ref{tab:app-modes} \\

\midrule

\multicolumn{2}{@{}l}{\textbf{Evidence and results}} \\[1pt]
Evidence for each failure mode
    & App.~\ref{app:evidence} \\
Original and standardized scores
    & App.~\ref{app:k2}; Table~\ref{tab:before_after} \\
Gold-label audit and verified cases
    & Apps.~\ref{app:e2}--\ref{app:e3} \\
Benchmark sizes and task inventory
    & Tables~\ref{tab:app-scale},~\ref{tab:app-inventory} \\

\midrule

\multicolumn{2}{@{}l}{\textbf{Repository-only artifacts}} \\[1pt]
Raw generations, run logs, plots code, and notebooks
    & Audit \href{https://github.com/qcri/cyberbench-audit}{repository} \\
\bottomrule
\end{tabular}
\caption{Roadmap to the artifacts supporting the benchmark audit. The appendix contains the information needed to assess the reported pipeline decisions; larger execution artifacts are provided in the audit repository.}
\label{tab:app-map}
\end{table}

SecBench has the largest substantive difference between the reported and released data. The paper reports 47{,}910 questions, but only 3{,}000 are publicly released. These comprise 2{,}730 MCQs and 270 SAQs. Only 661 of the released MCQs are in English, and none of the SAQs are. We therefore evaluate the 661 English MCQs.

The scored set is also smaller than the release when a benchmark contains
tasks or splits outside our evaluation scope. For SECURE, we evaluate three of
its six tasks: MAET, CWET, and KCV. For CTI-Bench, we evaluate five tasks and
exclude the separate 1{,}000 question CTI-RCM-2021 split. For RedSage-Bench,
we use the 30{,}000 closed-form question test split and exclude the 240 open-ended questions and the 50-question validation split. For CyberMetric, we use the separately released 500-question set that its authors describe as human-validated. Finally, SECURE-CWET contains one all-empty row with no prompt; the harness drops this row before evaluation.

\begin{table}[t]
\centering
\scriptsize
\renewcommand{\arraystretch}{1.15}
\setlength{\tabcolsep}{5pt}
\begin{tabular}{@{}lrrr@{}}
\toprule
\textbf{Benchmark} &
\textbf{Reported} &
\textbf{Released} &
\textbf{Scored} \\
\midrule
MMLU-CS       & 100      & 100      & 100 \\
SecEval       & 2{,}126  & 2{,}189  & 2{,}189 \\
SECURE        & 4{,}068  & 4{,}068  & 2{,}502 \\
CTI-Bench     & 4{,}947 & 5{,}610 & 4{,}610 \\
AthenaBench   & 8{,}100  & 8{,}100  & 8{,}100 \\
CyberMetric   & 10{,}000 & 10{,}180 & 500 \\
RedSage-Bench & 30{,}240 & 30{,}290 & 30{,}000 \\
SecBench      & 47{,}910 & 3{,}000  & 661 \\
\midrule
\textbf{Total scored} & & & \textbf{48{,}662} \\
\bottomrule
\end{tabular}
\caption{
Benchmark sizes used to define the audit scope. Per-task scored sizes are given in Table~\ref{tab:app-inventory}.
}
\label{tab:app-scale}
\end{table}

\subsection{Task Inventory}
\label{app:d}

The audit covers 23 tasks across the eight benchmarks. Table~\ref{tab:app-inventory} gives the task identifier (ID) used in the harness, task-level metric, denominator policy, and number of scored questions for each task. These counts sum to the 48{,}662 questions, as reported in Table~\ref{tab:app-scale}.

\begin{table*}[t]
\centering
\scriptsize
\renewcommand{\arraystretch}{1.12}
\setlength{\tabcolsep}{5pt}
\begin{tabular}{@{}llllr@{}}
\toprule
\textbf{Benchmark} &
\textbf{Task ID} &
\textbf{Metric} &
\textbf{Denominator policy} &
\textbf{Questions} \\
\midrule

MMLU-CS
& \texttt{mmlu\_cs}
& Accuracy
& Correct / total
& 100 \\

\midrule

SecEval
& \texttt{seceval}
& Set-exact-match accuracy
& Correct / total
& 2{,}189 \\

\midrule

SECURE
& \texttt{secure\_maet}
& Accuracy
& Correct / total
& 1{,}072 \\
& \texttt{secure\_cwet}
& Accuracy
& Correct / total
& 964 \\
& \texttt{secure\_kcv}
& Accuracy
& Correct / total
& 466 \\

\midrule

CTI-Bench
& \texttt{cti\_mcq}
& Accuracy
& Correct / total
& 2{,}500 \\
& \texttt{cti\_rcm}
& CWE accuracy
& Correct / total
& 1{,}000 \\
& \texttt{cti\_vsp}
& CVSS MAD
& All questions
& 1{,}000 \\
& \texttt{cti\_ate}
& Accuracy
& Correct / total
& 60 \\
& \texttt{cti\_taa}
& Binary / partial
& All questions
& 50 \\

\midrule

AthenaBench
& \texttt{ckt}
& Accuracy
& Correct / total
& 3{,}000 \\
& \texttt{rms}
& Question-level set-F1
& All questions
& 500 \\
& \texttt{athena\_taa}
& Binary / partial
& All questions
& 100 \\
& \texttt{athena\_ate}
& Accuracy
& Correct / total
& 500 \\
& \texttt{athena\_rcm}
& CWE accuracy
& Correct / total
& 2{,}000 \\
& \texttt{athena\_vsp}
& Normalized CVSS score
& All questions
& 2{,}000 \\

\midrule

CyberMetric
& \texttt{cybermetric}
& Accuracy
& Correct / total
& 500 \\

\midrule

RedSage-Bench
& \texttt{frameworks}
& Accuracy
& Correct / total
& 5{,}000 \\
& \texttt{generals}
& Accuracy
& Correct / total
& 5{,}000 \\
& \texttt{skills}
& Accuracy
& Correct / total
& 10{,}000 \\
& \texttt{cli}
& Accuracy
& Correct / total
& 5{,}000 \\
& \texttt{kali}
& Accuracy
& Correct / total
& 5{,}000 \\

\midrule

SecBench
& \texttt{secbench}
& Set-exact-match accuracy
& Correct / total
& 661 \\

\midrule
\multicolumn{4}{@{}l}{\textbf{Total}}
& \textbf{48{,}662} \\
\bottomrule
\end{tabular}

\caption{
Inventory of the 23 scored tasks. Metric names describe the task-level quantities recorded by the harness. The standardized denominator retains attempted questions rather than dropping unparseable model outputs. For TAA, both binary and partial-credit variants are retained for the audit, while standardized model comparisons use strict binary scoring (App.~\ref{app:i2}). Benchmark-level summaries include only comparable, bounded, higher-is-better metrics; raw MAD, set-F1, and partial-credit attribution are excluded where appropriate (App.~\ref{app:pipeline-table}).
}
\label{tab:app-inventory}
\end{table*}

Several task abbreviations are used throughout the appendix. Root-Cause Mapping (RCM) maps a CVE description to a CWE ID. Vulnerability Severity Prediction (VSP) predicts a CVSS vector. Attack-Technique Extraction (ATE) extracts MITRE ATT\&CK technique IDs, while Threat-Actor Attribution (TAA) identifies the actor associated with a threat-intelligence description. Response and Mitigation Selection (RMS) maps scenarios to ATT\&CK mitigation IDs. SECURE's MAET and CWET tasks are multiple-choice tasks based on MITRE ATT\&CK and CWE, respectively, while KCV is a true-or-false task over CVE records. AthenaBench's CKT is a five-option cybersecurity knowledge test.

The denominator column in Table~\ref{tab:app-inventory} describes the standardized scoring rule. Empty or unparseable model answers remain in the evaluation population rather than being discarded. For accuracy tasks, these answers therefore count as incorrect. Task-specific handling for structured metrics such as VSP is described in App.~\ref{app:h2}. For attacker attribution, the harness retains the alternative scoring variants needed for the audit, while standardized model comparisons use the strict binary convention described in App.~\ref{app:i2}.

\subsection{LLM Serving Stack}
\label{app:c2}

We serve all eight open-weight models locally with vLLM in
\texttt{bfloat16} and without quantization. Each evaluation job uses four
NVIDIA H200 141\,GB GPUs, with tensor parallelism set to the number of visible
devices. We set \texttt{gpu\_memory\_utilization} to 0.90 by default and to
0.85 for Qwen3.6 and RedSage-Qwen3, which require additional memory headroom.
Table~\ref{tab:app-serving} reports the maximum context length used for each
checkpoint.

The standardized inference configuration uses greedy decoding with temperature 0 and top-$p$ 1.0. Local vLLM runs additionally use \texttt{min\_tokens=50} to avoid empty end-of-sequence completions. No backend stop sequence is used in the standardized configuration. Maximum output length is calibrated by task, with 1{,}024 tokens as the default. These settings describe the standardized configuration. When reproducing an original benchmark pipeline or conducting a controlled failure-mode analysis, we instead use the benchmark-specific setting being studied, as documented in App.~\ref{app:pipeline-table}.

Two analyses use sampled decoding. For SECURE, we additionally evaluate the
documented temperature of $0.7$ and pin seed 42. For the CyberMetric
decoding-drift analysis, $\mathcal{F}_3(\mathcal{I})$, we compare the greedy
control with the documented temperature of $1.0$ and top-$p$ of $0.9$
(App.~\ref{app:g3}). These sampled runs pin a seed; the main greedy runs do not
depend on one.

GPT-5.4 and Sonnet 4.6 are evaluated through Azure-hosted endpoints (i.e., REST APIs). GPT-5.4 uses Azure OpenAI chat completion API with \texttt{api-version=2024-12-01-preview}. Sonnet 4.6 uses an Azure Anthropic-messages passthrough with \texttt{anthropic-version=2023-06-01}. Both are evaluated at temperature 0. Provider-side content filtering, such as guardrails, is disabled through the deployment configuration for both models. This avoids failed requests caused solely by the cybersecurity content of benchmark questions and makes pipeline behavior more reproducible.

The local runtime uses CUDA~12.1, Python~3.10, PyTorch, \texttt{transformers}, and vLLM. Open-weight checkpoints were retrieved at their then-current Hugging Face revisions, or from a fixed local checkpoint where applicable. We record the resolved checkpoint commit hashes with the released artifacts so that the evaluated weights can be recovered. Local inference was run from April 19--20, 2026, with two additional task runs from May 4--5, 2026. Hosted inference on Azure was run on May 5, and judge calls were performed between April 21 and May 5, 2026.

\begin{table}[t]
\centering
\scriptsize
\renewcommand{\arraystretch}{1.12}
\setlength{\tabcolsep}{4pt}
\begin{tabular}{@{}lrr@{}}
\toprule
\textbf{Checkpoint} &
\textbf{Context (token)} &
\textbf{Memory (\%)} \\
\midrule
Qwen3.6-35B-A3B            & 32{,}768 & 85 \\
RedSage-Qwen3-8B-DPO       & 16{,}384 & 85 \\
Llama-3.3-70B-Instruct     & 8{,}192  & 90 \\
Llama-Primus-Nemotron-70B  & 8{,}192  & 90 \\
Foundation-Sec-8B-Instruct & 8{,}192  & 90 \\
gpt-oss-20b                & 8{,}192  & 90 \\
gemma-4-31B-it             & 4{,}096  & 90 \\
Llama-Primus-Merged        & Default   & 90 \\
\bottomrule
\end{tabular}
\caption{
vLLM serving configuration for open-weight models. \emph{Context} is the configured \texttt{max\_model\_len}. \emph{Memory} is \texttt{gpu\_memory\_utilization}.
}
\label{tab:app-serving}
\end{table}

\subsection{Prompt Templates}
\label{app:a-prompts}

Prompt construction depends on both the benchmark and the task. The standardized harness preserves the benchmark question and intended answer semantics while making prompt and chat formatting explicit. Single-select multiple-choice tasks use the template shown in \promptref{prompt:single-select-mcq}. Multi-select and structured-output tasks use task-specific output instructions instead. Benchmark-specific system prompts are retained where they define the task presentation. Representative system prompts for CTI-Bench and SecEval are shown in \promptref{prompt:cti-system} and \promptref{prompt:seceval-system}, respectively. MMLU-CS and SecEval also prepend fixed few-shot exemplar blocks when required by the corresponding pipeline. The complete benchmark-by-benchmark prompt provenance is given in App.~\ref{app:pipeline-table}.

\begin{promptbox}[label={prompt:single-select-mcq}]{Shared single-select MCQ user template}
\begin{lstlisting}[style=sayf,aboveskip=0pt,belowskip=0pt]
You are given multiple choice questions. Answer with the
option letter (A, B, C, D) from the given choices
directly.

Question: {question}
{choices}
Answer:
\end{lstlisting}
\end{promptbox}

\begin{promptbox}[label={prompt:cti-system}]{CTI-Bench system prompt}
\begin{lstlisting}[style=sayf,aboveskip=0pt,belowskip=0pt]
You are a cybersecurity expert specializing in
cyberthreat intelligence.
\end{lstlisting}
\end{promptbox}

\begin{promptbox}[label={prompt:seceval-system}]{SecEval system prompt}
\begin{lstlisting}[style=sayf,aboveskip=0pt,belowskip=0pt]
Below are multiple-choice questions concerning
cybersecurity. Please select the correct answers
and respond with the letters ABCD only.
\end{lstlisting}
\end{promptbox}

\subsection{Reproducible Configuration}
\label{app:c3}

Each run stores the configuration needed to interpret its reported score. This includes the relevant pipeline configuration fields defined in Apps.~\ref{app:a}, together with model, judge, backend, task, and schema provenance. In particular, the record preserves the decoding configuration, token budget, stop-sequence handling, extraction and scoring rules, denominator policy, and prompt configuration associated with the run. This allows each reported score to be traced back to the pipeline that produced it. A representative configuration and policy record is shown in \coderef{code:harness-config}. The complete benchmark-specific values and their provenance are given in App.~\ref{app:pipeline-table}.

\begin{codebox}[label={code:harness-config}]{Harness config and per-run policy record}
\begin{lstlisting}[style=sayf,language=python,aboveskip=0pt,belowskip=0pt]
# decoding / token budget
temperature = 0.0
top_p       = 1.0
max_tokens  = per-task-calibrated
min_tokens  = 50    # local vLLM only
answer_stop = None

# scoring policies
denominator_policy =
  "accuracy = correct / total over all attempted questions;
   unparseable/empty model answers count as incorrect;
   judge-API failures are excluded from both numerator
   and denominator."

think_handling =
  "strip <think>...</think> before judging;
   if an answer-level stop is configured, apply it
   after removing reasoning."

scoring =
  "llm-as-judge: one call performs answer extraction
   and returns a CORRECT/INCORRECT verdict."

# provenance stored with each result:
# model {name, provider, endpoint}
# judge {name, provider}
# tasks []
# backend
# run_timestamp
# schema_version
\end{lstlisting}
\end{codebox}

\begin{codebox}[label={code:sayf-quickstart}]{Installing and running the harness}
\begin{lstlisting}[style=sayf,language=bash,aboveskip=0pt,belowskip=0pt]
pip install sayf-eval

# inference + judge
sayf-eval run \
  --tasks mcq seceval vsp taa \
  --model  openai/gpt-4o \
  --judge  anthropic/claude-sonnet-4-6 \
  --output-dir outputs/run1

# local vLLM endpoint as the model under test
vllm serve Qwen/Qwen3-8B --port 8000 --enforce-eager
sayf-eval run --tasks mcq \
  --model hosted_vllm/Qwen/Qwen3-8B \
  --base-url http://localhost:8000/v1 \
  --api-key EMPTY \
  --output-dir outputs/qwen
\end{lstlisting}
\end{codebox}

\subsection{Software}
\label{app:c4}

The released harness, Sayf-Eval,\footnote{\url{https://github.com/qcri/sayf-eval}} is distributed via PyPI and can run inference and judging through a common command-line interface. Hosted models and locally served OpenAI-compatible endpoints use the same evaluation workflow. \coderef{code:sayf-quickstart} shows the basic workflow. These commands illustrate package usage and are not intended to reproduce the exact experimental configuration used in this paper. Exact run configurations and provenance are provided with the released audit repository.\footnote{\url{https://github.com/qcri/cyberbench-audit}}

\section{Pipeline Specification Gaps}
\label{app:a-underspec}

Across eight benchmarks and nine pipeline configuration fields, we inspect 72 field--benchmark pairs. We classify each pair by comparing the benchmark documentation with its released implementation. A field is \emph{undefined} when the documentation does not specify it, \emph{contradicted} when the documented and released layers do not define the same executable behavior, and \emph{matched} when they agree. Of the 72 pairs, 36 are undefined, 8 are contradicted, and only 28 are specified and matched. Thus, 44 of 72 pipeline configuration fields (61\%) cannot be reproduced from benchmark documentation alone. Table~\ref{tab:spec-matrix} gives the complete field-level breakdown.

This underspecification matters because an unspecified field must be supplied by the evaluator. Different choices can produce different prompts, generations, extracted answers, scores, or aggregations even when the benchmark questions and evaluated model are unchanged. The same ambiguity also affects our reconstruction: when neither the documentation nor released implementation uniquely determines a field, our selected value is an explicit evaluation choice rather than a uniquely correct setting. We record these choices in the \textit{pipeline ledger}, defined in Table~\ref{tab:pipeline-ledger}, instead of treating them as part of benchmark specification.

The eight contradicted fields illustrate the different ways in which the documented and released layers can diverge:
\begin{itemize}[leftmargin=1.3em,topsep=3pt,parsep=1pt,itemsep=1pt]
\item \textbf{CTI-Bench scoring rule:} the documentation describes exact-match scoring, while the released TAA scorer also credits alias-connected and related threat actors.
\item \textbf{AthenaBench scoring rule:} the VSP normalization constant $R$=7.7 appears in \texttt{config.yaml} but is absent in the paper, so the reported metric cannot be reconstructed from the paper alone.
\item \textbf{SECURE decoding:} the paper specifies temperature $T$=0.7, but no inference implementation is released to enact that setting.
\item \textbf{CyberMetric prompt template:} the prompt in the README differs from the prompt hard-coded in the released evaluator.
\item \textbf{CyberMetric decoding:} the paper specifies $T$=1.0, top-$p$~0.9, and top-$k$~50, while the released evaluator sets none of these parameters and therefore inherits backend defaults.
\item \textbf{RedSage-Bench chat formatting:} the two released inference examples use incompatible chat-formatting configurations.
\item \textbf{RedSage-Bench decoding:} the two released inference examples specify different temperatures.
\item \textbf{RedSage-Bench maximum output tokens:} the two released inference examples specify different token budgets.
\end{itemize}

\newcommand{\specMatch}{\cellcolor{green!8}\textcolor{green!45!black}{\scriptsize\ding{51}}}
\newcommand{\specUndefined}{\cellcolor{gray!7}\textcolor{black!30}{$\cdot$}}
\newcommand{\specConflict}{\cellcolor{red!10}\textcolor{red!45!black}{\scriptsize$\times$}}

\begin{table*}[t]
\centering
\scriptsize
\renewcommand{\arraystretch}{1.18}
\setlength{\tabcolsep}{4pt}

\begin{tabularx}{\textwidth}{@{}l*{9}{>{\centering\arraybackslash}X}>{\centering\arraybackslash}X@{}}
\toprule
& \multicolumn{2}{c}{\textbf{Prompt}}
& \multicolumn{3}{c}{\textbf{Inference}}
& \multicolumn{3}{c}{\textbf{Extraction}}
& \multicolumn{1}{c}{\textbf{Aggregation}}
& \\
\cmidrule(lr){2-3}
\cmidrule(lr){4-6}
\cmidrule(lr){7-9}
\cmidrule(lr){10-10}

\textbf{Benchmark}
& Template
& Chat
& Decoding
& Max tokens
& Stop
& Extractor
& Scoring
& Denominator
& Rule
& \textbf{Gaps} \\
\midrule

MMLU-CS
& \specMatch
& \specMatch
& \specMatch
& \specMatch
& \specUndefined
& \specMatch
& \specMatch
& \specMatch
& \specMatch
& 1 \\

SecEval
& \specUndefined
& \specUndefined
& \specUndefined
& \specUndefined
& \specUndefined
& \specMatch
& \specMatch
& \specUndefined
& \specMatch
& 6 \\

SECURE
& \specMatch
& \specUndefined
& \specConflict
& \specUndefined
& \specUndefined
& \specMatch
& \specMatch
& \specUndefined
& \specUndefined
& 6 \\

CTI-Bench
& \specMatch
& \specUndefined
& \specMatch
& \specUndefined
& \specUndefined
& \specMatch
& \specConflict
& \specUndefined
& \specMatch
& 5 \\

AthenaBench
& \specMatch
& \specUndefined
& \specMatch
& \specUndefined
& \specUndefined
& \specMatch
& \specConflict
& \specUndefined
& \specMatch
& 5 \\

CyberMetric
& \specConflict
& \specMatch
& \specConflict
& \specUndefined
& \specUndefined
& \specMatch
& \specMatch
& \specUndefined
& \specMatch
& 5 \\

RedSage-Bench
& \specUndefined
& \specConflict
& \specConflict
& \specConflict
& \specUndefined
& \specUndefined
& \specMatch
& \specUndefined
& \specUndefined
& 8 \\

SecBench
& \specUndefined
& \specUndefined
& \specUndefined
& \specUndefined
& \specUndefined
& \specUndefined
& \specMatch
& \specUndefined
& \specUndefined
& 8 \\
\bottomrule
\end{tabularx}

\caption{Specification status of the nine pipeline configuration fields across the eight benchmarks. A green check indicates that a field is specified in the documentation and \textit{matched} by the released implementation. A gray dot indicates an \textit{undefined} field, and a red cross indicates that the documented and released layers do not define the same executable behavior, so one has \textit{contradicted} the other. \emph{Gaps} counts undefined and contradicted fields. Overall, 44 of 72 fields (61\%) are gaps; stop sequences are undefined for all eight benchmarks.}
\label{tab:spec-matrix}
\end{table*}

The pipeline ledger also reports measured effects for fields that we perturb. These are \emph{isolated single-field counterfactuals}: we vary one pipeline configuration field on the same questions while holding the remaining fields fixed. Downstream changes, such as extraction or denominator policy, can be evaluated by re-scoring stored outputs. Upstream changes, such as prompts or inference settings, require re-generation (App.~\ref{app:b2}). These effects are therefore distinct from the original-to-standardized differences in Table~\ref{tab:before_after}, which reflect the combined change from the original pipeline to the standardized pipeline.

For each original pipeline, the ledger records the extractor used to reproduce its released behavior. When no extractor is released, we explicitly document the one we supply. The standardized pipeline instead uses the pinned LLM-based extraction and scoring rules described in App.~\ref{app:a-extract}. Some failures cannot be isolated with a controlled counterfactual because the original pipeline cannot be executed without the failure or because the effect depends on an interaction between model output style and the evaluator. In these cases, we report a clearly labeled \emph{peer-gap estimate}, defined as the median score of unaffected models minus the affected model's score. A ledger entry marked \emph{not isolated} indicates that the field was supplied or changed but was not independently ablated.

\section{Benchmark Pipelines}
\label{app:a}

This section documents how each benchmark pipeline is reconstructed and standardized. We distinguish three sources of information. The \emph{documented} layer is the behavior stated in the benchmark paper or documentation. The \emph{released} layer is the behavior implemented by the public artifact. When these layers are incomplete or inconsistent, we explicitly record the resolution used in our audit. We then identify the settings used by the standardized harness. This separation is important because a supplied or normalized setting is an evaluation choice, not necessarily a uniquely correct interpretation of the benchmark.

\subsection{Pipeline Summary}

Table~\ref{tab:pipeline-summary} summarizes the released artifacts and the resulting benchmark-level scores. The artifact columns indicate whether the corresponding component is released and usable as specified. These indicators describe artifact availability and consistency; they are not failure-mode observations. Binary failure-mode incidence is reported in Table~\ref{tab:failure_coverage}.

The original and standardized columns report the mean and range across the 10 evaluated models. Benchmark-level means include only comparable, bounded, higher-is-better task scores. We exclude CTI-Bench VSP because it reports raw MAD, CTI-Bench TAA because its ``correct+plausible'' score includes partial credit, and AthenaBench RMS because it reports set-F1. AthenaBench VSP is retained because its normalization, $\max(0,1-\mathrm{MAD}/7.7)\times100$, produces a bounded higher-is-better score.

\newcommand{\artifactYes}{\cellcolor{green!8}\textcolor{green!45!black}{\scriptsize\ding{51}}}
\newcommand{\artifactNo}{\cellcolor{red!8}\textcolor{red!45!black}{\scriptsize\ding{55}}}

\begin{table*}[t]
\centering
\scriptsize
\renewcommand{\arraystretch}{1.16}
\setlength{\tabcolsep}{3.5pt}
\begin{tabularx}{\textwidth}{@{}lccccr*{6}{>{\centering\arraybackslash}X}r@{}}
\toprule
& \multicolumn{4}{c}{\textbf{Released artifacts}} & & \multicolumn{3}{c}{\textbf{Original score (\%)}} & \multicolumn{3}{c}{\textbf{Standardized score (\%)}} & \\
\cmidrule(lr){2-5}
\cmidrule(lr){7-9}
\cmidrule(lr){10-12}
\textbf{Benchmark} & Prompt & Inference & Evaluator & Params & \textbf{Questions} & Min & Mean & Max & Min & Mean & Max & \textbf{$\Delta$(mean)} \\
\midrule
MMLU-CS       & \artifactYes & \artifactYes & \artifactYes & \artifactYes & 100      & 36.0 & 68.8 & 87.0 & 74.0 & 83.0 & 90.0 & +14.2 \\
SecEval       & \artifactYes & \artifactYes & \artifactYes & \artifactNo  & 2{,}189  & 0.3  & 39.3 & 78.0 & 57.0 & 71.4 & 82.0 & +32.1 \\
SECURE        & \artifactYes & \artifactNo  & \artifactNo  & \artifactNo  & 2{,}502  & 25.8 & 72.7 & 92.4 & 79.7 & 88.2 & 92.5 & +15.5 \\
CTI-Bench     & \artifactYes & \artifactNo  & \artifactYes & \artifactNo  & 4{,}610  & 27.3 & 45.7 & 55.8 & 37.5 & 47.0 & 55.4 & +1.3 \\
AthenaBench   & \artifactYes & \artifactYes & \artifactYes & \artifactNo  & 8{,}100  & 21.0 & 48.0 & 76.8 & 50.6 & 58.3 & 75.3 & +10.3 \\
CyberMetric   & \artifactYes & \artifactYes & \artifactYes & \artifactNo  & 500      & 1.0  & 59.4 & 96.8 & 85.2 & 92.3 & 96.2 & +32.9 \\
RedSage-Bench & \artifactNo  & \artifactYes & \artifactYes & \artifactYes & 30{,}000 & 25.5 & 75.2 & 90.9 & 75.1 & 84.6 & 91.1 & +9.4 \\
SecBench      & \artifactNo  & \artifactNo  & \artifactNo  & \artifactNo  & 661      & 33.9 & 59.8 & 85.2 & 66.3 & 81.9 & 89.7 & +22.1 \\
\bottomrule
\end{tabularx}
\caption{Original and standardized pipeline summary. A green check indicates that the artifact is released and usable as specified; a red cross indicates that it is absent, contradictory, or not executable as written. For each pipeline, \emph{min}, \emph{mean}, and \emph{max} summarize benchmark scores across the 10 evaluated models. $\Delta$(mean) is the standardized mean minus the original mean, in percentage points.}
\label{tab:pipeline-summary}
\end{table*}

\subsection{Standardization Choices}
\label{app:d2}

The standardized harness uses the common interface and configuration described in App.~\ref{app:c}. It records the prompt and inference configuration, raw model response, extracted prediction, question-level score, invalid-response status, and aggregate score for every run. Standardized extraction uses the pinned LLM-based policy described in App.~\ref{app:a-extract}. We reviewed all eight benchmarks and found that none defines output-format compliance as the capability being evaluated. The judge can therefore recognize semantically equivalent answers without changing the intended task. A benchmark that explicitly evaluated output-format compliance would require a different policy.

Most pipeline resolutions follow documented behavior or supply a missing mechanical setting. Four choices admit meaningful alternatives and are therefore judgment-dependent. Table~\ref{tab:app-changes} makes these choices explicit. They should not be interpreted as uniquely correct fixes. Instead, the measured differences show that comparative conclusions can depend on defensible evaluation conventions.

\newcommand{\ledgerWhite}{}
\newcommand{\ledgerGray}{\rowcolor{gray!8}}
\newcommand{\ledgerpad}{\rule{0pt}{2.6ex}}

\begin{table*}[t]
\centering
\scriptsize
\renewcommand{\arraystretch}{1.10}
\setlength{\extrarowheight}{0.8pt}
\setlength{\tabcolsep}{4pt}

\begin{tabularx}{\textwidth}{@{}>{\raggedright\arraybackslash}p{0.15\textwidth}>{\raggedright\arraybackslash}p{0.13\textwidth}>{\raggedright\arraybackslash}p{0.21\textwidth}>{\raggedright\arraybackslash}X@{}}
\toprule
\textbf{Standardized choice} & \textbf{Scope} & \textbf{Alternative} & \textbf{Measured effect} \\
\midrule

\ledgerWhite \ledgerpad Semantic LLM extraction & All benchmarks & Benchmark-specific released or reconstructed extractors & Released extraction rules differ by up to 79.7\,pp on identical RedSage-Bench generations. CTI-Bench VSP extractors disagree on 30.4\% of contested model--question pairs ($\mathcal{F}_1(\mathcal{E})$). \\

\ledgerGray \ledgerpad Correct-over-total denominator & CTI-Bench, SECURE, AthenaBench & Exclude unparseable predictions and score correct over valid & CTI-RCM changes from 100.0\% to 0.2\% for Gemma-4, corresponding to $500\times$ inflation under correct-over-valid. AthenaBench VSP shifts by up to 85.4\,pp ($\mathcal{F}_2(\mathcal{E})$). \\

\ledgerWhite \ledgerpad Strict binary attribution & CTI-Bench, AthenaBench & Award partial credit to related threat actors & Scoring rules differ by up to 70\,pp across the attribution tasks. In CTI-Bench, ``Correct+Plausible'' increases scores by up to 30\,pp ($\mathcal{F}_2(\mathcal{A})$). \\

\ledgerGray \ledgerpad Generative response scoring & MMLU-CS, RedSage-Bench & Rank answer choices using log probabilities & Differences reach 40.9\,pp and can favor either scoring method depending on the model ($\mathcal{F}_1(\mathcal{A})$). \\

\bottomrule
\end{tabularx}

\caption{Standardization choices for which a defensible alternative materially changes the measurement. The measured effects quantify sensitivity to these choices. They do not imply that one convention is uniquely correct.}
\label{tab:app-changes}
\end{table*}

\subsection{Prompt Sources}
\label{app:a1}

Prompt provenance differs substantially across benchmarks. CTI-Bench, AthenaBench, and SECURE provide task prompts with the released data. SecEval and CyberMetric encode their prompts in evaluation code. MMLU-CS relies on its established evaluation convention, while RedSage-Bench constructs its prompt in benchmark harness code rather than storing it with each question. SecBench releases the question data but no evaluation prompt, so we reconstruct one from the released fields.

Having a prompt somewhere in the release does not by itself pin the experiment. CyberMetric's README prompt differs from the evaluator prompt. RedSage-Bench also ships inference examples with conflicting chat-formatting and generation settings. As a result, following the documentation and executing the released artifact can produce different pipelines under the same benchmark name. Table~\ref{tab:spec-matrix} summarizes this distinction across all 72 pipeline configuration fields.

\subsection{Pipeline Ledger}
\label{app:pipeline-table}

Table~\ref{tab:pipeline-ledger} records the benchmark-specific pipeline resolutions. The \textit{Released} column describes executable behavior where an implementation exists. \textit{Resolution} records the setting used to reconstruct or standardize the pipeline. The \textit{Action} column uses four labels: \textit{retain} preserves released behavior, \textit{fix} changes behavior that prevents valid execution, \textit{normalize} applies a consistent evaluation convention, and \textit{supply} fills an unspecified field. Common settings already described in App.~\ref{app:c} are not repeated unless they resolve a benchmark-specific ambiguity or contribute to a measured effect.

When possible, the \textit{Measured effect} column reports an isolated single-field counterfactual on the same questions. Downstream changes can be evaluated by re-scoring stored responses, while upstream changes require re-generation (App.~\ref{app:b2}). \textit{Not isolated} means that the field was changed or supplied but was not independently ablated. A peer-gap estimate, marked with $^\dagger$, is used only when a controlled counterfactual is unavailable; it is not treated as a controlled ablation.

\newcommand{\actRetain}{Retain}
\newcommand{\actFix}{Fix}
\newcommand{\actNorm}{Normalize}
\newcommand{\actSupply}{Supply}
\newcommand{\ledgerbench}[1]{\midrule\multicolumn{5}{@{}l}{\textbf{#1}}\\*}

\onecolumn

\begingroup
\scriptsize
\setlength{\tabcolsep}{3pt}
\renewcommand{\arraystretch}{1.10}
\setlength{\extrarowheight}{0.8pt}
\setlength{\LTleft}{0pt}
\setlength{\LTright}{0pt}

\begin{longtable}{@{}>{\raggedright\arraybackslash}p{0.13\textwidth}>{\raggedright\arraybackslash}p{0.22\textwidth}>{\raggedright\arraybackslash}p{0.25\textwidth}>{\raggedright\arraybackslash}p{0.09\textwidth}>{\raggedright\arraybackslash}p{0.24\textwidth}@{}}

\caption{Benchmark pipeline ledger. \emph{Released} describes the behavior implemented by the public artifact. \emph{Resolution} records the choice used to reconstruct or standardize the pipeline. \emph{Action} indicates whether we retain released behavior, fix a broken setting, normalize a working but inconsistent convention, or supply an unspecified field. \emph{Measured effect} reports an isolated counterfactual where available; ``Not isolated'' indicates that the field was not independently ablated. The $^\dagger$ symbol denotes a peer-gap estimate rather than a controlled counterfactual.}
\label{tab:pipeline-ledger}\\

\toprule
\textbf{Field} & \textbf{Released} & \textbf{Resolution} & \textbf{Action} & \textbf{Measured effect} \\
\midrule
\endfirsthead

\multicolumn{5}{@{}l}{\footnotesize\itshape Pipeline ledger, continued}\\
\toprule
\textbf{Field} & \textbf{Released} & \textbf{Resolution} & \textbf{Action} & \textbf{Measured effect} \\
\midrule
\endhead

\addlinespace[2pt]
\multicolumn{5}{r@{}}{\footnotesize\itshape Continued on next page}\\
\endfoot

\endlastfoot

\multicolumn{5}{@{}l}{\textbf{MMLU-CS}}\\*
\ledgerWhite \ledgerpad Chat formatting & Official evaluation uses raw completion without a chat template. & Use each model's native chat template for the generative path; preserve raw completion for the logprob reproduction. & \actNorm & Not isolated. \\
\ledgerGray \ledgerpad Max output tokens & Official logprob evaluation uses \texttt{max\_tokens=1}. & Retain one token for logprob scoring and use a sufficient budget for generative scoring; reasoning prompt-mode runs use 4{,}096 tokens. & \actSupply & Not isolated. \\
\ledgerWhite \ledgerpad Extraction rule & The official logprob path generates no free-form answer and therefore has no response extractor. & Use the pinned LLM judge for the generative path. & \actSupply & Not isolated. \\
\ledgerGray \ledgerpad Scoring rule & Rank A--D using next-token log probability. & Use generative response scoring for standardized comparisons; retain logprob scoring as an audit alternative. & \actNorm & Qwen3.6 differs by 23.0\,pp: 57.0\% generative versus 80.0\% logprob. \\

\ledgerbench{SecEval}
\ledgerWhite \ledgerpad Decoding & Not specified; the evaluator inherits backend defaults. & Use $T{=}0$ and top-$p{=}1.0$. & \actSupply & Not isolated. \\
\ledgerGray \ledgerpad Max output tokens & The evaluator sets \texttt{max\_new\_tokens=5}. & Retain five tokens where supported; use the 16-token minimum on API backends that reject smaller values. & \actFix & GPT-5.4 changes from 0.3\% to 81.4\%, a +81.1\,pp shift. At five tokens, all 2{,}189 requests fail before producing valid model output. \\

\ledgerbench{SECURE}
\ledgerWhite \ledgerpad Chat formatting & No inference implementation is released. & Use each model's native chat template with no additional system prompt. & \actSupply & Not isolated. \\
\ledgerGray \ledgerpad Decoding & The paper specifies $T{=}0.7$, but no released implementation enacts it. & Enact $T{=}0.7$ with top-$p{=}1.0$; sampled runs pin seed 42. & \actSupply & Not isolated. \\
\ledgerWhite \ledgerpad Max output tokens & Not specified in a released implementation. & Use 1{,}024 tokens. & \actSupply & Not isolated. \\
\ledgerGray \ledgerpad Stop sequences & Not specified in a released implementation. & Use no stop sequence. & \actSupply & Not isolated. \\
\ledgerWhite \ledgerpad Extraction rule & No reference extractor is released. & Reconstruct the original path using a final-answer letter extractor, with True/False forms mapped to T/F; standardized scoring uses the pinned judge. & \actSupply & Not isolated. \\
\ledgerGray \ledgerpad Scoring rule & No executable scorer is released. & Use exact match for the reconstructed original path and the standardized judge verdict for standardized scoring. & \actSupply & Not isolated. \\
\ledgerWhite \ledgerpad Denominator policy & No implementation is released; reported scores are consistent with excluding invalid predictions. & Retain every attempted question and count unparseable model answers as incorrect. & \actNorm & Primus-Nemotron on CWET changes from 100.0\% correct-over-valid to 9.4\% correct-over-total, a 90.6\,pp difference. \\
\ledgerGray \ledgerpad Aggregation & No benchmark-level aggregation implementation is released. & Report task scores separately; do not introduce an additional cross-task mean. & \actSupply & Not isolated. \\

\ledgerbench{CTI-Bench}
\ledgerWhite \ledgerpad Chat formatting & Released notebooks target hosted chat APIs and do not define a local-model chat template. & Use each model's native chat template together with the CTI system prompt. & \actSupply & Not isolated. \\
\ledgerGray \ledgerpad Max output tokens & The released notebook uses \texttt{max\_tokens=2048}. & Retain 2{,}048 for MCQ, RCM, and VSP; use calibrated budgets of 8{,}192 for ATE and 4{,}096 for TAA. & \actNorm & Not isolated. \\
\ledgerWhite \ledgerpad Extraction rule & MCQ takes the final line; RCM and VSP take the last matching expression anywhere in the response, although their prompts request a final-line answer. & Reproduce the released rules for the original pipeline; use the pinned judge for standardized scoring. & \actNorm & On VSP, the final-line and anywhere extractors disagree on 30.4\% of model--question pairs for which at least one rule extracts a vector; disagreement reaches 96.5\% for Primus-Merged. \\
\ledgerGray \ledgerpad Scoring rule & TAA additionally awards Correct+Plausible credit to alias-connected or related actors. & Use strict binary scoring for standardized comparisons while recognizing true semantic aliases. Retain Correct+Plausible as an audit alternative. & \actNorm & Correct+Plausible raises scores by up to 30.0\,pp relative to strict scoring (Qwen3.6: 32.0\% to 62.0\%). \\
\ledgerWhite \ledgerpad Denominator policy & Unparseable predictions are excluded from the denominator on affected tasks. & Use correct-over-total and report invalid-response rates separately. & \actNorm & Gemma-4 on RCM changes from 100.0\% to 0.2\%, a 99.8\,pp difference and $500\times$ inflation under correct-over-valid. \\

\ledgerbench{AthenaBench}
\ledgerWhite \ledgerpad Decoding & Uses $T{=}0$; top-$p$ and no seed. & Retain $T{=}0$ and set top-$p{=}1.0$. & \actSupply & Not isolated. \\
\ledgerGray \ledgerpad Scoring rule & VSP uses $\max(0,1-\mathrm{MAD}/R)\times100$, with $R$ read from \texttt{config.yaml}. & Retain the released metric and make $R{=}7.7$ explicit whenever VSP is reported. & \actRetain & Relative to CTI-Bench's raw-MAD convention, the metric direction and scale shift model ranks by up to five positions. \\
\ledgerWhite \ledgerpad Denominator policy & Five tasks retain all questions; VSP excludes predictions whose CVSS vectors cannot be parsed. & For VSP, retain failed extractions and assign the maximum deviation of 10; other tasks are unchanged. & \actNorm & Gemma-4 changes from 85.4\% to 0.0\%; the median decrease across the 10 models is 55.9\,pp. \\

\ledgerbench{CyberMetric}
\ledgerWhite \ledgerpad Prompt template & The released evaluator contains the literal answer placeholder \texttt{ANSWER: X}; its prompt also differs from the README. & Use the evaluator prompt as the original reference, but replace the literal placeholder with an unambiguous answer-format instruction. & \actFix & 90.9\,pp peer-gap estimate$^\dagger$ for Foundation-Sec: 1.0\% versus a 91.9\% peer median. \\
\ledgerGray \ledgerpad Chat formatting & The evaluator targets a hosted chat API and defines no local-model path. & Use each model's native chat template while preserving the evaluator's system instruction. & \actSupply & Not isolated. \\
\ledgerWhite \ledgerpad Decoding & The evaluator sets no decoding parameters and inherits backend defaults, although the paper specifies $T{=}1.0$, top-$p{=}0.9$, and top-$k{=}50$. & Enact the documented sampling configuration for the corresponding analysis; use greedy decoding as the deterministic control. & \actFix & Primus-Merged changes from 17.2\% to 57.2\%, a 40.0\,pp shift; correctness changes on 274 of 500 questions. \\
\ledgerGray \ledgerpad Max output tokens & Not specified. & Use 1{,}024 tokens. & \actSupply & Not isolated. \\

\ledgerbench{RedSage-Bench}
\ledgerWhite \ledgerpad Prompt template & The prompt is constructed by \texttt{cybersec\_prompt\_fn}; its \texttt{include\_context} setting differs across call sites. & Use the released prompt layout with \texttt{include\_context=False}. & \actSupply & Not isolated. \\
\ledgerGray \ledgerpad Chat formatting & Released run configurations override chat-template handling. & Use each model's native chat template. & \actNorm & Not isolated. \\
\ledgerWhite \ledgerpad Decoding & The generative configuration uses $T{=}0$, with top-$p$ and seed unset. & Retain greedy decoding and set top-$p{=}1.0$; no sampling seed is required for the deterministic run. & \actNorm & Not isolated. \\
\ledgerGray \ledgerpad Max output tokens & The generative MCQ task uses \texttt{generation\_size=100}; the logprob path does not generate a response. & Use 2{,}048 tokens for the generative path so reasoning spans can complete before answer extraction. & \actNorm & Evaluated jointly with stop-sequence handling below. \\
\ledgerWhite \ledgerpad Stop sequences & The generative MCQ task uses \texttt{stop\_sequence=["\textbackslash n"]}. & Generate without a backend newline stop, remove the completed reasoning span, and apply answer handling only after reasoning is stripped. & \actFix & Qwen3.6's benchmark mean changes from 0.0\% to 85.9\%; per-task recovery ranges from 81.4 to 90.1\,pp. \\
\ledgerGray \ledgerpad Extraction rule & Three metrics are registered on the same generations: exact match, prefix exact match, and regex MCQ accuracy; none is designated canonical. & Use the pinned judge for standardized scoring and retain the released metrics for sensitivity analysis. & \actNorm & Exact versus prefix match differs by up to 79.7\,pp on identical generations (Primus-Merged: 0.3\% versus 80.0\%). \\
\ledgerWhite \ledgerpad Scoring rule & Both logprob and generative scoring are available. & Use generative response scoring for standardized comparisons; retain logprob scoring as an audit alternative. & \actNorm & Differences reach 40.9\,pp for Gemma-4; Qwen3.6 moves in the opposite direction by 26.7\,pp. \\

\ledgerbench{SecBench}
\ledgerWhite \ledgerpad Prompt template & No evaluation prompt is released. & Reconstruct the prompt from the released question and answer-choice fields, requiring only the selected option letter or letters. & \actSupply & Not isolated; no released prompt provides a controlled baseline. \\
\ledgerGray \ledgerpad Chat formatting & Not specified. & Use each model's native chat template with no additional system prompt. & \actSupply & Not isolated. \\
\ledgerWhite \ledgerpad Decoding & Not specified. & Use greedy decoding with $T{=}0$ and top-$p{=}1.0$. & \actSupply & Not isolated. \\
\ledgerGray \ledgerpad Max output tokens & Not specified. & Use 16 tokens for the short answer format. & \actSupply & Not isolated. \\
\ledgerWhite \ledgerpad Stop sequences & Not specified. & Use no stop sequence. & \actSupply & Not isolated. \\
\ledgerGray \ledgerpad Extraction rule & No extractor is released. & Reconstruct the original path by extracting standalone A--D letters and normalizing the selected set; standardized scoring uses the pinned judge. & \actSupply & Not isolated. \\
\ledgerWhite \ledgerpad Scoring rule & No executable scorer is released. & Use exact match on the normalized answer set. & \actSupply & Not isolated. \\
\ledgerGray \ledgerpad Denominator policy & Not specified. & Use correct-over-total; unparseable answers count as incorrect. & \actSupply & Not isolated. \\
\ledgerWhite \ledgerpad Aggregation & Not specified. & Report one accuracy over the 661 English multiple-choice questions. & \actSupply & Not isolated. \\

\bottomrule

\end{longtable}
\endgroup

\twocolumn

\section{Extraction and Judging}
\label{app:judge}

This section describes the standardized extraction and judging procedure used by the harness and evaluates its reliability. We first specify the pinned judge policy, then validate its output behavior and decisions through human and targeted stress-test analyses, and finally assess sensitivity to the choice of judge using an independent model.

\subsection{Judge Policy}
\label{app:a-extract}

The standardized scoring path uses a single LLM judge call to extract the model's stated answer and grade it. This reduces dependence on benchmark-specific surface-form extractors while preserving the task's intended answer semantics. For reference, \coderef{code:mcq-extractor} shows the deterministic multiple-choice extractor used in our surface-form comparisons. It prioritizes an explicit final answer and otherwise falls back through progressively weaker surface-form cues. Unlike the standardized judge, it cannot distinguish a committed answer from a letter or identifier mentioned only during reasoning. This distinction contributes to the extractor divergence measured under $\mathcal{F}_1(\mathcal{E})$.
\begin{codebox}[label={code:mcq-extractor}]{Deterministic MCQ extractor}
\begin{lstlisting}[style=sayf,language=python,aboveskip=0pt,belowskip=0pt]
# for task_type=="mcq"; else return last line
extract_mcq(response):
  lines = nonempty_lines(response)

  # explicit line, "**Answer:** X" then "Answer: X"
  # case-insensitive, scan last->first
  for pat in [r"\*\*(?:Final )?Answer:\*\*\s*([A-D])",
              r"(?:Final )?Answer:\s*([A-D])"]:
    for line in reversed(lines):
      if m = search(pat, line, IGNORECASE):
        return upper(m)

  # a line that is exactly one letter A-D
  for line in reversed(lines):
    if upper(line) in {A,B,C,D}:
      return upper(line)

  # a line starting with a letter not followed by ")"
  for line in reversed(lines):
    if upper(line[0]) in {A..D} and line[1] != ")":
      return upper(line[0])

  # any A-D token, case-insensitive, scan last->first
  for line in reversed(lines):
    if m = search(r"\b([A-D])\b", upper(line)):
      return m

  return last_line
\end{lstlisting}
\end{codebox}
The pinned judge is GPT-5.4 served through Azure OpenAI at temperature~0. Before judging, reasoning enclosed in \texttt{<think>} tags is removed. A judge-side failure, such as an API error or empty judge response, excludes the corresponding question from both numerator and denominator. By contrast, an empty or unparseable \emph{model} answer remains in the denominator and is scored incorrect. No judge-side failures occurred in the evaluation.

The judge prompt is shown in \promptref{prompt:judge-policy}. Each task supplies a \texttt{format\_hint} and \texttt{compare\_rule} that define the expected answer representation and grading rule. Examples include a single multiple-choice letter, a normalized \texttt{CVSS:3.1/} vector for VSP, and a sorted set of parent ATT\&CK technique identifiers for ATE. The judge must first extract what the model actually states and then issue a binary verdict. It is explicitly instructed not to infer or repair an answer.
\begin{promptbox}[label={prompt:judge-policy}]{Pinned judge prompt}
\begin{lstlisting}[style=sayf,aboveskip=0pt,belowskip=0pt]
You are a strict evaluator for cybersecurity benchmarks.
Your role has TWO steps.

{question_block}
Correct Answer: {ground_truth}
Model's Response:
{model_answer}

STEP 1 - EXTRACT the model's answer.
- Look for the model's explicit final answer (e.g.,
  "Answer:", concluding line, single bolded line).
- IGNORE letters/IDs/text that appear only inside
  explanations of other options or thinking-aloud prose.
- If the model gives multiple inconsistent answers, use
  its most prominent/final selection.
- DO NOT correct, infer, or improve the answer - extract
  verbatim what it actually said.
- Format the extracted answer as: {format_hint}

STEP 2 - VERDICT.
{compare_rule}

Output ONLY this JSON, nothing else:
{
  "extracted_answer": "<your formatted extraction>",
  "verdict": "CORRECT" or "INCORRECT",
  "justification": "<one short sentence>"
}
\end{lstlisting}
\end{promptbox}

\subsection{Judge Validation}
\label{app:a-judgestats}

We audit the pinned judge at three levels: output-format compliance, human validation, and targeted stress testing. Across all 48{,}662 questions and 10 evaluated models, the judge makes 486{,}620 grading calls. Re-parsing every raw judge response shows near-perfect adherence to the requested schema. JSON conformance is 100.0\% for multiple-choice calls and 99.99\% for open-ended calls; fewer than 0.01\% require the tolerant parsing fallback, and every successfully parsed verdict is either \texttt{CORRECT} or \texttt{INCORRECT}. Table~\ref{tab:judge-behavior} summarizes extraction behavior by answer type.

Multiple-choice outputs are almost always directly extractable. Open-ended tasks are more difficult because models can mention several candidate identifiers or entities without committing to one. This occurs particularly in free-form ID-set and attribution tasks. The judge is instructed not to synthesize an answer from such transient mentions. \exampleref{example:judge-none} illustrates this behavior: the model mentions the correct CWE during its reasoning but later contradicts itself and never commits to a final CWE, so the judge correctly records \texttt{NONE} rather than crediting the earlier mention.

\evbox{example:judge-none}
{No committed final answer\\(CTI-Bench RCM, Llama-3.3)}
{Map the CVE to a CWE; The last line of your response must contain only the CWE ID.\\Correct answer: CWE-79.}
{Opens correctly, quoting ``\dots a classic example of a CWE-79 \dots Cross-site Scripting \dots'', then degenerates into a contradictory loop (``\dots the closest match is actually CWE-184, no, \dots CWE-707, no \dots'') and never emits a final CWE line.}
{Judge JSON: \texttt{extracted\_answer}=``NONE'', \texttt{verdict}=INCORRECT, \texttt{justification}=``never provides a clear final CWE selection \dots unresolved contradictory reasoning.''}
{The correct answer appears during reasoning but is never committed as the final answer. The judge therefore records \texttt{NONE}, and the question is scored incorrect.}

\begin{table}[t]
\centering
\scriptsize
\renewcommand{\arraystretch}{1.12}
\setlength{\tabcolsep}{5pt}
\begin{tabular}{@{}lrrrr@{}}
\toprule
\textbf{Answer type} & \textbf{Calls} & \textbf{JSON (\%)} & \textbf{Extract (\%)} & \textbf{Format (\%)} \\
\midrule
Multiple-choice & 414{,}520 & 100.00 & 99.98 & 99.92 \\
Open-ended & 72{,}100 & 99.99 & 97.87 & 97.87 \\
\bottomrule
\end{tabular}
\caption{Pinned-judge behavior by answer type over 486{,}620 grading calls. \textit{Calls} is the number of grading calls made by the judge. \textit{JSON} is the rate of valid three-field JSON responses, \textit{Extract} is the rate of non-empty extracted answers, and \textit{Format} is the rate at which the extraction satisfies the task's expected answer format.}
\label{tab:judge-behavior}
\end{table}

We next manually validate a random stratified sample of 80 judge decisions, 10 from each benchmark and distributed across its tasks so that structured and open-ended outputs are represented. Human verification agrees with the judge's extraction and verdict on all 80 questions, including 64 multiple-choice and 16 open-ended cases.

Because a random sample contains relatively few difficult extractions, we additionally target two hard strata: the 1{,}733 calls for which the pinned judge returns \texttt{NONE}, and the 50{,}456 multiple-choice-family calls for which its verdict differs from a deterministic final-answer regex on the same response. On the \texttt{NONE} stratum, the independent Sonnet 4.6 judge corroborates the absence of a committed answer on 94.6\% of calls. On the regex-disagreement stratum, the two judges agree on 98.1\% of verdicts, with Cohen's $\kappa$=0.83. Agreement across the two hard strata is 98.0\%, compared with 99.6\% over the full independent-judge evaluation given in App.~\ref{app:a-judgeindep}.

\begin{table}[t]
\centering
\scriptsize
\renewcommand{\arraystretch}{1.12}
\setlength{\tabcolsep}{4pt}
\begin{tabular}{@{}lcccc@{}}
\toprule
\textbf{Evaluated model} & \textbf{Calls} & \textbf{Agree (\%)} & \textbf{$\kappa$} & \textbf{$\Delta$ (pp)} \\
\midrule
GPT-5.4 & 48{,}612 & 99.68 & 0.988 & +0.22 \\
Gemma-4 & 48{,}612 & 99.44 & 0.981 & $-0.18$ \\
Qwen3.6 & 48{,}612 & 99.60 & 0.988 & $-0.19$ \\
Llama-3.3 & 48{,}612 & 99.68 & 0.991 & $-0.21$ \\
GPT-OSS & 48{,}612 & 98.43 & 0.957 & +0.20 \\
Primus-Nemotron & 48{,}612 & 99.81 & 0.995 & $-0.10$ \\
Primus-Merged & 48{,}612 & 99.93 & 0.998 & $-0.00$ \\
Foundation-Sec & 48{,}612 & 99.84 & 0.996 & +0.08 \\
RedSage-Qwen3 & 48{,}612 & 99.91 & 0.997 & $-0.01$ \\
\midrule
All & 437{,}508 & 99.59 & 0.989 & $-0.02$ \\
\bottomrule
\end{tabular}
\caption{Agreement between the pinned judge (GPT-5.4) and the independent judge (Sonnet 4.6) on identical stored model outputs. \emph{Agree} is verdict agreement. $\kappa$ is Cohen's kappa. $\Delta$ is the evaluated model's accuracy under the pinned judge minus its accuracy under the independent judge. The \emph{All} row pools all 437{,}508 verdict pairs when computing agreement and Cohen's $\kappa$; $\Delta$ is the population-weighted mean score difference across models, which equals the unweighted mean here because each model contributes 48{,}612 calls.}
\label{tab:judge-independence}
\end{table}

We also manually inspect an adversarial sample of 120 questions, split evenly between these two hard strata. Two annotators independently label every question using a written guideline and adjudication procedure released with the artifacts. Inter-annotator agreement is Cohen's $\kappa$=0.93, with 99.2\% raw agreement. After adjudication, the pinned judge is correct on 113 of 120 questions (94.2\%; Wilson 95\% CI $[88.4,97.1]$). The seven raw judge errors include both over-crediting an uncommitted answer and missing a committed answer obscured by model-output artifacts. Two correspond to the known SECURE true-or-false format-hint mismatch and are corrected before benchmark scores are computed.

\subsection{Independent Judge}
\label{app:a-judgeindep}

Because GPT-5.4 is both the pinned judge and one of the evaluated models, we test whether the choice of judge materially favors it or changes model comparisons. Sonnet 4.6 independently re-grades the stored outputs of the other nine evaluated models on 22 of the 23 tasks, excluding CTI-Bench attacker attribution. Model outputs are held fixed; only the judge changes. This produces 48{,}612 re-grading calls per model and 437{,}508 calls in total.

As listed in Table~\ref{tab:judge-independence}, the judges agree on 99.6\% of verdicts, with Cohen's $\kappa$=0.99. Per-model score differences are also small. GPT-5.4 scores 0.22\,pp higher under the pinned judge than under Sonnet 4.6, the largest absolute difference among the nine re-graded models. The population-weighted mean difference is -0.02\,pp. Recomputing the nine-model ranking under the independent judge yields Spearman $\rho$=0.98, with GPT-5.4 ranked first under both judges. Agreement falls below $\kappa$=0.9 on only two tasks: CyberMetric ($\kappa$=0.83) and SECURE-KCV ($\kappa$=0.73). The SECURE-KCV disagreement is driven by a true-or-false vs. letter format-hint mismatch in the pinned judge's raw outputs. This mismatch is detected and corrected before score computation, so it does not affect the reported SECURE-KCV scores.

\section{Audit Protocol}
\label{app:b}

This section details the audit procedure, evidence types, the affected models, benchmarks, or questions, and stage coverage for the 15 failure modes.

\begin{table*}[t]
\centering
\scriptsize
\setlength{\tabcolsep}{4pt}
\renewcommand{\arraystretch}{1.12}
\begin{tabularx}{\textwidth}{@{}l>{\raggedright\arraybackslash}p{0.19\textwidth}>{\raggedright\arraybackslash}p{0.15\textwidth}>{\centering\arraybackslash}p{0.07\textwidth}>{\raggedright\arraybackslash}X@{}}
\toprule
\textbf{ID} & \textbf{Failure mode} & \textbf{Analysis} & \textbf{Held} & \textbf{Effect evidence} \\
\midrule
\rule{0pt}{2.8ex}$\mathcal{F}_1(\mathcal{D})$ & Limited capability coverage & Dataset audit & --- & Distribution of question types within each benchmark. \\
\rowcolor{gray!8}
\rule{0pt}{2.8ex}$\mathcal{F}_2(\mathcal{D})$ & Gold-label correctness & Label audit & --- & Search-grounded verification of disagreement flags. \\
\midrule
\rule{0pt}{2.8ex}$\mathcal{F}_1(\mathcal{P})$ & Format-token leakage & Output audit / peer gap & Yes & Invalid-response behavior and peer gap; standardized scoring requires re-generation with the corrected prompt. \\
\rowcolor{gray!8}
\rule{0pt}{2.8ex}$\mathcal{F}_2(\mathcal{P})$ & Prompt--question conflict & Output audit & Yes & Single-letter response rate on questions whose gold answer requires multiple selections. \\
\rule{0pt}{2.8ex}$\mathcal{F}_3(\mathcal{P})$ & Template incompatibility & Output audit / peer gap & Yes & Output-format behavior and peer-gap estimate. \\
\midrule
\rowcolor{gray!8}
\rule{0pt}{2.8ex}$\mathcal{F}_1(\mathcal{I})$ & Stop-sequence mismatch & Re-generation & No & Paired runs with and without the conflicting stop sequence. \\
\rule{0pt}{2.8ex}$\mathcal{F}_2(\mathcal{I})$ & Token-budget filter & Re-generation & No & Runs under the rejected and valid token budgets. \\
\rowcolor{gray!8}
\rule{0pt}{2.8ex}$\mathcal{F}_3(\mathcal{I})$ & Decoding drift & Re-generation & No & Paired runs under released/default and documented decoding settings. \\
\midrule
\rule{0pt}{2.8ex}$\mathcal{F}_1(\mathcal{E})$ & Extractor divergence & Re-scoring & Yes & Alternative extractors applied to identical responses. \\
\rowcolor{gray!8}
\rule{0pt}{2.8ex}$\mathcal{F}_2(\mathcal{E})$ & Denominator inflation & Re-scoring & Yes & Correct-over-valid and correct-over-total applied to identical predictions. \\
\rule{0pt}{2.8ex}$\mathcal{F}_3(\mathcal{E})$ & Metric-direction mismatch & Re-scoring & Yes & Identical task results ranked under opposite metric directions. \\
\rowcolor{gray!8}
\rule{0pt}{2.8ex}$\mathcal{F}_4(\mathcal{E})$ & Prompt-mode sensitivity & Re-generation & No & Same questions evaluated under zero-shot, few-shot, and CoT prompts. \\
\midrule
\rule{0pt}{2.8ex}$\mathcal{F}_1(\mathcal{A})$ & Logprob vs.\ generative scoring & Paired evaluation & No & Same questions evaluated through logprob and generative scoring paths. \\
\rowcolor{gray!8}
\rule{0pt}{2.8ex}$\mathcal{F}_2(\mathcal{A})$ & Task-level metric drift & Re-scoring & Yes & Alternative credit rules applied to identical extracted predictions. \\
\rule{0pt}{2.8ex}$\mathcal{F}_3(\mathcal{A})$ & Aggregation inconsistency & Re-scoring & Yes & Alternative denominator and aggregation conventions applied to the corresponding question-level results. \\
\bottomrule
\end{tabularx}
\caption{Primary analysis used to characterize each failure mode. \emph{Held} indicates whether model outputs are held fixed, so the comparison operates on identical stored model responses. ``---'' indicates the analysis does not involve model outputs. Failure incidence is determined independently by the binary inspection reported in Table~\ref{tab:failure_coverage}.}
\label{tab:app-modes}
\end{table*}

\begin{table}[t]
\centering
\scriptsize
\setlength{\tabcolsep}{3pt}
\renewcommand{\arraystretch}{1.12}
\begin{tabularx}{\columnwidth}{@{}l>{\raggedright\arraybackslash}X>{\raggedright\arraybackslash}p{0.28\columnwidth}@{}}
\toprule
\textbf{Mode} & \textbf{Witness} & \textbf{Reference} \\
\midrule
\rule{0pt}{2.8ex}$\mathcal{F}_1(\mathcal{D})$ & Question-type distribution & Fig.~\ref{fig:coverage}; App.~\ref{app:e1} \\
\rowcolor{gray!8}
\rule{0pt}{2.8ex}$\mathcal{F}_2(\mathcal{D})$ & Search-grounded label verification and verified cases & App.~\ref{app:e2}; \caserange{case:redsage-gold}{case:seceval-gold} \\
\midrule
\rule{0pt}{2.8ex}$\mathcal{F}_1(\mathcal{P})$ & Two CyberMetric responses & App.~\ref{app:f2}; \examplerange{example:p1a-cybermetric-foundation}{example:p1b-cybermetric-gemma} \\
\rowcolor{gray!8}
\rule{0pt}{2.8ex}$\mathcal{F}_2(\mathcal{P})$ & SecEval multi-select responses & Table~\ref{tab:ablation-multianswer}; \exampleref{example:p2-seceval-gemma} \\
\rule{0pt}{2.8ex}$\mathcal{F}_3(\mathcal{P})$ & SECURE template response & App.~\ref{app:f4}; \exampleref{example:p3-secure-maet-primus} \\
\midrule
\rowcolor{gray!8}
\rule{0pt}{2.8ex}$\mathcal{F}_1(\mathcal{I})$ & RedSage stop-sequence ablation & Table~\ref{tab:ablation-stopseq}; \exampleref{example:i1-redsage-qwen} \\
\rule{0pt}{2.8ex}$\mathcal{F}_2(\mathcal{I})$ & SecEval API rejection & App.~\ref{app:g2}; \exampleref{example:i2-seceval-gpt54} \\
\rowcolor{gray!8}
\rule{0pt}{2.8ex}$\mathcal{F}_3(\mathcal{I})$ & CyberMetric decoding ablation & Table~\ref{tab:ablation-decoding}; \exampleref{example:i3-cybermetric-primus} \\
\midrule
\rule{0pt}{2.8ex}$\mathcal{F}_1(\mathcal{E})$ & Same-generation extractor comparisons & Tables~\ref{tab:extractor_disagreement},~\ref{tab:ablation-redsage-extract}; \exampleref{example:e1-cti-vsp-primus} \\
\rowcolor{gray!8}
\rule{0pt}{2.8ex}$\mathcal{F}_2(\mathcal{E})$ & Alternative denominator policies & Table~\ref{tab:ablation-denominator}; \exampleref{example:e2-cti-rcm-gemma} \\
\rule{0pt}{2.8ex}$\mathcal{F}_3(\mathcal{E})$ & Same predictions under opposite metric directions & App.~\ref{app:h3} \\
\rowcolor{gray!8}
\rule{0pt}{2.8ex}$\mathcal{F}_4(\mathcal{E})$ & Three prompt modes on the same questions & App.~\ref{app:h4} \\
\midrule
\rule{0pt}{2.8ex}$\mathcal{F}_1(\mathcal{A})$ & Logprob and generative scoring & Table~\ref{tab:ablation-logprob}; App.~\ref{app:i1} \\
\rowcolor{gray!8}
\rule{0pt}{2.8ex}$\mathcal{F}_2(\mathcal{A})$ & Alternative attacker-attribution credit rules & Table~\ref{tab:ablation-taadrift}; \exampleref{example:a2-cti-taa-llama} \\
\rule{0pt}{2.8ex}$\mathcal{F}_3(\mathcal{A})$ & Cross-benchmark aggregation comparison & App.~\ref{app:i3}; Table~\ref{tab:before_after} \\
\bottomrule
\end{tabularx}
\caption{Evidence index for the 15 failure modes. Question-level examples provide illustrative witnesses, while tables and appendix sections report the corresponding aggregate or comparative evidence.}
\label{tab:app-evidence}
\end{table}

\subsection{Procedure}
\label{app:b1}

The meta-evaluation methodology defined in~\S\ref{sec:methodology} describes the audit protocol and treats controlled perturbation as opportunistic. Here we clarify how the audit was executed and how we handle cases where an effect cannot be isolated. The authors performed the audit. The harness automates configuration logging, output collection, parsing diagnostics, re-scoring, and aggregate comparisons. Inspection of benchmark documentation and released implementations, anomaly tracing, and interpretation of observed discrepancies combine automated diagnostics with manual review. Gold-label auditing additionally uses the search-grounded verification procedure in App.~\ref{app:e2}.

Failure incidence and effect estimation are separate. We inspect every benchmark for every failure mode, yielding 120 benchmark--failure interactions. Each interaction is recorded as observed or not observed, independent of whether a controlled perturbation is possible. The resulting binary incidence matrix is reported in Table~\ref{tab:failure_coverage}. After a failure is identified, we estimate its effect when the pipeline admits a meaningful comparison. A controlled perturbation may be unavailable because the original configuration cannot be reconstructed, the relevant setting is not exposed, or isolating the field would require generations that were not run. In such cases, we report the available inspection evidence or a clearly identified peer-gap estimate rather than treating it as a controlled ablation.

\subsection{Perturbations}
\label{app:b2}

The audit uses four forms of evidence. \emph{Dataset audits} inspect benchmark questions or labels without perturbing model outputs. \emph{Re-scoring} applies an alternative extraction, scoring, denominator, metric, or aggregation rule to stored responses, holding the model generations fixed. \emph{Re-generation} changes a prompt or inference configuration and generates new responses for the same questions. Finally, \emph{paired evaluation} compares evaluation interfaces that cannot be reduced to re-scoring identical generated text, such as logprob and generative multiple-choice scoring.

The distinction matters for reproducibility. Re-scoring is deterministic given the stored responses and evaluation configuration. Re-generation reproduces the pinned experimental configuration rather than guaranteeing byte-identical outputs. For open-weight models, we pin checkpoint revisions and the serving stack. For GPT-5.4 and Sonnet 4.6, we additionally record the hosted model identifier and run timestamp because a hosted endpoint may change while retaining the same API-facing name. Table~\ref{tab:app-modes} records the primary evidence used for each failure mode and whether the corresponding comparison holds model outputs fixed.

\subsection{Affected Units}
\label{app:b4}

Failure incidence is defined at the benchmark--failure level, whereas the effect in Table~\ref{tab:failure_modes} can have different affected units. Dataset-stage failures are summarized across benchmarks. Most prompt, inference, extraction, and scoring failures are summarized across affected models. Aggregation failures can instead be benchmark-level. So, \textit{affected units} should be read as the population over which the reported effect is observed and summarized, not as a second failure-incidence matrix. Per-model outputs and logs are retained in the audit repository.

\subsection{Evidence Index}
\label{app:b6}

Table~\ref{tab:app-evidence} indexes the evidence supporting each failure mode. Nine modes admit question-level examples showing the prompt, raw model output, extraction behavior, and scoring consequence; $\mathcal{F}_1(\mathcal{P})$ has two such examples. These boxes are illustrative witnesses rather than the basis for the aggregate effect estimates. Other modes are inherently distributional or comparative: capability coverage and label correctness require dataset-level evidence, metric direction and aggregation operate over collections of predictions, and prompt-mode and logprob sensitivity require paired evaluations. We therefore report each failure at the level at which its mechanism can be demonstrated.

\subsection{Stage Coverage}
\label{app:b5}

Table~\ref{tab:app-coverage} aggregates the binary failure-incidence matrix by pipeline stage. Every benchmark is inspected for every failure mode associated with each stage. Consequently, all 40 benchmark--stage combinations are covered, and the denominators sum to the 120 benchmark--failure interactions in Table~\ref{tab:failure_coverage}. The numerators sum to the 34 observed interactions. An unobserved interaction therefore means that the failure was inspected but not found; it does not mean that the corresponding stage was omitted from the audit.

\begin{table}[t]
\centering
\scriptsize
\setlength{\tabcolsep}{3pt}
\renewcommand{\arraystretch}{1.12}
\begin{tabular}{@{}lccccc@{}}
\toprule
\textbf{Benchmark} & $\boldsymbol{\mathcal{D}}$ & $\boldsymbol{\mathcal{P}}$ & $\boldsymbol{\mathcal{I}}$ & $\boldsymbol{\mathcal{E}}$ & $\boldsymbol{\mathcal{A}}$ \\
\midrule
MMLU-CS       & 2/2 & 0/3 & 0/3 & 0/4 & 1/3 \\
SecEval       & 1/2 & 1/3 & 1/3 & 0/4 & 1/3 \\
SECURE        & 1/2 & 1/3 & 0/3 & 1/4 & 0/3 \\
CTI-Bench     & 1/2 & 1/3 & 1/3 & 3/4 & 2/3 \\
AthenaBench   & 2/2 & 0/3 & 0/3 & 3/4 & 1/3 \\
CyberMetric   & 1/2 & 1/3 & 1/3 & 0/4 & 0/3 \\
RedSage-Bench & 2/2 & 0/3 & 1/3 & 1/4 & 1/3 \\
SecBench      & 2/2 & 0/3 & 0/3 & 0/4 & 0/3 \\
\midrule
\textbf{Total} & 12/16 & 4/24 & 4/24 & 8/32 & 6/24 \\
\bottomrule
\end{tabular}
\caption{Stage-level audit coverage. Each cell reports observed failure interactions over failure modes examined at that stage. All 120 benchmark--failure interactions were inspected; 34 were observed.}
\label{tab:app-coverage}
\end{table}

\begin{figure*}[t]
\begin{promptbox}[label={prompt:knowledge-analytical}]{Knowledge-vs.-analytical classifier}
\begin{lstlisting}[style=sayf,aboveskip=0pt,belowskip=0pt]
SYSTEM
You are a careful annotator classifying cybersecurity benchmark questions into one of two classes:

K = Knowledge-oriented. Answered by retrieving a fact, definition, standard parameter, or procedural step from cybersecurity domain knowledge. Single-step factual recall. Surface form is usually a short MCQ on definitions, properties, configuration values, standards, or named entities.

A = Analytical. Requires reasoning over a scenario, structured extraction from text, multi-step inference, scoring, or attribution. Surface form usually involves a free-form description that the answerer must analyze (CVSS vector composition, MITRE ATT&CK technique extraction, CWE root-cause classification, threat-actor attribution).

Decision rules:
- If a single fact or definition produces the answer, choose K.
- If the answer requires reading a scenario / report / description and inferring or extracting structured information from it, choose A.
- Multi-choice format alone does not make a question K - what matters is whether the *answer* requires recall or reasoning.
- Choose A if the question presents a non-trivial input (code, log, CVE description, attack narrative) that must be analyzed before answering.

Always reply with strict JSON in a fenced ```json``` block, with exactly these keys: "class" (K or A), "confidence" (low/medium/high), "rationale" (one short sentence).

USER
Here are six worked examples; classify the final question in the same format.

Question: Which of the following is a desirable property of a biometric system?
Options: A) Permanent, B) Transferability, C) Uniformity, D) Forgiveness
```json {"class": "K", "confidence": "high", "rationale": "Single-step factual recall about biometric-system design properties."} ```

[Five additional worked examples omitted here; included in the released artifact.]

Question: {question}
\end{lstlisting}
\end{promptbox}
\end{figure*}

\subsection{Reading the Failure Table}
\label{app:b7}

Table~\ref{tab:failure_modes} reports four quantities for each failure. \emph{Impact measure} defines the quantity being summarized. \emph{Affected units} identifies the population over which the effect is observed, such as models or benchmarks. \emph{Maximum effect} is the largest observed value over that population, and \emph{Median effect} is its median when enough affected units exist to summarize a distribution. A dash indicates that a meaningful median is unavailable. The two dataset-stage rows instead report a single aggregate quantity spanning the final two columns.

The impact measures are interpreted as follows. \emph{Dominant question-type share} is the fraction of sampled questions assigned to the most common capability type. \emph{Flag precision} is the fraction of checked disagreement flags confirmed as gold-label errors. \emph{Peer score gap} is the score difference between an affected model and the median of unaffected peers under the same benchmark condition. \emph{Single-letter response rate} is the fraction of multi-answer questions for which a model returns only one answer letter. \emph{Score gap} is the score difference induced by the compared pipeline settings. \emph{Extractor score gap} applies alternative extraction rules to the same generations. \emph{Rank shift} is the displacement in model rank under the compared metric conventions. \emph{Score spread} is the range across the evaluated prompt modes. \emph{Convention gap} measures the difference between alternative task-level credit rules. \emph{Denominator gap} measures the score difference associated with inconsistent denominator or aggregation conventions.

\section{Failure Evidence}
\label{app:evidence}

This section provides evidence for the 15 failure modes in Table~\ref{tab:failure_modes}, organized by pipeline stage with aggregate results and question-level examples.

\subsection{Dataset}
\label{app:e}

\subsubsection{Capability classification ($\mathcal{F}_1(\mathcal{D})$)}
\label{app:e1}
As presented in~\S\ref{sec:pipeline_failures}, we classify each question in each benchmark as knowledge-oriented or analytical by majority vote of four LLM classifiers: GPT-5.4, Qwen3.6, Llama-3.3, and RedSage-Qwen3. Each classifier uses the same six-shot prompt at temperature~0 and returns a strict JSON label. Across a stratified sample of 2{,}155 questions from all tasks, agreement is substantial, with Fleiss's $\kappa$=0.753 and pairwise Cohen's $\kappa$ ranging from 0.66 to 0.87. Because the classifiers overlap with the evaluated models, we treat these labels as a coverage estimate rather than gold annotations. \promptref{prompt:knowledge-analytical} shows the classification instruction and output schema; the complete six-shot prompt is released with the audit artifacts.

\subsubsection{Gold-label correctness ($\mathcal{F}_2(\mathcal{D})$)}
\label{app:e2}
The gold-label audit has two automated stages followed by human validation. First, for each question we consider models with parseable predictions and flag the question when at least 50\% select the same non-gold answer, effectively acting as a \emph{disagreement filter}. The threshold applies to the most common non-gold answer, that is, questions with a split non-gold vote below 50\% or a majority matching the gold label are not flagged. This stage is a triage mechanism, not a label-quality judgment.

\begin{table}[t]
\centering
\scriptsize
\renewcommand{\arraystretch}{1.12}
\setlength{\tabcolsep}{5pt}
\begin{tabular}{@{}lr@{}}
\toprule
\textbf{Benchmark} & \textbf{Flagged questions} \\
\midrule
MMLU-CS       & 2 \\
SecEval       & 216 \\
SECURE        & 33 \\
CTI-Bench     & 37 \\
AthenaBench   & 795 \\
CyberMetric   & 0 \\
RedSage-Bench & 53 \\
SecBench      & 4 \\
\midrule
\textbf{Total} & 1{,}140 \\
\bottomrule
\end{tabular}
\caption{Distribution of the 1{,}140 disagreement flags across benchmarks. Of these, 998 are checked by the search-grounded verifier}
\label{tab:app-label-perbench}
\end{table}

Flagged questions are then checked by a GPT-5.4 search-grounded verifier restricted to multi-tiered, authoritative cybersecurity sources. The verifier returns one of four outcomes: \emph{gold correct}, \emph{gold mislabel}, \emph{both wrong}, or \emph{uncertain}. Tier~1 sources include CVE, CWE, CVSS, MITRE ATT\&CK, NVD, CISA, and relevant RFC and NIST documents~\citep{sikos2023cybersecurity}. Tier~2 includes coordinated-disclosure and vendor advisories. Tier~1 evidence takes precedence when sources conflict; unresolved or insufficient evidence yields \emph{uncertain} rather than a forced binary decision.

The disagreement filter flags 1{,}140 questions, of which 998 are checked by the grounded verifier. Among these, 238 are confirmed gold mislabels (23.8\%), 653 are false-positive flags for which the gold label is correct (65.4\%), 16 are both wrong (1.6\%), and 91 are uncertain (9.1\%). Thus, 23.8\% is the flag precision on the checked set, not a benchmark-wide label-error rate. Its Wilson 95\% CI is $[21.3\%,26.6\%]$. As the flagged set is selected through model disagreement rather than random sampling, we do not extrapolate this rate to the full benchmarks. Of the 1{,}140 flags, 405 reach at least 75\% model agreement and 84 are unanimous. Table~\ref{tab:app-label-perbench} shows that the flags concentrate heavily in AthenaBench and SecEval.

Two human annotators independently validate a stratified 50-question sample spanning all four verifier outcomes. Inter-annotator agreement is Cohen's $\kappa=0.68$ with 94\% raw agreement. After adjudication, the human check confirms 46 of 50 verifier decisions (92\%; Wilson 95\% CI $[81.2,96.8]$). Of the remaining four cases, two are verifier errors and two cannot be resolved from the cited evidence. We therefore describe the verifier results as \emph{automatically verified}, so \emph{manual validation} refers only to judgments made by the human annotators.

\paragraph{Gold-label cases.}
\label{app:e3}
For each benchmark with a verifier-confirmed mislabel, we give one representative case showing the benchmark question, published gold answer, model-majority prediction, verifier reasoning, and supporting source (Cases~\ref{case:redsage-gold}--\ref{case:seceval-gold}).

\mlcase{case:redsage-gold}
{RedSage-Bench / \texttt{skills} (idx 2149)}
{What is the total length of an IP packet, including the header and payload, when the payload size is 32 bytes and the IP header size is 20 bytes? (A)~84 (B)~52 (C)~60 (D)~72}
{C (60 bytes)}
{B (52 bytes)}
{The IPv4 total-length field counts the whole datagram, including header and data, so $20+32=52$ bytes. Option~B is correct and the published gold label~C is wrong.}
{\item \url{https://www.rfc-editor.org/rfc/rfc791} \quad{\textit{``Total Length~\ldots~is the length of the datagram, measured in octets, including internet header and data.''}}}

\mlcase{case:cti-vsp-gold}
{CTI-Bench / \texttt{vsp} (idx 125)}
{Analyze the following CVE description and calculate the CVSS v3.1 Base Score~\ldots~Summarize each metric's value and provide the final CVSS v3.1 vector string.}
{CVSS:3.1/\brk AV:N/\brk AC:L/\brk PR:N/\brk UI:N/\brk S:U/\brk C:H/\brk I:H/\brk A:H}
{CVSS:3.1/\brk AV:N/\brk AC:L/\brk PR:N/\brk UI:N/\brk S:U/\brk C:N/\brk I:N/\brk A:H}
{The NVD record for CVE-2021-42141 contains two assessments. NVD's enrichment gives C:H/I:H/A:H, while CISA-ADP gives C:N/I:N/A:H. The description states only that the flaw may cause denial of service, supporting availability impact alone. This means the published gold overstates confidentiality and integrity impact.}
{\item \url{https://nvd.nist.gov/vuln/detail/CVE-2021-42141}\\{\textit{``ADP: CISA-ADP, Base Score: 7.5 HIGH Vector: CVSS:3.1/AV:N/AC:L/PR:N/UI:N/S:U/C:N/I:N/A:H.''}}}

\mlcase{case:athena-vsp-gold}
{AthenaBench / \texttt{athena\_vsp} (idx 1642)}
{Analyze the following CVE description and calculate the CVSS v3.1 Base Score\ldots}
{CVSS:3.1/\brk AV:L/\brk AC:L/\brk PR:N/\brk UI:R/\brk S:U/\brk C:H/\brk I:H/\brk A:H}
{CVSS:3.1/\brk AV:N/\brk AC:L/\brk PR:N/\brk UI:N/\brk S:U/\brk C:H/\brk I:H/\brk A:H}
{The CVE description states that remote attackers can execute arbitrary code without authentication, supporting AV:N and PR:N. NVD and the Zero Day Initiative both assign the network-accessible vector; the published AV:L/UI:R values conflict with those sources.}
{\item \url{https://nvd.nist.gov/vuln/detail/CVE-2025-3484}\\
{\textit{``CNA: ZDI, Base Score: 9.8 CRITICAL Vector: CVSS:3.0/AV:N/AC:L/PR:N/UI:N/S:U/C:H/I:H/A:H.''}}}

\mlcase{case:secure-kcv-gold}
{SECURE / \texttt{secure\_kcv} (idx 84)}
{Given the CVE-2024-5048 record, a SourceCodester Student Management System SQL-injection issue~\ldots~True or False: remote code execution is a potential impact.}
{T}
{F}
{The CVE/NVD record classifies the issue as CWE-89 with C:L/I:L/A:L impacts and provides no evidence of code execution. The published gold appears incorrect and the model majority is supported by the authoritative record.}
{\item \url{https://nvd.nist.gov/vuln/detail/CVE-2024-5048} \quad{\textit{``CVSS:3.1/AV:N/AC:L/PR:L/UI:N/S:U/C:L/I:L/A:L.''}}}

\mlcase{case:seceval-gold}
{SecEval / \texttt{seceval} (idx 98)}
{A web application uses GET requests with query parameters to perform state-changing actions~\ldots~What are the security implications?\\
A.\ Sensitive data may be exposed in server logs\\
B.\ GET requests can't have a body\\
C.~\ldots \quad D.~\ldots}
{AD}
{ABCD}
{OWASP warns that query-string data can appear in logs, history, caches, and referrers, supporting~A; state-changing GET requests also create additional security concerns. The published gold omits at least one supported option on this multiple-answer question.}
{\item \url{https://owasp.org/www-community/vulnerabilities/Information_exposure_through_query_strings_in_url} \quad{\textit{``The parameter values~\ldots~will be exposed in: Referer Header, Web Logs, Shared Systems, Browser History, Browser Cache~\ldots''}}}

\subsubsection{Prompt}
\label{app:f}

\subsubsection{Format-token leakage ($\mathcal{F}_1(\mathcal{P})$)}
\label{app:f2}
CyberMetric's released prompt ends with the literal template \texttt{Always return in this format: `ANSWER:~X'}, where \texttt{X} is intended as a placeholder. This token triggers systematic response failures in four models. Foundation-Sec produces an empty response on 490 of 500 questions, Primus-Nemotron on 401, Gemma-4 reproduces the placeholder on 404, and Primus-Merged continues the instruction rather than answering on 343. The released extractor silently maps these outputs to \texttt{None}, which the scorer counts as an incorrect answer. No rejected response contains a recoverable answer, so the failure originates in the prompt rather than extraction. Across the four affected models, the peer score gap reaches 90.9\,pp, with a median of 81.3\,pp. \examplerange{example:p1a-cybermetric-foundation}{example:p1b-cybermetric-gemma} show the two dominant behaviors.

\evbox{example:p1a-cybermetric-foundation}
{CyberMetric / Foundation-Sec ($\mathcal{F}_1(\mathcal{P})$)}
{\emph{``\ldots Always return in this format: `ANSWER: X'\,''} (prompt tail); gold is \texttt{A}}
{\texttt{""} (empty; the model emits nothing at all on 490 of 500 questions)}
{released regex finds no \texttt{ANSWER:\brk{} [A-D]} match in an empty string $\to$ \texttt{None}, which the scorer compares directly against the gold letter}
{Foundation-Sec scores 1.0\% on these questions against a peer median of 91.9\%, a 90.9\,pp peer gap. The empty responses come from the template, not from a failed generation run.}

\evbox{example:p1b-cybermetric-gemma}
{CyberMetric / Gemma-4 ($\mathcal{F}_1(\mathcal{P})$)}
{\emph{``\ldots Choose the correct answer (A, B, C, or D) only. Always return in this format: `ANSWER: X'\,''}}
{\texttt{'ANSWER: X'\brk{} 'ANSWER: X'\brk{} 'ANSWER: X'\brk{} 'ANSWER: X'\ldots} (the placeholder reproduced verbatim and repeated to the token budget)}
{released regex \texttt{re.\brk{}search(\brk{}r"ANSWER:\brk{}?\textbackslash s*(\brk{}[\brk{}A-D])",\brk{} resp,\brk{} re.\brk{}I)} requires a letter in \texttt{[A-D]}; the literal \texttt{X} is not, so the match fails $\to$ \texttt{None}}
{473 of the 500 outputs are non-empty yet return no prediction, giving 4.0\% against a peer median of 91.9\%. No re-extraction over these stored generations recovers an answer because the model never emits one, so the defect is in the prompt rather than the extractor. The standardized score comes from re-generating with the placeholder removed, so we report a peer-gap estimate rather than a controlled ablation.}

\subsubsection{Prompt--question conflict ($\mathcal{F}_2(\mathcal{P})$)}
\label{app:f3}
SecEval's benchmark-level instruction says to ``select the correct answers,'' but individual questions can encourage a singular response. On the 927 questions with multiple correct answers, the eight models with parseable responses produce a single-letter answer on 10.1\%--33.0\% of questions, with a median of 26.2\% (Table~\ref{tab:ablation-multianswer}). Set-exact-match scores these responses as incorrect even when the selected letter is part of the gold set. \exampleref{example:p2-seceval-gemma} illustrates the failure.

\evbox{example:p2-seceval-gemma}
{SecEval / Gemma-4 ($\mathcal{F}_2(\mathcal{P})$)}
{\emph{``You are evaluating the network security of a mobile application. Select the controls that should be implemented to ensure secure communication between the mobile application and its backend servers.''}}
{\texttt{D: Ensuring backend services}}
{single standalone letter $\to$ \texttt{D}; gold is \texttt{AD}}
{Set-exact-match scores 0 despite \texttt{D} being one of the two correct options. 306 of the 2{,}189 questions (14.0\%) pair a multi-letter gold with a single-letter prediction; over the 927 multi-answer questions alone this is 33.0\%.}

\subsubsection{Template incompatibility ($\mathcal{F}_3(\mathcal{P})$)}
\label{app:f4}
On SECURE, the failure arises from a mismatch between the task template, Primus-Merged's response style, and the reconstructed extractor. Rather than returning only the requested option, Primus-Merged frequently produces explanations or echoes the answer choices; on MAET, this occurs on 950 of 1,072 questions. The extractor then reads the first answer-choice letter it encounters, which can come from the echoed list rather than the model's intended selection. Consequently, many responses are scored according to incidental output structure rather than task correctness, producing a 48\,pp peer score gap. \exampleref{example:p3-secure-maet-primus} shows a representative case.

\evbox{example:p3-secure-maet-primus}
{SECURE-MAET / Primus-Merged ($\mathcal{F}_3(\mathcal{P})$)}
{\emph{``For the given question: Which incident involved adversaries using a rogue master to control emergency systems?, and four options: A) Maroochy Water Breach, B) 2017 Dallas Siren incident, C) Stuxnet Attack, D) BlackEnergy Attack\ldots''}}
{\texttt{Choose your answer from the list below:\brk{} A) Maroochy Water Breach\brk{} B) 2017 Dallas Siren incident\brk{} C) Stuxnet Attack\brk{} D) BlackEnergy Attack\brk{} X) None of the above\brk{} This question requires knowledge of specific incidents\ldots}}
{first answer-choice letter in the response $\to$ \texttt{A}, which is the first line of the model's echoed option list, not an answer}
{Gold is \texttt{B}, so the question is scored wrong for a reason unrelated to difficulty. 950 of the 1{,}072 MAET questions return an explanation or an echoed list rather than an option, so the score largely measures template compliance.}

\begin{table}[t]
\centering
\scriptsize
\renewcommand{\arraystretch}{1.12}
\setlength{\tabcolsep}{5pt}
\begin{tabular}{@{}lr@{}}
\toprule
\textbf{Model} & \textbf{Single-letter (\%)} \\
\midrule
Sonnet 4.6       & 10.8 \\
Gemma-4          & 33.0 \\
Qwen3.6          & 10.1 \\
Llama-3.3        & 23.7 \\
Primus-Nemotron  & 26.1 \\
Primus-Merged    & 31.6 \\
Foundation-Sec   & 26.4 \\
RedSage-Qwen3    & 28.7 \\
\midrule
\textbf{Median}  & 26.2 \\
\bottomrule
\end{tabular}
\caption{Single-letter response rate on SecEval's 927 multi-answer questions for the eight models with parseable outputs on this subset.}
\label{tab:ablation-multianswer}
\end{table}

\subsection{Inference}
\label{app:g}

\subsubsection{Stop-sequence mismatch ($\mathcal{F}_1(\mathcal{I})$)}
\label{app:g1}
RedSage-Bench's released generative path stops on a newline. For Qwen3.6, the newline appears inside the reasoning preamble before any answer token, producing empty responses. Removing the backend stop, allowing reasoning to complete, and applying answer handling afterward raises the benchmark mean from 0.0\% to 85.9\%. Per-task recovery ranges from 81.4 to 90.1\,pp. \exampleref{example:i1-redsage-qwen} shows the mechanism.

\evbox{example:i1-redsage-qwen}
{RedSage-Bench / Qwen3.6 ($\mathcal{F}_1(\mathcal{I})$)}
{the released generative MCQ prompt, unchanged; only the backend stop sequence differs between the two runs}
{\emph{with} \texttt{stop=[\brk{}"\textbackslash n"]}: \texttt{""} (the stop fires inside the \texttt{<think>} preamble before any answer token) \quad\textbar\quad \emph{stop removed}: the answer is produced normally}
{unchanged in both runs; with the stop in place, there is no text to extract}
{Qwen3.6 scores 0.0\% with the released stop sequence and 85.9\% without it, an 85.9\,pp difference on identical questions with the same extractor (Table~\ref{tab:ablation-stopseq}).}

\begin{table}[t]
\centering
\scriptsize
\renewcommand{\arraystretch}{1.12}
\setlength{\tabcolsep}{5pt}
\begin{tabular}{@{}lrrr@{}}
\toprule
& \multicolumn{2}{c}{\textbf{Accuracy (\%)}} & \\
\cmidrule(lr){2-3}
\textbf{Task} & Released & Stop-free & $\boldsymbol{\Delta}$ \textbf{(pp)} \\
\midrule
\texttt{cli}        & 0.0 & 88.8 & +88.8 \\
\texttt{frameworks} & 0.0 & 83.7 & +83.7 \\
\texttt{generals}   & 0.0 & 85.4 & +85.4 \\
\texttt{kali}       & 0.0 & 81.4 & +81.4 \\
\texttt{skills}     & 0.0 & 90.1 & +90.1 \\
\midrule
\textbf{Mean} & 0.0 & 85.9 & +85.9 \\
\bottomrule
\end{tabular}
\caption{RedSage-Bench stop-sequence ablation for Qwen3.6. The benchmark-mean recovery is 85.9\,pp.}
\label{tab:ablation-stopseq}
\end{table}

\subsubsection{Token-budget filter ($\mathcal{F}_2(\mathcal{I})$)}
\label{app:g2}
SecEval specifies a five-token output budget, below the 16-token minimum accepted by the GPT-5.4 Azure OpenAI endpoint used in our evaluation. All 2{,}189 requests therefore return HTTP~400 rather than a model generation. The evaluator stores these error payloads as responses, allowing the answer extractor to match occasional A--D characters inside the error text and report 0.3\% accuracy. Raising the budget to 16 tokens produces valid generations and raises accuracy to 81.4\%, a difference of 81.1\,pp. \exampleref{example:i2-seceval-gpt54} shows the failure.

\evbox{example:i2-seceval-gpt54}
{SecEval / GPT-5.4 ($\mathcal{F}_2(\mathcal{I})$)}
{system \emph{``Below are multiple-choice questions concerning cybersecurity. Please select the correct answers and respond with the letters ABCD only.''}; user \emph{``Question: Which mitigation prevent stack overflow bug? A: Stack Canary. B: ALSR. C: CFI. D: Code Sig\ldots''}}
{\texttt{ERROR: HTTP 400: \{"error": \{"message": "Invalid `max\_output\_tokens': integer below minimum value. Expected a value >= 16, but got 5 instead.", "type": "invalid\_request\_error", "param": "max\_output\_tokens"\ldots\}\}}}
{the evaluator stores the error string as if it were a response, and the A--D letter scan runs over the JSON payload}
{No generation ever occurred. All 2{,}189 questions carry a payload of this form, so the reported 0.3\% is entirely due to accidental letter matches inside error text. A backend constraint is being reported as a capability score.}

\subsubsection{Decoding drift ($\mathcal{F}_3(\mathcal{I})$)}
\label{app:g3}
CyberMetric documents temperature~1.0, top-$p$~0.9, and top-$k$~50, while its released evaluator sets none of them and therefore inherits backend defaults. Comparing the documented configuration with a greedy control on the same 500 questions changes Primus-Merged from 17.2\% to 57.2\%, a 40.0\,pp difference; correctness changes on 274 questions (Table~\ref{tab:ablation-decoding}). \exampleref{example:i3-cybermetric-primus} illustrates the format-compliance mechanism.

\evbox{example:i3-cybermetric-primus}
{I3 --- CyberMetric / Primus-Merged ($\mathcal{F}_3(\mathcal{I})$)}
{identical question and prompt under both configurations; only decoding differs}
{\emph{greedy} ($T{=}0$, released default): \texttt{(where X is the letter of the correct answer). Do not leave a space between the colon and the letter. Do not use quotes\ldots} \quad \emph{documented} ($T{=}1.0$, top-$p{=}0.9$): \texttt{ANSWER: A You must know: To provide a security solution\ldots}}
{greedy output continues the instruction text and yields no answer letter ($\to$ \texttt{None}); the sampled output emits the required \texttt{ANSWER:\brk{} A} form ($\to$ \texttt{A})}
{Gold is \texttt{A}. Correctness changes on 274 of the 500 questions between the two configurations, moving Primus-Merged from 17.2\% to 57.2\%. The difference is driven by decoding-dependent format compliance rather than a change in the underlying question.}

\begin{table}[t]
\centering
\scriptsize
\renewcommand{\arraystretch}{1.12}
\setlength{\tabcolsep}{4pt}
\begin{tabular}{@{}lrrr@{}}
\toprule
& \multicolumn{2}{c}{\textbf{Accuracy (\%)}} & \\
\cmidrule(lr){2-3}
\textbf{Model} & Greedy & Documented & $\boldsymbol{\Delta}$ \textbf{(pp)} \\
\midrule
GPT-5.4          & 96.0 & 96.2 & +0.2 \\
Sonnet 4.6       & 96.8 & 96.6 & -0.2 \\
Gemma-4          & 4.0  & 4.8  & +0.8 \\
Qwen3.6          & 92.2 & 91.0 & -1.2 \\
Llama-3.3        & 91.6 & 89.6 & -2.0 \\
GPT-OSS          & 86.4 & 83.4 & -3.0 \\
Primus-Nemotron  & 19.2 & 30.6 & +11.4 \\
Primus-Merged    & 17.2 & 57.2 & +40.0 \\
Foundation-Sec   & 1.0  & 2.8  & +1.8 \\
RedSage-Qwen3    & 89.8 & 86.8 & -3.0 \\
\bottomrule
\end{tabular}
\caption{CyberMetric decoding sensitivity on the same 500 questions. \emph{Greedy} uses $T=0$; \emph{Documented} uses the benchmark's stated sampling configuration.}
\label{tab:ablation-decoding}
\end{table}

\subsection{Extraction}
\label{app:h}

\subsubsection{Extractor divergence ($\mathcal{F}_1(\mathcal{E})$)}
\label{app:h1}
Semantically equivalent output can receive different scores under different extraction rules. On CTI-Bench VSP, we apply two released conventions to identical stored generations. AthenaBench's \emph{final-line} rule searches the final answer line for a \texttt{CVSS:3.1/} vector, whereas CTI-Bench's \emph{anywhere} rule takes the last vector appearing anywhere in the response. Among model--question pairs for which at least one rule extracts a vector, the two disagree on 30.4\%; disagreement reaches 96.5\% for Primus-Merged (Table~\ref{tab:extractor_disagreement}). \exampleref{example:e1-cti-vsp-primus} shows a representative response.

\evbox{example:e1-cti-vsp-primus}
{CTI-Bench VSP / Primus-Merged ($\mathcal{F}_1(\mathcal{E})$)}
{\emph{``\ldots provide the final CVSS v3.1 vector string.''}}
{\texttt{\ldots The final answer is:\brk{} CVSS:3.1/AV:N/AC:L/PR:N/UI:N/S:U/C:H/I:H/A:H\brk{}This course unit focuses on the CVE-2017-1000480 vulnerability\ldots} (the correct vector, followed by unrelated text for the rest of the 32-line response)}
{\emph{final-line} reads the last non-empty line and returns \texttt{None}; \emph{anywhere} returns the last CVSS vector found in the response, which matches the gold vector}
{The same generation is correct under one extraction rule and unparseable under the other. The two rules disagree on 96.5\% of Primus-Merged's 513 contested questions, even though both are plausible interpretations of the released pipeline.}

RedSage-Bench provides another comparison within the same generation. Its released evaluator computes \emph{Exact Match} (EM), \emph{Prefix Exact Match} (PEM), and \emph{Regex MCQ eXtraction} (RX) on the same generated responses without designating one canonical. Primus-Merged changes from 0.3\% under EM to 80.0\% under PEM, a 79.7\,pp difference, because the model often gives the correct letter first and then continues in prose. The full per-model comparison is reported in Table~\ref{tab:ablation-redsage-extract}; the 79.7\,pp difference is the maximum extractor-divergence effect summarized in Table~\ref{tab:failure_modes}.

\begin{table}[t]
\centering
\scriptsize
\renewcommand{\arraystretch}{1.12}
\setlength{\tabcolsep}{3.5pt}
\begin{tabular}{@{}lrrr@{}}
\toprule
& \multicolumn{2}{c}{\textbf{Disagreement (\%)}} & \\
\cmidrule(lr){2-3}
\textbf{Model} & All & Contested only & Contested ($n$) \\
\midrule
GPT-5.4          & 0.0  & 0.0  & 1{,}000 \\
Sonnet 4.6       & 0.1  & 0.1  & 999 \\
Gemma-4          & 0.2  & 28.6 & 7 \\
Qwen3.6          & 42.8 & 52.8 & 810 \\
Llama-3.3        & 42.1 & 44.4 & 948 \\
GPT-OSS          & 46.0 & 60.6 & 759 \\
Primus-Nemotron  & 0.3  & 1.3  & 234 \\
Primus-Merged    & 49.5 & 96.5 & 513 \\
Foundation-Sec   & 19.2 & 19.9 & 964 \\
RedSage-Qwen3    & 16.9 & 18.6 & 908 \\
\midrule
\textbf{Pooled}  & 21.7 & 30.4 & 7{,}142 \\
\bottomrule
\end{tabular}
\caption{
Disagreement between final-line and anywhere CVSS extraction on CTI-Bench VSP. \emph{All} is the percentage of all 1{,}000 questions per model for which the two extractors return different results. \emph{Contested only} is the disagreement percentage restricted to model--question pairs for which at least one extractor returns a CVSS vector. \emph{Contested ($n$)} is the number of such pairs. The pooled row aggregates all 10 models.}
\label{tab:extractor_disagreement}
\end{table}

\begin{table}[t]
\centering
\scriptsize
\renewcommand{\arraystretch}{1.12}
\setlength{\tabcolsep}{4pt}
\begin{tabular}{@{}lrrr@{}}
\toprule
& \multicolumn{3}{c}{\textbf{Accuracy (\%)}} \\
\cmidrule(lr){2-4}
\textbf{Model} & EM & PEM & RX \\
\midrule
GPT-5.4          & 90.7 & 90.7 & 90.2 \\
Sonnet 4.6       & 90.5 & 91.3 & 91.0 \\
Gemma-4          & 2.1  & 45.6 & 45.7 \\
Qwen3.6          & 86.6 & 86.6 & 85.9 \\
Llama-3.3        & 61.3 & 85.8 & 85.1 \\
GPT-OSS          & 0.1  & 0.1  & 25.5 \\
Primus-Nemotron  & 84.2 & 86.0 & 85.9 \\
Primus-Merged    & 0.3  & 80.0 & 79.0 \\
Foundation-Sec   & 79.1 & 79.1 & 78.2 \\
RedSage-Qwen3    & 10.8 & 86.0 & 85.3 \\
\bottomrule
\end{tabular}
\caption{Three released RedSage-Bench extraction rules applied to identical stored generations. \textit{EM} is exact match, \textit{PEM} is prefix exact match, and \textit{RX} is regex MCQ extraction.}
\label{tab:ablation-redsage-extract}
\end{table}

\subsubsection{Denominator inflation ($\mathcal{F}_2(\mathcal{E})$)}
\label{app:h2}
Let $c$ denote correct predictions, $v$ parseable predictions, and $n$ all attempted questions. Correct-over-valid reports $c/v$, whereas correct-over-total reports $c/n$. On CTI-RCM, Gemma-4 has only two parseable outputs among 1{,}000 questions, both correct. The released convention therefore reports 100.0\%, whereas correct-over-total reports 0.2\%, a 99.8\,pp difference and $500\times$ inflation. \exampleref{example:e2-cti-rcm-gemma} shows representative outputs. Table~\ref{tab:ablation-denominator} reports all accuracy-type model--task pairs for which the two denominator policies differ by more than 5\,pp. AthenaBench VSP exhibits the same mechanism: excluding unparseable CVSS vectors affects nine of 10 models, and retaining them with maximum deviation lowers scores by up to 85.4\,pp, with a median decrease of 55.9\,pp.

\evbox{example:e2-cti-rcm-gemma}
{CTI-RCM / Gemma-4 ($\mathcal{F}_2(\mathcal{E})$)}
{\emph{``Analyze the following CVE description and map it to the appropriate CWE\ldots Ensure the last line of your response contains only the CWE ID. CVE Description: In the Linux kernel through 6.7.1, there is a use-after-free in \texttt{cec\_queue\_msg\_fh}\ldots''}}
{three representative outputs: \texttt{L} \quad\textbar\quad \texttt{262594. 262594. 262594. 262594.\ldots} \quad\textbar\quad \texttt{""}}
{last-line \texttt{CWE-\textbackslash d+} matching returns \texttt{None} on all three. Across the task, the extractor matches only 2 of 1{,}000 outputs (\texttt{CWE-79} and \texttt{CWE-287}), both correct; 945 outputs are non-empty but contain no CWE ID}
{Here $c{=}2$, $v{=}2$, and $n{=}1000$. Correct-over-valid scoring reports $c/v=100.0\%$, while correct-over-total reports $c/n=0.2\%$ on the same generations. The resulting $500\times$ inflation is entirely due to the denominator convention.}

\begin{table}[t]
\centering
\scriptsize
\renewcommand{\arraystretch}{1.10}
\setlength{\tabcolsep}{3.5pt}
\begin{tabular}{@{}llrrr@{}}
\toprule
& & \multicolumn{2}{c}{\textbf{Accuracy (\%)}} & \\
\cmidrule(lr){3-4}
\textbf{Model} & \textbf{Task} & $c/v$ & $c/n$ & $\boldsymbol{\Delta}$ \textbf{(pp)} \\
\midrule
Gemma-4         & \texttt{secure-cwet} & 42.9  & 0.6  & -42.2 \\
Primus-Nemotron & \texttt{secure-cwet} & 100.0 & 9.4  & -90.6 \\
Foundation-Sec  & \texttt{secure-cwet} & 82.8  & 43.4 & -39.4 \\
\midrule
Gemma-4         & \texttt{secure-maet} & 58.3  & 0.7  & -57.7 \\
Primus-Nemotron & \texttt{secure-maet} & 91.1  & 8.6  & -82.5 \\
Foundation-Sec  & \texttt{secure-maet} & 81.5  & 35.0 & -46.5 \\
\midrule
Gemma-4         & \texttt{secure-kcv}  & 84.8  & 12.0 & -72.8 \\
Foundation-Sec  & \texttt{secure-kcv}  & 77.8  & 1.5  & -76.3 \\
Primus-Merged   & \texttt{secure-kcv}  & 58.0  & 43.8 & -14.2 \\
\midrule
Gemma-4         & \texttt{cti-rcm} & 100.0 & 0.2  & -99.8 \\
Primus-Nemotron & \texttt{cti-rcm} & 71.7  & 9.1  & -62.6 \\
Foundation-Sec  & \texttt{cti-rcm} & 69.1  & 63.6 & -5.5 \\
\midrule
Gemma-4         & \texttt{cti-mcq} & 52.9  & 12.9 & -39.9 \\
Qwen3.6         & \texttt{cti-mcq} & 74.9  & 67.2 & -7.7 \\
Llama-3.3       & \texttt{cti-mcq} & 66.1  & 24.7 & -41.4 \\
GPT-OSS         & \texttt{cti-mcq} & 70.2  & 60.4 & -9.9 \\
Primus-Merged   & \texttt{cti-mcq} & 17.6  & 0.2  & -17.4 \\
Foundation-Sec  & \texttt{cti-mcq} & 32.9  & 27.8 & -5.1 \\
RedSage-Qwen3   & \texttt{cti-mcq} & 67.6  & 49.8 & -17.8 \\
\bottomrule
\end{tabular}
\caption{Accuracy for different denominator policies. Correct-over-valid ($c/v$) and correct-over-total ($c/n$) differ by more than 5\,pp on identical stored outputs. AthenaBench VSP is excluded because its score is a normalized deviation rather than a correct-count accuracy.}
\label{tab:ablation-denominator}
\end{table}

\subsubsection{Metric-direction mismatch ($\mathcal{F}_3(\mathcal{E})$)}
\label{app:h3}
CTI-Bench and AthenaBench score the same CVSS prediction problem in opposite directions. CTI-Bench reports mean absolute deviation (MAD), for which lower is better, whereas AthenaBench reports $100\times\max(0,1-\mathrm{MAD}/7.7)$, for which higher is better. Without explicitly normalizing direction, the same model performances can therefore induce different orderings. Across the 10 models, the corresponding rank displacement reaches five positions.

\subsubsection{Prompt-mode sensitivity ($\mathcal{F}_4(\mathcal{E})$)}
\label{app:h4}
AthenaBench's ATE evaluator extracts only the final answer line. Changing the prompt mode can therefore change whether a correctly identified technique appears in the extractable position. We evaluate zero-shot, four-shot, and chain-of-thought prompts on the same questions while holding the task and extractor fixed. Across the 10 models, the score spread reaches 40\,pp, with a median of 7\,pp. Because the prompt modes produce different generations, this is a prompt--extractor sensitivity rather than a same-generation extraction ablation.

\subsection{Aggregation}
\label{app:i}

\subsubsection{Logprob vs.\ generative scoring ($\mathcal{F}_1(\mathcal{A})$)}
\label{app:i1}
Multiple-choice performance can also depend on whether the evaluator ranks answer choices by token log probability or generates a textual answer and scores the resulting response (Table~\ref{tab:ablation-logprob}). On RedSage-Bench, Gemma-4 scores 86.6\% under logprob scoring but 45.7\% under generative scoring, a 40.9\,pp difference. Qwen3.6 moves in the opposite direction, from 59.2\% logprob to 85.9\% generative. The generative results here use the stop-free configuration, so this comparison is separate from $\mathcal{F}_1(\mathcal{I})$. Hosted GPT-5.4 and Sonnet 4.6 are omitted from the logprob comparison because their evaluated APIs do not expose the required token log probabilities.

\begin{table}[t]
\centering
\scriptsize
\renewcommand{\arraystretch}{1.10}
\setlength{\tabcolsep}{4pt}
\begin{tabular}{@{}lrrrr@{}}
\toprule
& \multicolumn{4}{c}{\textbf{Accuracy (\%)}} \\
\cmidrule(lr){2-5}
& \multicolumn{2}{c}{\textbf{MMLU-CS}} & \multicolumn{2}{c}{\textbf{RedSage-Bench}} \\
\cmidrule(lr){2-3}\cmidrule(lr){4-5}
\textbf{Model} & Generative & Logprob & Generative & Logprob \\
\midrule
Gemma-4          & 59.0 & 66.0 & 45.7 & 86.6 \\
Qwen3.6          & 57.0 & 80.0 & 85.9 & 59.2 \\
Llama-3.3        & 78.0 & 82.0 & 85.1 & 84.8 \\
GPT-OSS          & 36.0 & 49.0 & 25.5 & 26.7 \\
Primus-Nemotron  & 87.0 & 87.0 & 85.9 & 84.1 \\
Primus-Merged    & 66.0 & 83.0 & 79.0 & 73.8 \\
Foundation-Sec   & 80.0 & 80.0 & 78.2 & 74.4 \\
RedSage-Qwen3    & 81.0 & 85.0 & 85.3 & 84.2 \\
\bottomrule
\end{tabular}
\caption{Generative vs logprob multiple-choice scoring for the eight open-weight models on two benchmarks.}
\label{tab:ablation-logprob}
\end{table}

\subsubsection{Task-level metric drift ($\mathcal{F}_2(\mathcal{A})$)}
\label{app:i2}
Attacker-attribution tasks use materially different credit rules. The standardized harness uses strict binary grading while recognizing true semantic aliases, such as APT28 and Fancy Bear, as equivalent. CTI-Bench additionally awards \emph{plausible} credit to related actors, while AthenaBench uses a strict binary verdict. Within CTI-Bench, moving from strict to \emph{correct+plausible} scoring rule raises a model's score by up to 30\,pp. Across CTI-Bench and AthenaBench, these scoring conventions produce cross-task score gaps of up to 70\,pp (Table~\ref{tab:ablation-taadrift}). \exampleref{example:a2-cti-taa-llama} illustrates how the same CTI prediction receives different credit under the two released CTI conventions.

\evbox{example:a2-cti-taa-llama}
{CTI-TAA / Llama-3.3 ($\mathcal{F}_2(\mathcal{A})$)}
{\emph{``Analysis of Campaign-2. The second campaign of [PLACEHOLDER] observed is spread via a phishing link that downloads an archive file named `Indian Army Recruitment\ldots'\,''}}
{prediction extracted as \texttt{transparent tribe}; gold is \texttt{sidecopy}}
{identical prediction, two verdicts from the same released evaluator: strict exact match $\to$ \textit{incorrect}; the shipped alias/related-actor graph records \texttt{connection = P} (related actor) $\to$ \textit{credited} under Correct+Plausible}
{Across the 50 questions, 22 are strictly correct (44.0\%) and 35 are Correct+Plausible (70.0\%); 13 questions receive different credit under the two conventions. Neither convention is designated canonical in the release, producing a 26.0\,pp spread on identical outputs.}

\begin{table}[t]
\centering
\scriptsize
\renewcommand{\arraystretch}{1.10}
\setlength{\tabcolsep}{3.5pt}
\begin{tabular}{@{}lrrrr@{}}
\toprule
& \multicolumn{3}{c}{\textbf{Accuracy (\%)}} & \\
\cmidrule(lr){2-4}
& \multicolumn{2}{c}{\textbf{CTI-Bench}} & \textbf{AthenaBench} & \\
\cmidrule(lr){2-3}\cmidrule(lr){4-4}
\textbf{Model} & Strict & C+P & Binary & $\boldsymbol{\Delta}$ \textbf{(pp)} \\
\midrule
GPT-5.4          & 74.0 & 86.0 & 33.0 & +53.0 \\
Sonnet 4.6       & 86.0 & 94.0 & 47.0 & +47.0 \\
Gemma-4          & 20.0 & 26.0 & 7.0  & +19.0 \\
Qwen3.6          & 32.0 & 62.0 & 14.0 & +48.0 \\
Llama-3.3        & 44.0 & 70.0 & 0.0  & +70.0 \\
GPT-OSS          & 42.0 & 64.0 & 5.0  & +59.0 \\
Primus-Nemotron  & 16.0 & 32.0 & 12.0 & +20.0 \\
Primus-Merged    & 6.0  & 14.0 & 8.0  & +6.0 \\
Foundation-Sec   & 20.0 & 42.0 & 25.0 & +17.0 \\
RedSage-Qwen3    & 12.0 & 20.0 & 21.0 & -1.0 \\
\bottomrule
\end{tabular}
\caption{Attacker-attribution accuracy under different aggregation conventions. CTI-Bench reports \emph{Strict} and \emph{Correct+Plausible} (C+P) scoring, while AthenaBench uses \emph{Binary} scoring. The column $\Delta$ is the cross-benchmark score gap in pp between CTI-Bench C+P and AthenaBench binary accuracy.}
\label{tab:ablation-taadrift}
\end{table}

\subsubsection{Aggregation inconsistency ($\mathcal{F}_3(\mathcal{A})$)}
\label{app:i3}
Gemma-4 illustrates how nominally comparable accuracy percentages can diverge under incompatible aggregation rules. CTI-RCM reports 100.0\% because only its two parseable predictions enter the denominator, whereas SecEval reports 9.5\% under correct-over-total with set-exact-match scoring, an apparent 90.5\,pp gap. Re-scoring CTI-RCM as correct-over-total reduces it to 0.2\%, showing that denominator alignment alone can reverse the apparent comparison. As reported in Table~\ref{tab:before_after}, under the full standardized pipeline, which additionally aligns prompting, extraction, and scoring where semantics permit, Gemma-4 scores 70.9\% on CTI-RCM and 78.3\% on SecEval. The example illustrates why benchmark-level comparisons require compatible aggregation rules rather than percentages alone.

\section{Cross-Benchmark Analysis}
\label{app:j}

This section provides additional analyses of how the audited tasks relate to one another and whether they support consistent model comparisons. We examine effective dimensionality, task-level rank agreement and pairwise order reversals, and the relationship between semantic overlap and ranking similarity.

\subsection{Effective Dimensionality}
\label{app:j1}

We column-standardize the $23\times10$ strict-verdict task-score matrix and apply PCA. The first principal component explains 95.25\% of the variance. Horn's parallel analysis with 500 random matrices at the 95th percentile finds that only this component exceeds the corresponding null eigenvalue, indicating that most score variation lies along a single dominant axis.

\subsection{Rank Agreement}
\label{app:j2}

For task-to-task comparisons, we measure rank agreement with Kendall's $\tau$-b, which accounts for ties. With only 10 models, small values may not be distinguishable from zero, so we interpret the coefficients as measures of correspondence rather than precise population estimates. For original-versus-standardized benchmark rankings in App.~\ref{app:k4}, we use Spearman's $\rho$.

\subsection{Pairwise Rank Changes}
\label{app:j3}

A pairwise order reversal occurs when two models are ordered one way by one task and the opposite way by another. We enumerate these reversals over all $\binom{10}{2}=45$ model pairs for each task pair and release the full catalog with the evaluation artifacts. The clearest examples arise between semantically similar tasks with different evaluation conventions. CTI-Bench and AthenaBench vulnerability scoring evaluate the same CVSS quantity but use opposite metric directions (App.~\ref{app:h3}); their model rankings agree only weakly under Kendall's $\tau$-b ($0.29$), with rank displacement reaching five positions. Their attacker-attribution tasks likewise show weak agreement ($\tau$-b $=0.24$).

\subsection{Semantic Overlap}
\label{app:j4}

We embed benchmark questions after removing boilerplate, represent each task by the centroid of its question embeddings, and compute pairwise task-content similarity using cosine similarity. We compare this matrix with a task-ranking agreement matrix based on pairwise Kendall's $\tau$ using a Mantel permutation test. Across the $\binom{23}{2}=253$ task pairs, content similarity and ranking agreement are moderately associated ($r=0.516$, $p<0.002$), with a squared correlation of $r^2\approx0.27$. Thus, semantic similarity explains only part of the variation in task-induced rankings. Because each $\tau$ is estimated from rankings over only 10 models, sampling variability may attenuate the observed association, so we avoid interpreting $r^2$ as a precise fraction of variance explained. Independently, PCA of the task-score matrix recovers a single dominant component (App.~\ref{app:j1}), consistent with a shared model-strength axis producing similar score patterns across tasks with substantially different content. Together, these results suggest that the strong score-level redundancy across tasks cannot be attributed primarily to duplicated or semantically overlapping questions.

\section{Standardization and Rank Stability}
\label{app:k}

This section records the benchmark-specific standardization choices, the resulting score changes, and the stability of the observed rank shifts. It also separates changes attributable to generation from those attributable to extraction.

\subsection{Standardized Components}
\label{app:k1}

The pipeline ledger (Table~\ref{tab:pipeline-ledger}) records the action taken for every benchmark-field pair. \emph{Retain} preserves released behavior, whereas \emph{Fix}, \emph{Normalize}, and \emph{Supply} identify fields changed or supplied by the standardized pipeline. Read together with Table~\ref{tab:spec-matrix}, the two tables distinguish what the benchmark specifies from what the standardized evaluation executes.

\subsection{Original vs.\ Standardized Scores}
\label{app:k2}

Table~\ref{tab:before_after} reports the per-model task scores before and after pipeline standardization. CyberMetric \texttt{det}/\texttt{samp} and MMLU-CS \texttt{gen}/\texttt{logp} are alternative configurations of one task each, not additional tasks, so the inventory in Table~\ref{tab:app-inventory} still contains 23 tasks. MMLU-CS logprob scoring is undefined for GPT-5.4 and Sonnet 4.6 because the evaluated hosted APIs do not expose the required token log probabilities.

The symbol $^\dagger$ marks an original score dominated by a denominator artifact: the released evaluator scores only parseable predictions, and the valid set is a small fraction of the attempted questions. GPT-5.4's original SecEval score requires a different caution. The released five-token budget is rejected by the backend, so the resulting 0.3\% reflects accidental extraction from API error payloads rather than model capability (App.~\ref{app:g2}). Finally, standardized scores use the pinned extraction and grading policy of App.~\ref{app:a-extract}, so an original-to-standardized difference can reflect changes in prompting, inference, extraction, scoring, or aggregation as specified in the pipeline ledger.

\begin{sidewaystable*}[p]
\centering
\scriptsize
\renewcommand{\arraystretch}{1.18}
\setlength{\extrarowheight}{0.5pt}
\setlength{\tabcolsep}{1.4pt}
\begin{tabularx}{\linewidth}{@{}>{\raggedright\arraybackslash}p{0.080\linewidth}>{\raggedright\arraybackslash}p{0.052\linewidth}>{\raggedright\arraybackslash}p{0.060\linewidth}*{10}{>{\centering\arraybackslash}X>{\columncolor{gray!8}\centering\arraybackslash}X}@{}}
\toprule
& & & \multicolumn{2}{c}{\textbf{GPT-5.4\textsuperscript{*}}} & \multicolumn{2}{c}{\textbf{Sonnet 4.6\textsuperscript{*}}} & \multicolumn{2}{c}{\textbf{Gemma-4}} & \multicolumn{2}{c}{\textbf{Qwen3.6}} & \multicolumn{2}{c}{\textbf{Llama-3.3}} & \multicolumn{2}{c}{\textbf{GPT-OSS}} & \multicolumn{2}{c}{\textbf{Primus-Nemotron}} & \multicolumn{2}{c}{\textbf{Primus-Merged}} & \multicolumn{2}{c}{\textbf{Foundation-Sec}} & \multicolumn{2}{c}{\textbf{RedSage-Qwen3}} \\
\cmidrule(lr){4-5}\cmidrule(lr){6-7}\cmidrule(lr){8-9}\cmidrule(lr){10-11}\cmidrule(lr){12-13}\cmidrule(lr){14-15}\cmidrule(lr){16-17}\cmidrule(lr){18-19}\cmidrule(lr){20-21}\cmidrule(lr){22-23}
\textbf{Benchmark} & \textbf{Task} & \textbf{Metric} & Before & After & Before & After & Before & After & Before & After & Before & After & Before & After & Before & After & Before & After & Before & After & Before & After \\
\midrule
\textbf{MMLU-CS} & \texttt{gen} & Accuracy & 74.0 & 87.0 & 70.0 & 90.0 & 59.0 & 90.0 & 57.0 & 88.0 & 78.0 & 81.0 & 36.0 & 86.0 & 87.0 & 76.0 & 66.0 & 74.0 & 80.0 & 76.0 & 81.0 & 82.0 \\
& \texttt{logp} & Accuracy & -- & -- & -- & -- & 66.0 & 90.0 & 80.0 & 88.0 & 82.0 & 81.0 & 49.0 & 86.0 & 87.0 & 76.0 & 83.0 & 74.0 & 80.0 & 76.0 & 85.0 & 82.0 \\
\midrule
\textbf{SecEval} & \texttt{seceval} & Accuracy & 0.3 & 82.0 & 78.0 & 71.8 & 9.5 & 78.3 & 54.3 & 74.2 & 71.0 & 69.7 & 0.4 & 72.6 & 61.0 & 72.8 & 12.6 & 61.7 & 58.5 & 57.0 & 47.2 & 74.2 \\
\midrule
\textbf{SECURE} & \texttt{maet} & Accuracy & 93.3 & 93.1 & 94.2 & 94.1 & 58.3 & 92.3 & 88.8 & 90.2 & 85.2 & 86.7 & 71.4 & 87.3 & 91.1\ensuremath{^\dagger} & 91.0 & 11.4 & 78.5 & 81.5 & 84.6 & 59.0 & 89.6 \\
& \texttt{cwet} & Accuracy & 94.3 & 95.0 & 94.6 & 95.0 & 42.9 & 92.3 & 89.3 & 92.0 & 87.7 & 90.0 & 71.6 & 88.9 & 100.0\ensuremath{^\dagger} & 94.1 & 8.0 & 78.4 & 82.8 & 83.4 & 60.1 & 91.5 \\
& \texttt{kcv} & Accuracy & 87.8 & 88.4 & 88.4 & 88.4 & 84.8 & 84.8 & 88.4 & 89.5 & 81.5 & 87.5 & 81.9 & 88.2 & 0.0\ensuremath{^\dagger} & 86.9 & 58.0 & 82.2 & 77.8 & 81.1 & 67.2 & 81.5 \\
\midrule
\textbf{CTI-Bench} & \texttt{mcq} & Accuracy & 78.3 & 78.5 & 83.6 & 84.1 & 52.9 & 74.4 & 74.9 & 73.0 & 66.1 & 65.6 & 70.2 & 69.3 & 71.6 & 69.2 & 17.6 & 45.0 & 32.9 & 56.4 & 67.6 & 65.2 \\
& \texttt{rcm} & Accuracy & 74.0 & 74.2 & 75.5 & 75.4 & 100.0\ensuremath{^\dagger} & 70.9 & 68.0 & 71.8 & 66.1 & 63.1 & 66.2 & 65.9 & 71.7 & 65.3 & 64.2 & 67.5 & 69.1 & 69.4 & 78.0 & 75.7 \\
& \texttt{vsp} & MAD$\downarrow$ & 0.98 & 1.02 & 0.80 & 0.81 & 1.38 & 0.93 & 1.38 & 1.34 & 1.59 & 1.49 & 1.27 & 1.95 & 1.43 & 1.34 & 1.67 & 1.91 & 1.56 & 1.39 & 1.15 & 1.42 \\
& \texttt{ate} & Accuracy & 5.0 & 5.0 & 8.3 & 6.7 & 5.0 & 8.3 & 0.0 & 3.3 & 0.0 & 1.7 & 3.3 & 3.3 & 0.0 & 1.7 & 0.0 & 0.0 & 1.7 & 1.7 & 0.0 & 0.0 \\
& \texttt{taa} & Accuracy & 86.0 & 70.0 & 94.0 & 82.0 & 26.0 & 50.0 & 62.0 & 20.0 & 70.0 & 38.0 & 64.0 & 26.0 & 32.0 & 40.0 & 14.0 & 14.0 & 42.0 & 20.0 & 20.0 & 30.0 \\
\midrule
\textbf{AthenaBench} & \texttt{ckt} & Accuracy & 91.3 & 91.0 & 92.5 & 92.8 & 8.4 & 86.1 & 79.4 & 84.9 & 70.5 & 82.2 & 69.7 & 81.1 & 66.6 & 81.3 & 19.5 & 76.6 & 68.9 & 78.5 & 80.9 & 79.2 \\
& \texttt{rms} & Set-F1 & 40.7 & 41.4 & 60.1 & 59.6 & 0.0 & 21.5 & 5.0 & 3.4 & 3.5 & 10.9 & 2.5 & 2.3 & 4.1 & 14.9 & 0.1 & 8.4 & 0.1 & 24.8 & 15.6 & 23.7 \\
& \texttt{taa} & Accuracy & 33.0 & 30.0 & 47.0 & 42.0 & 7.0 & 22.0 & 14.0 & 16.0 & 0.0 & 18.0 & 5.0 & 11.0 & 12.0 & 19.0 & 8.0 & 19.0 & 25.0 & 21.0 & 21.0 & 19.0 \\
& \texttt{ate} & Accuracy & 68.6 & 66.4 & 82.2 & 79.2 & 2.4 & 49.4 & 40.6 & 49.6 & 27.0 & 29.4 & 24.2 & 26.8 & 39.0 & 53.2 & 27.2 & 33.6 & 0.0 & 38.0 & 51.8 & 51.0 \\
& \texttt{rcm} & Accuracy & 71.0 & 71.5 & 73.7 & 73.6 & 1.9 & 64.8 & 59.9 & 65.7 & 51.0 & 61.0 & 57.3 & 58.2 & 21.2 & 57.8 & 38.8 & 55.7 & 9.2 & 60.5 & 68.4 & 68.3 \\
& \texttt{vsp} & MAD-norm & 85.8 & 85.7 & 88.6 & 88.7 & 85.4 & 87.8 & 83.1 & 58.9 & 74.4 & 71.6 & 78.8 & 75.7 & 73.6 & 74.6 & 53.3 & 72.3 & 70.2 & 65.7 & 73.5 & 72.0 \\
\midrule
\textbf{CyberMetric} & \texttt{det} & Accuracy & 96.0 & 96.0 & 96.8 & 96.2 & 4.0 & 95.2 & 92.2 & 95.6 & 91.6 & 93.0 & 86.4 & 92.0 & 19.2 & 93.4 & 17.2 & 86.0 & 1.0 & 85.2 & 89.8 & 90.2 \\
& \texttt{samp} & Accuracy & 96.2 & 96.0 & 96.6 & 96.2 & 4.8 & 95.2 & 91.0 & 95.6 & 89.6 & 93.0 & 83.4 & 92.0 & 30.6 & 93.4 & 57.2 & 86.0 & 2.8 & 85.2 & 86.8 & 90.2 \\
\midrule
\textbf{RedSage-Bench} & \texttt{cli} & Accuracy & 93.0 & 92.7 & 93.7 & 93.6 & 43.5 & 88.8 & 88.8 & 90.1 & 87.3 & 86.9 & 24.5 & 89.7 & 87.9 & 87.1 & 78.5 & 75.9 & 78.3 & 75.2 & 86.7 & 86.6 \\
& \texttt{frameworks} & Accuracy & 87.9 & 88.1 & 90.0 & 90.2 & 46.4 & 86.2 & 83.7 & 84.6 & 83.2 & 83.0 & 24.6 & 81.1 & 84.8 & 84.6 & 78.9 & 73.4 & 79.6 & 77.4 & 85.2 & 84.4 \\
& \texttt{generals} & Accuracy & 90.6 & 90.2 & 90.4 & 90.7 & 47.1 & 87.1 & 85.4 & 86.7 & 85.8 & 85.8 & 26.5 & 82.1 & 86.6 & 86.8 & 78.6 & 74.9 & 77.7 & 74.5 & 84.1 & 83.7 \\
& \texttt{kali} & Accuracy & 86.3 & 86.4 & 87.3 & 87.2 & 39.6 & 81.2 & 81.4 & 83.3 & 79.5 & 80.3 & 25.3 & 80.5 & 80.2 & 80.3 & 74.0 & 70.4 & 71.9 & 68.7 & 80.8 & 80.5 \\
& \texttt{skills} & Accuracy & 93.2 & 93.2 & 93.3 & 93.6 & 51.8 & 91.4 & 90.1 & 91.6 & 89.7 & 89.4 & 26.4 & 90.2 & 90.1 & 89.5 & 84.9 & 80.7 & 83.5 & 81.1 & 89.5 & 88.8 \\
\midrule
\textbf{SecBench} & \texttt{secbench} & Accuracy & 84.6 & 87.7 & 85.2 & 89.7 & 33.9 & 86.7 & 62.3 & 89.1 & 71.4 & 82.1 & 41.1 & 81.6 & 69.6 & 84.8 & 38.1 & 66.3 & 67.0 & 70.9 & 44.3 & 80.2 \\
\bottomrule
\end{tabularx}
\caption{Per-task scores before and after pipeline standardization for all 10 models. Gray columns show standardized scores. The Metric column identifies the reported quantity: accuracy (\%), question-level Set-F1 (\%), mean absolute CVSS-score deviation (MAD, lower is better), or normalized MAD (\%). CTI-Bench \texttt{taa} uses Correct+Plausible accuracy before standardization and strict binary accuracy after standardization. CyberMetric \texttt{det}/\texttt{samp} and MMLU-CS \texttt{gen}/\texttt{logp} are alternative configurations of one task each. \textsuperscript{*}~hosted/API model; \ensuremath{^\dagger}~denominator artifact.}
\label{tab:before_after}
\end{sidewaystable*}

\subsection{Tie Breaking}
\label{app:k3}

When two models have identical standardized scores on a benchmark, we break the tie by stable sorting in descending score, preserving a fixed model order. The same rule is applied to original and standardized rankings, making every rank difference in Table~\ref{tab:rank_shifts} deterministic. Exact ties are rare at the reported precision, so this rule affects only tied cells and does not change the reported count of large rank shifts.

\subsection{Bootstrap Stability}
\label{app:k4}

The reported standardized scores are point estimates from one stored evaluation run. We quantify sampling variability with a paired question-level bootstrap that requires no new generation. Within each task, we resample stored question outcomes with replacement for 5{,}000 replicates, applying the same resample to every model so that comparisons retain a shared question basis. For each replicate, we recompute the benchmark score, re-rank the models, and calculate Spearman's $\rho$ against the fixed original ranking.

The observed samples reproduce the $\rho$ values in Table~\ref{tab:rank_shifts} to two decimal places. Every 95\% bootstrap interval in Table~\ref{tab:rank-shift-ci} excludes $1$, indicating that the observed reorderings are not explained by question sampling alone. MMLU-CS remains negatively correlated with its original ranking, while SecEval's interval includes zero, indicating an almost complete reshuffling in both cases.

The individual large shifts are also stable. Each of the 35 benchmark-level movements of at least three positions retains its direction in at least 97.7\% of bootstrap replicates; 33 do so in at least 98\%, and 24 retain their direction in all 5{,}000 replicates. The two least stable cases are three-position drops on MMLU-CS. Its larger shifts are substantially more stable: Gemma-4 moves $+6$ positions with a bootstrap interval of $[+3,+7]$, Primus-Nemotron moves $-7$ with $[-9,-5]$, and Foundation-Sec moves $-6$ with $[-7,-3]$.

Score uncertainty is smaller than the resulting ranking changes but still limits fine-grained comparisons. Bootstrap half-widths range from approximately $\pm0.4$\,pp on RedSage-Bench to $\pm7.5$\,pp on MMLU-CS, and adjacent models often have overlapping intervals. On SecBench, for example, Sonnet 4.6 scores 89.7 $[87.4,92.0]$, Qwen3.6 scores 89.1 $[86.7,91.4]$, and GPT-5.4 scores 87.7 $[85.3,90.2]$. Thus, the effect of pipeline standardization on the rankings is robust to resampling even when the precise ordering of nearby models is not statistically resolvable.

\begin{table}[t]
\centering
\scriptsize
\setlength{\tabcolsep}{6pt}
\renewcommand{\arraystretch}{1.10}
\begin{tabular}{@{}lrrr@{}}
\toprule
& \multicolumn{3}{c}{\textbf{Spearman $\boldsymbol{\rho}$}} \\
\cmidrule(lr){2-4}
\textbf{Benchmark} & Lower & Estimate & Upper \\
\midrule
MMLU-CS       & -0.71 & -0.47 & -0.25 \\
SecEval       & -0.08 & +0.03 & +0.15 \\
SECURE        & +0.56 & +0.65 & +0.72 \\
CTI-Bench     & +0.56 & +0.64 & +0.76 \\
AthenaBench   & +0.39 & +0.42 & +0.47 \\
CyberMetric   & +0.47 & +0.73 & +0.81 \\
RedSage-Bench & +0.60 & +0.71 & +0.72 \\
SecBench      & +0.32 & +0.50 & +0.58 \\
\bottomrule
\end{tabular}
\caption{Spearman correlation between original and standardized model rankings with question-level bootstrap uncertainty. \emph{Estimate} is the observed correlation; \emph{Lower} and \emph{Upper} are the endpoints of the 95\% confidence interval from 5{,}000 paired bootstrap resamples. Every interval excludes $1$; only SecEval includes $0$.}
\label{tab:rank-shift-ci}
\end{table}

\subsection{Generation vs.\ Extraction}
\label{app:k5}

The original-to-standardized comparison changes both how responses are generated and how they are interpreted. We extend the pipeline notation of \S\ref{sec:pipeline} to separate these effects. Let
\begin{equation}
\mathcal{G}_b^{x}(m)=\mathcal{I}_b^{x}\!\left(m,\mathcal{P}_b^{x}(\mathcal{D}_b)\right), x\in\{o,s\},
\label{eq:decomp-generation}
\end{equation}
denote the generations produced under the original ($o$) or standardized ($s$) prompt and inference configuration. We further distinguish three extraction-and-scoring rules: the original benchmark rule $\mathcal{E}_b^{o}$, a deterministic final-answer regex $\mathcal{E}_b^{r}$, and the pinned judge $\mathcal{E}_b^{j}$. The resulting pipeline score is
\begin{equation}
\mathcal{S}_b^{x,y}(m)=\mathcal{A}_b^{y}\!\left(\mathcal{E}_b^{y}\!\left(\mathcal{G}_b^{x}(m)\right)\right), y\in\{o,r,j\},
\label{eq:decomp-score}
\end{equation}
where $\mathcal{A}_b^{y}$ denotes the aggregation associated with that evaluation rule. The four comparison cases are: $\mathcal{S}_b^{o,o}$, the original benchmark score; $\mathcal{S}_b^{o,r}$, regex evaluation of the original generations; $\mathcal{S}_b^{s,r}$, the same regex evaluation of the standardized generations; and $\mathcal{S}_b^{s,j}$, judge evaluation of the standardized generations.

For benchmark $b$, let $\mathbf{S}_b^{x,y}$ denote the vector of $\mathcal{S}_b^{x,y}(m)$ over the 10 models. We measure the generation contribution by
\begin{equation}
\rho_{\mathrm{gen}}=\rho\!\left(\mathbf{S}_b^{o,r},\mathbf{S}_b^{s,r}\right),
\end{equation}
which changes $\mathcal{P}_b$ and $\mathcal{I}_b$ while holding the regex evaluation fixed. We measure the extraction contribution by
\begin{equation}
\rho_{\mathrm{ext}}=\rho\!\left(\mathbf{S}_b^{s,r},\mathbf{S}_b^{s,j}\right),
\end{equation}
which holds standardized generations fixed while changing their interpretation. Finally,
\begin{equation}
\rho_{\mathrm{orig}}=\rho\!\left(\mathbf{S}_b^{o,r},\mathbf{S}_b^{o,o}\right)
\end{equation}
checks how closely the deterministic regex reconstruction preserves the original benchmark ranking. Lower $\rho$ indicates greater ranking change.

The comparison includes tasks that admit deterministic extraction of single- or multi-letter answers, true-or-false responses, CWE and ATT\&CK identifiers, or CVSS vectors. The two attacker-attribution tasks are excluded because semantic alias handling cannot be represented reliably by a generic regex. The reconstruction check is strong on seven benchmarks, where $\rho_{\mathrm{orig}}$ ranges from 0.78 to 0.95. CTI-Bench is lower at 0.53 because its CWE and CVSS extraction rules diverge more substantially from the generic regex.

Table~\ref{tab:decomp} shows that generation changes produce more ranking reordering than extraction on five of the eight benchmarks: SecEval, CTI-Bench, AthenaBench, RedSage-Bench, and SecBench. The contrast is strongest on AthenaBench, where $\rho_{\mathrm{gen}}=0.34$ and $\rho_{\mathrm{ext}}=0.99$, and CTI-Bench, where they are 0.59 and 0.89. Extraction contributes more on SECURE and CyberMetric, although both CyberMetric correlations remain high. MMLU-CS is sensitive to both, with $\rho_{\mathrm{gen}}=0.25$ and $\rho_{\mathrm{ext}}=0.20$. Thus, the rank shifts cannot be attributed simply to replacing benchmark extractors with an LLM judge: substantial reordering is already present when extraction is held fixed. The extraction effect is also reproduced by Sonnet 4.6 (App.~\ref{app:a-judgeindep}). CTI-Bench differs from the full comparison in Table~\ref{tab:rank_shifts} as its attacker-attribution task is excluded here.

\begin{table}[t]
\centering
\scriptsize
\setlength{\tabcolsep}{5pt}
\renewcommand{\arraystretch}{1.10}
\begin{tabular}{@{}lccc@{}}
\toprule
\textbf{Benchmark} & $\boldsymbol{\rho_{\mathrm{gen}}}$ & $\boldsymbol{\rho_{\mathrm{ext}}}$ & $\boldsymbol{\rho_{\mathrm{orig}}}$ \\
\midrule
MMLU-CS       & +0.25 & +0.20 & +0.91 \\
SecEval       & -0.19 & +0.27 & +0.82 \\
SECURE        & +0.80 & +0.58 & +0.78 \\
CTI-Bench     & +0.59 & +0.89 & +0.53 \\
AthenaBench   & +0.34 & +0.99 & +0.91 \\
CyberMetric   & +0.93 & +0.82 & +0.94 \\
RedSage-Bench & +0.55 & +0.72 & +0.84 \\
SecBench      & +0.38 & +0.61 & +0.95 \\
\bottomrule
\end{tabular}
\caption{Decomposition of the original-to-standardized rank shift. $\rho_{\mathrm{gen}}=\rho(\mathbf{S}_b^{o,r},\mathbf{S}_b^{s,r})$ holds extraction fixed while changing generation; $\rho_{\mathrm{ext}}=\rho(\mathbf{S}_b^{s,r},\mathbf{S}_b^{s,j})$ holds standardized generations fixed while changing extraction; and $\rho_{\mathrm{orig}}=\rho(\mathbf{S}_b^{o,r},\mathbf{S}_b^{o,o})$ checks the regex reconstruction against the original ranking. Lower $\rho$ indicates greater ranking change. All correlations use stored original and standardized generations; this decomposition requires no additional inference.}
\label{tab:decomp}
\end{table}

\section{Recommendation Details}
\label{app:recs}

This section expands on \S\ref{sec:recommendations} by outlining automation limits and the information benchmark releases should report for reliable, reproducible evaluation.

\subsection{Automation Scope}
\label{app:m}

The harness automates configuration logging, model-output collection, response validation, parsing diagnostics, deterministic re-scoring, and aggregate comparisons. It also makes controlled pipeline changes reproducible when the relevant configuration is exposed. Automation does not, however, determine benchmark intent or establish semantic ground truth. Decisions such as whether related threat actors deserve partial credit, which aliases should be equivalent, which capabilities a benchmark should cover, or whether a published label is correct require externally justified policies or evidence. The gold-label audit therefore uses an automated search-grounded verifier followed by human validation (App.~\ref{app:e2}) rather than treating automated agreement as authoritative ground truth.

\subsection{Evaluation Pipeline Card}
\label{app:card}

We recommend that benchmark releases provide an \emph{evaluation pipeline card} for each scored task rather than only one specification for the benchmark as a whole. As shown in this paper, different tasks within the same benchmark can use different prompts, inference settings, extraction procedures, metrics, denominator policies, and aggregation rules, so each reported task score should resolve to the exact pipeline that produced it.

For practical reuse, the card should be provided both as human-readable documentation and as a versioned machine-readable record, such as JSON conforming to a public schema. This would allow evaluation repositories and reporting systems to compare results only when their task and pipeline specifications are compatible. This recommendation complements Evaluation Cards~\citep{ghosh2026evaluation}, which replaces flat model--benchmark--score reporting with structured evaluation records that resolve results to their underlying benchmark, split, and metric configuration.

For each task, the pipeline card should report:

\begin{itemize}[leftmargin=1.3em,topsep=2pt,parsep=1pt,itemsep=1pt]
\item \textbf{Identity:} benchmark, task, split, metric, pipeline version, and stable identifiers needed to distinguish the evaluation from related variants.
\item \textbf{Dataset:} scored question population, question and label sources, label provenance, inclusion or exclusion rules, and the date or snapshot of evolving references such as CVE and ATT\&CK.
\item \textbf{Prompt:} exact task prompt, system prompt, demonstrations, chat-formatting policy, and answer-format instructions.
\item \textbf{Inference:} decoding parameters, output-token budget, stop sequences, seed policy, serving behavior, and relevant backend constraints.
\item \textbf{Extraction:} the executable extraction procedure or judge, including its prompt and decoding configuration when LLM-based.
\item \textbf{Scoring:} metrics, credit rule, alias and invalid-response handling, and denominator policy.
\item \textbf{Aggregation:} how question-level measurements are combined into the reported task score.
\item \textbf{Reliability:} known invalid-response, extraction, scoring, or configuration sensitivities and the diagnostics used to detect them.
\item \textbf{Reproducibility:} what to record in each evaluation run and which analyses require re-scoring or re-generation.
\end{itemize}

Model identity, model-specific serving details, timestamps, and numerical results belong to the corresponding evaluation-run record rather than the task-level pipeline card. A result can be interpreted as a run-specific score linked to a versioned task-level pipeline specification. \pipelinecardref{card:cti-vsp} shows this distinction using only the CTI-Bench VSP task.

\begin{pipelinecard}[label={card:cti-vsp}]{CTI-Bench / VSP / CVSS MAD}
\footnotesize

\noindent\textbf{Identity.}
Benchmark: CTI-Bench. Task: Vulnerability Severity Prediction (VSP). Scored split: 1{,}000 released VSP questions. Reported metric: CVSS mean absolute deviation (MAD), lower is better.

\medskip
\noindent\textbf{Dataset.}
Each question provides a CVE description and asks for a CVSS v3.1 vector. All 1{,}000 released VSP questions are scored. The benchmark does not specify a snapshot date for the underlying CVE/CVSS references.

\medskip
\noindent\textbf{Prompt.}
The released CTI-Bench task prompt is preserved. The system instruction is \emph{``You are a cybersecurity expert specializing in cyberthreat intelligence.''} The task prompt asks for analysis of the supplied CVE description and requires the final CVSS v3.1 vector string. The standardized pipeline uses the task system prompt and the evaluated system's native chat formatting.

\medskip
\noindent\textbf{Inference.}
The task uses a maximum output budget of 2{,}048 tokens. The standardized pipeline uses greedy decoding with temperature $0$, top-$p=1.0$, and no backend stop sequence. No task-specific sampling seed is required under deterministic decoding.

\medskip
\noindent\textbf{Extraction.}
The released pipeline returns the last matching \texttt{CVSS:3.1/} vector found anywhere in the response, although the prompt requests the final vector in the answer. The standardized pipeline uses the pinned semantic extraction policy and expects a normalized \texttt{CVSS:3.1/} vector. Reasoning enclosed in \texttt{<think>...</think>} is removed before extraction, and the extractor must not repair or infer an unstated answer.

\medskip
\noindent\textbf{Scoring.}
The extracted vector is scored using mean absolute deviation between the predicted and gold CVSS values. Lower MAD indicates better performance. An empty or unparseable response remains part of the attempted-question population and is not discarded.

\medskip
\noindent\textbf{Aggregation.}
Question-level CVSS deviations are aggregated over the 1{,}000 scored questions into one task-level MAD value. No cross-task aggregation is part of this task specification.

\medskip
\noindent\textbf{Reliability.}
The extraction rule is a known sensitivity point. Applying final-line and anywhere CVSS extraction to identical stored responses yields 30.4\% disagreement among cases for which at least one rule extracts a vector. Invalid and unparseable responses should therefore be reported explicitly alongside the task score.

\medskip
\noindent\textbf{Reproducibility.}
A run using this task should record the pipeline version, exact prompt configuration, decoding parameters, token budget, stop-sequence configuration, extraction rule, scoring rule, denominator policy, and raw response. Changes to extraction, scoring, or denominator policy can then be evaluated by re-scoring stored responses; changes to prompting or inference require new generations.

\end{pipelinecard}

In practice, each task card should have an equivalent machine-readable representation with stable field names and versioned identifiers, allowing a reported result to link unambiguously to the task, metric, and pipeline configuration that produced it.

The same task-level specification of~\pipelinecardref{card:cti-vsp} can be represented in a machine-readable form, as illustrated in \coderef{code:pipeline-card-json}.

\begin{codebox}[label={code:pipeline-card-json}]{Machine-readable version of~\pipelinecardrefwhite{card:cti-vsp}}
\begin{lstlisting}[style=sayf,aboveskip=0pt,belowskip=0pt]
{
  "identity": {
    "benchmark": "CTI-Bench",
    "task": "vsp",
    "pipeline_version": "standardized"
  },
  "dataset": {
    "questions": 1000,
    "input": "CVE description",
    "target": "CVSS v3.1 vector",
    "reference_snapshot": null
  },
  "prompt": {
    "system": "cyberthreat-intelligence expert",
    "chat_format": "native",
    "answer_format": "CVSS:3.1 vector"
  },
  "inference": {
    "temperature": 0,
    "top_p": 1.0,
    "max_tokens": 2048,
    "stop": null
  },
  "extraction": {
    "method": "pinned semantic extractor",
    "output": "normalized CVSS:3.1 vector"
  },
  "scoring": {
    "metric": "CVSS MAD",
    "direction": "lower",
    "invalid_response": "retained"
  },
  "aggregation": {
    "level": "task",
    "questions": 1000
  },
  "reliability": {
    "extractor_disagreement_contested_pct": 30.4
  },
  "reproducibility": {
    "store_raw_outputs": true,
    "rescoring_supported": true,
    "prompt_or_inference_change_requires_regeneration": true
  }
}
\end{lstlisting}
\end{codebox}

\end{document}